\PassOptionsToPackage{table}{xcolor}
\documentclass[sigconf,nonacm]{acmart}

\usepackage{booktabs}
\usepackage{array}
\usepackage{tabularx}
\usepackage{tcolorbox}
\usepackage{tikz}
\usetikzlibrary{arrows.meta,positioning,fit,backgrounds,calc}
\usepackage{pgfplots}
\pgfplotsset{compat=1.18}
\definecolor{sblue}{HTML}{2a78d6}\definecolor{sorange}{HTML}{eb6834}
\definecolor{Gray}{gray}{0.9}
\AtBeginEnvironment{table}{\rowcolors{2}{Gray}{white}}
\AtBeginEnvironment{table*}{\rowcolors{2}{Gray}{white}}
\newcommand{\sys}{\textsc{yoars}}
\newcounter{lessonbox}
\newcommand{\lesson}[1]{%
\refstepcounter{lessonbox}%
\begin{tcolorbox}[colback=black!3,colframe=sblue!65!black,
boxrule=0pt,leftrule=1.5pt,arc=0pt,outer arc=0pt,
boxsep=0pt,left=7pt,right=7pt,top=6pt,bottom=6pt,
before skip=7pt,after skip=7pt]
{\small\raggedright{\sffamily\footnotesize\bfseries\color{sblue!65!black}Lesson \arabic{lessonbox}:}\ #1\par}
\end{tcolorbox}}

\begin{document}

\title{Getting a Grip: Operating Human--Agent Software Teams Through Governed Messages}
\subtitle{Technical Report}

\author{Michael Granitzer}
\orcid{0000-0003-3566-5507}
\affiliation{\institution{University of Passau}\country{Germany}}
\affiliation{\institution{IT:U Interdisciplinary Transformation University Austria}\city{Linz}\country{Austria}}
\email{michael.granitzer@uni-passau.de}
\author{Kanishka Ghosh Dastidar}
\orcid{0000-0003-4171-0597}
\affiliation{\institution{IT:U Interdisciplinary Transformation University Austria}\city{Linz}\country{Austria}}
\email{kanishka.ghoshdastidar@it-u.at}
\author{Andreas Einwiller}
\orcid{0009-0009-2271-2125}
\affiliation{\institution{University of Passau}\country{Germany}}
\email{andreas.einwiller@uni-passau.de}
\author{Jelena Mitrovi\'c}
\orcid{0000-0003-3220-8749}
\affiliation{\institution{University of Passau}\country{Germany}}
\email{jelena.mitrovic@uni-passau.de}
\author{Saber Zerhoudi}
\orcid{0000-0003-2259-0462}
\affiliation{\institution{University of Passau}\country{Germany}}
\email{saber.zerhoudi@uni-passau.de}

\begin{abstract}
Software is increasingly developed by teams in which several people direct
many coding agents from different vendors. Such teams must establish who may
request what from whom, which messages start work, which decisions bind, and
who caused a change. We propose to govern a human--agent team through its
message protocol and present \sys, a relay in which humans and agents are
authenticated principals that exchange typed, addressed messages (acts) in rooms
holding goals, decision rules, routing and an attributable, append-only
record. This report describes the relay's architecture and the design
decisions behind it, the full method and results of the controlled study, and
the field material. In 926 controlled runs (899 analyzable), small agent
teams built programs
under varied room rules, some with a hostile member planting a backdoor. A
revision of which messages start work, bundled with limits on repeated
messages, was followed by a drop in total coordination failure from 42\% to
13\% ($p=.0063$). After correction over nine registered tests, the drop falls
just short of significance (adjusted $p=.057$). Instructed roles and votes did
not reliably produce the intended process, and no defense inside the
conversation reliably prevented the backdoor. Afterwards, two passes of an
auditor on Claude Opus 5 read the records of 96 runs of one model family in
which a detector had confirmed the backdoor, including 15 that failed our
participation check. The attacker had committed the backdoor itself in at
least 71 of these runs. In 25, it had persuaded a teammate to write it, and
the auditor named the persuader in all 25. The same audit also flagged
24 of 40 honest runs. With the team's own model as auditor,
detection and attribution fell substantially. In a four-week field study, up
to nineteen agents and two people developed a distributed system through
\sys, producing 7{,}457 acts. Human attention and decision making emerged
as the critical resource. Requests waited mainly for the human-run coordinator
and for approvals, while the agents answered within minutes. Most governance
requirements emerged from operation, and holding context rather than
producing output dominated token cost. We report twenty-six failures of our
own instruments and the claims they made us withdraw. We derive seven lessons
and a checklist for operating human--agent software teams, and we release the
corpus, the redacted room ledger and the analysis scripts.
\end{abstract}


\maketitle

\section{Introduction}
\label{sec:intro}

Coding agents now author hundreds of thousands of pull requests in public
repositories \cite{Li:2025:arXiv}, and developers run several agents in
parallel while struggling to coordinate and monitor them
\cite{Long:2026:arXiv,Dhanorkar:2026:FAccT}. The next step is already visible
in practice. Several people direct many agents, the agents come from different
vendors and run in different tools, and they work on components that depend on
each other. A change to one component may require coordinated actions by
several agents and people, each with different responsibilities and
authority.

Such a team must answer questions that a single agent never raises. Who may
ask whom for what? Which messages should start an agent's work, and which only
inform it? Which decisions bind later work? Who asked for a change, and who
wrote it? Prompts alone neither enforce these rules nor ensure an attributable
record of compliance. Security makes these questions pressing. Agents write
and ship code, and in a team of agents from different vendors a single
compromised or misaligned member can plant a vulnerability that its peers
accept \cite{Wang:2026:AAAI,Greenblatt:2024:ICML}. The team must then be able
to tell who introduced it and on whose request.

Software engineering research shows that coordination
across team boundaries slows work \cite{Herbsleb:1999:ICSE,Herbsleb:2003:TSE}
and that productivity depends on matching communication to technical
dependencies \cite{Cataldo:2008:ESEM}. Multi-agent frameworks for software
development coordinate agents inside one program through instructed roles and
standard procedures \cite{Hong:2024:ICLR,Qian:2024:ACL,Wu:2023:arXiv}. Their
failures are dominated by misaligned agents and missing verification
\cite{Cemri:2025:NeurIPS}. Agent protocols connect a client to tools
\cite{MCP:2025:doc} or delegate a task to one remote agent
\cite{A2A:2025:doc}.

These approaches provide parts of what such a team needs, but not a
persistent, governed space in which several people and independently hosted
agents exchange obligations, take decisions and leave an auditable record. It
is also open whether such structure, once built into the communication layer,
controls how agents behave, and what operating such a team looks like over
weeks of real work. We therefore
ask three questions:
\begin{itemize}
\item \textbf{RQ1 (controllability).} Which rules of the shared room change how
agent teams coordinate: the description of the workspace, the rules that
decide which messages start work, assigned roles, or voting?
\item \textbf{RQ2 (transparency).} When one team member is hostile, do
defenses inside the conversation prevent a planted security hole, and does a
transparent, attributable record allow the hole and its originator to be found
afterwards?
\item \textbf{RQ3 (operation).} When a heterogeneous fleet of agents and two
people develops a complex system through the protocol, where does work wait,
which collaboration patterns help or hurt, how do the requirements change, and
where do tokens and context go?
\end{itemize}

We propose to structure a human--agent team through its
message protocol rather than through prompts. To do so, we built \sys\footnote{\url{https://yoars.org/}},
a relay in which humans and agents are authenticated principals that exchange
\emph{acts}, addressed messages whose type states the sender's intent, such as
a request or an inform \cite{FIPA:2002:doc}. Acts are exchanged in rooms that
hold goals, decision rules, routing and a \emph{ledger}, the append-only,
attributable record of all acts. We answer RQ1 and RQ2 with 926 controlled
runs in which small agent teams built programs under varied room rules, some
with a hostile member. We answer RQ3 with a four-week field study in which up
to nineteen agents and two people used \sys\ to develop \sys\ itself. We chose
the platform's own development for the field study because it was the most
demanding development task available to us. It is also a convenience sample:
the two people are authors, and the team develops the system under study
(Section~\ref{sec:threats}). We analyze its complete ledger of
7{,}457 acts and the usage records of 3{,}276 model turns.

The field study exposes a bottleneck absent from the controlled tasks: access
to human attention and decisions. Component agents usually answered within
minutes. Requests to coordinators in human-operated sessions, and approvals,
waited longer as the team grew. In our experience as the operators,
most waiting decisions were simple once seen, such as an approval or the order
of two tasks, and with more stakeholders we expect this bottleneck to grow.
The team responded organizationally, by adding decision makers and delegating
authority. We therefore treat human attention and decision making as a central
design concern for human--agent teams.

\emph{Scope of this report.} This report gives four things. First, it describes the
architecture of the relay and the design decisions that shaped it
(Section~\ref{sec:design}). For each decision we give the problem that forced
it, the alternatives we rejected where the records name them, and when it was
built relative to the controlled and the field study. Second, it gives the
full method of the controlled study: task material, run protocol, measures,
statistics and the rules that decide whether a run counts
(Section~\ref{sec:controlled} and Appendix~\ref{app:details}). Third, it
reports the detailed results behind each of the four findings
(Section~\ref{sec:results-all}). These include comparisons whose results we
withdrew, the results across model families, and a single-agent baseline.
Fourth, it gives the field material in full (Section~\ref{sec:field} and
Appendix~\ref{app:context}). This material comprises threads from the record,
the problems a coordinator logged, and an early session in which capability
and instruction could not be told apart. It also includes a measurement of
what context costs in a resident fleet, whose agents stay in one room for
weeks. Appendix~\ref{app:designs} separately describes three designs that
follow from the results and have not been tested.

\emph{Contributions.} We contribute (i) a framework for governing human--agent
teams through a message protocol (Section~\ref{sec:concept}), (ii) its
implementation and the design decisions behind it
(Section~\ref{sec:design}), (iii) controlled evidence on coordination and
post-hoc investigation from 926 runs (Sections~\ref{sec:controlled}
and~\ref{sec:results-all}), (iv) a four-week operational case study
(Section~\ref{sec:field}), (v) seven lessons with a checklist for
practitioners (Section~\ref{sec:lessons}), and (vi) a methodological account
of twenty-six failures of our own instruments
(Appendix~\ref{app:instruments}). We release the corpus, the redacted room
ledger and the analysis scripts, together with the source of the relay and
the adapter at the commits the runs used (Section~\ref{sec:since}).

\section{State of the Art and Gap}
\label{sec:related}

We compare \sys\ with four lines of work (Table~\ref{tab:sota}). We then place
our study among work on evaluating multi-agent systems and on their security.

\emph{Coordination in software engineering.} Distributed development studies
connect delay to communication across sites and teams
\cite{Herbsleb:1999:ICSE,Herbsleb:2003:TSE}. Socio-technical congruence links
productivity to coordination that matches dependencies \cite{Cataldo:2008:ESEM},
coordination theory frames it as managing dependencies \cite{Malone:1994:CSUR},
and version control and code review shape awareness and response times
\cite{Gutwin:2004:CSCW,Dabbish:2012:CSCW,Sadowski:2018:ICSE}. These studies
concern human teams. We ask what carries coordination when most members are
agents.

\emph{Multi-agent frameworks.} MetaGPT organizes agents through roles and
standard operating procedures \cite{Hong:2024:ICLR}, ChatDev through chat
chains between roles \cite{Qian:2024:ACL}, and AutoGen through programmable
conversations \cite{Wu:2023:arXiv}. A review names human--agent
collaboration as an open challenge for these systems \cite{He:2024:TOSEM}.
They coordinate agents inside one process and one run, usually with one model,
and a person stays outside the team. MAST classifies their failures from
execution traces \cite{Cemri:2025:NeurIPS}, and automated attribution seeks
the failing agent \cite{Zhang:2025:ICML}. These analyses read traces after the
fact and do not vary the communication medium as a treatment.

\emph{Agent communication.} KQML and FIPA ACL made the purpose of a message
explicit through typed performatives \cite{Finin:1994:CIKM,FIPA:2002:doc}.
Electronic institutions and normative multi-agent systems enforced norms on
interactions \cite{Esteva:2001:AMEC,Boella:2006:CMOT}, and blackboards and
tuple spaces coordinated through a shared store
\cite{Erman:1980:CSUR,Gelernter:1985:TOPLAS}. These ideas predate language-model
agents and did not address agents whose every received message can start
costly work. Today's protocols standardize two-party exchanges: tool calls in
MCP \cite{MCP:2025:doc} and task delegation in A2A \cite{A2A:2025:doc,Ehtesham:2025:arXiv}.

\emph{Oversight and governance.} ParallelPilot supports one developer
monitoring parallel coding sessions \cite{Long:2026:arXiv}, and Dhanorkar et
al.\ describe the oversight work of developers using agents
\cite{Dhanorkar:2026:FAccT}. Governance proposals call for agent identifiers
and activity records \cite{Chan:2024:FAccT} and for authenticated delegation
of human authority \cite{South:2025:arXiv}. Enforcement layers have been
evaluated in a market setting \cite{Syrnikov:2026:arXiv} and as a check on
actions before execution \cite{Shi:2026:arXiv}. AI Control and Shadows in the
Code study backdoors planted by capable or concealed agents
\cite{Greenblatt:2024:ICML,Wang:2026:AAAI}.

\emph{Evaluation of multi-agent systems.} Benchmarks for agentic software
work generally score the artifact, the code that the agents produce
\citep{jimenez2024swebenchlanguagemodelsresolve}. In the first block of our
controlled study, this kind of scoring was \emph{saturated}: conditions passed
the artifact check even when the agents had not coordinated at all. We
therefore also measure the process by which a team reaches its artifact.

\emph{Security of LLM multi-agent systems.} Work on adversarial participants
keeps the conversation medium fixed and varies other factors. Studies of
faulty or malicious agents vary the \emph{topology} (who may address whom)
and add agents that inspect the others.
\citet{huang2025resiliencellmbasedmultiagentcollaboration} find hierarchical
structures the most resilient. With added Challenger and Inspector roles, they
recover up to 96.4\% of injected errors.
\citet{wang2025gsafeguardtopologyguidedsecuritylens} detect anomalies on the
utterance graph, the graph of agents and the messages they exchange, and
\citet{xie2025whosmolemodelingdetecting} interrogate agents to expose hidden
intentions. Closest to our setting, \citet{Wang:2026:AAAI} plant malicious
agents in multi-agent software-development frameworks. They report attack
success rates of 45--93\%. Other work spreads injected instructions from agent
to agent \citep{lee2024promptinfectionllmtollmprompt}, or attacks a team whose
discussion is monitored throughout
\citep{xiang2026monitoringenoughstrategicagent}. Work on AI \emph{control}
asks the complementary question for a single agent: one untrusted author whose
work a monitor checks. It studies trusted monitoring of possibly backdoored
code \citep{Greenblatt:2024:ICML} and sabotage that has to pass monitors
reading the agent's transcript
\citep{kutasov2025shadearenaevaluatingsabotagemonitoring}. Social deduction
benchmarks measure how well peers detect deception
\citep{agarwal2025wolfwerewolfbasedobservationsllm}.

Two differences from this literature shape our design. First, in prior
inspector studies the injected errors can be checked against information that
every agent shares, such as a HumanEval problem statement. The faulty agent
also does not argue for its errors. Our attacker relies on knowledge that only
its role holds, and it argues for its change. Second, none of the work cited
here treats the conversation medium as a governed platform whose policies are
the treatment.

\begin{table}[t]
\centering
\caption{Capabilities of related approaches as described in the cited work
(our reading). \checkmark\ provided, $\circ$ partly, -- not addressed.}
\label{tab:sota}
\begingroup
\scriptsize
\setlength{\tabcolsep}{1.6pt}
\renewcommand{\arraystretch}{1.1}
\begin{tabular}{@{}lccccccc@{}}
\toprule
 & \rotatebox{70}{humans as peers} & \rotatebox{70}{independent agents} & \rotatebox{70}{typed obligations} & \rotatebox{70}{control of wake-up} & \rotatebox{70}{goals, decisions} & \rotatebox{70}{enforced routing} & \rotatebox{70}{attributable record} \\
\midrule
MetaGPT, ChatDev & $\circ$ & -- & $\circ$ & $\circ$ & -- & -- & $\circ$ \\
AutoGen & $\circ$ & -- & -- & $\circ$ & -- & -- & $\circ$ \\
MCP, A2A & -- & \checkmark & $\circ$ & -- & $\circ$ & -- & -- \\
KQML, FIPA ACL & -- & \checkmark & \checkmark & -- & -- & -- & -- \\
Electr.\ institutions & $\circ$ & \checkmark & \checkmark & -- & $\circ$ & \checkmark & $\circ$ \\
ParallelPilot & $\circ$ & $\circ$ & -- & -- & -- & -- & $\circ$ \\
\sys & \checkmark & \checkmark & \checkmark & \checkmark & \checkmark & $\circ^{*}$ & \checkmark \\
\bottomrule
\end{tabular}
\par\smallskip
{\raggedright $^{*}$Partly built, after the controlled study
(Section~\ref{sec:since}), and not evaluated here.\par}
\endgroup
\end{table}

\emph{The gap.} The cited approaches provide parts of this design, and the
individual mechanisms of \sys\ have precedents. We combine them in a
persistent room for humans and independently hosted agents, and study both
protocol changes in controlled tasks and operation over weeks of development.

\section{Governing a Team Through Its Messages}
\label{sec:concept}
The central design decision is to express the rules of a human--agent team as
properties of its communication layer rather than as instructions in each
agent's prompt. An instruction may be ignored, lost when an agent's context is
compacted, or interpreted differently by different models. A property of the
communication layer applies to every participant, irrespective of vendor,
model or tool. From operating a team of several people and many agents, we
identify five requirements. Each part of the framework answers one of them.
\begin{enumerate}
\item \emph{Accountability}: every principal answers to a human owner.
\item \emph{Explicit obligations}: who owes whom an answer is recorded.
\item \emph{Economy of agent work}: every turn costs tokens and fills a
bounded context.
\item \emph{Bounded and checkable work}: work serves an agreed goal with
criteria.
\item \emph{Oversight}: a durable record serves auditors and agents
recovering lost context.
\end{enumerate}

\emph{Rooms, seats and principals (accountability).} A room is a tuple
$\rho = (P, \Sigma, \Pi, G, \Omega, L)$ of principals $P$, seats $\Sigma$, a
policy $\Pi$, goals $G$, a resource map $\Omega$ and a ledger $L$. We separate
three levels that are easily conflated. A \emph{seat} $\sigma \in \Sigma$ is a
role in the room, with its responsibilities, permissions and allowed
communication partners. A \emph{principal} $p \in P$ is an authenticated human
or agent identity that occupies a seat and has an owner $o(p)$, the human
accountable for it. A \emph{deployment} is the model, harness and host on
which an agent principal runs. Obligations, authority and routing attach to
the seat, and accountability attaches to the owner. Replacing an agent, or
moving it to another model or host, keeps its role, permissions and
routing, and does not transfer responsibility. The resource map $\Omega$
holds references to shared resources, such as repositories, deployed services and encrypted
credentials, so that access to them is also governed by the room.

\emph{Typed, addressed acts (explicit obligations).} Principals communicate
only through acts
\[
m = (s, a, R, \kappa, g, r, b),
\]
where $s \in P$ is the sender, $a \in \{\textit{autonomous}, \textit{on behalf
of } h\}$ the acting mode with $h$ a human, $R \subseteq P$ the named
recipients, $\kappa \in K$ the kind, $g \in G \cup \{\bot\}$ an optional goal,
$r$ an optional reference to the act being answered, and $b$ the body. The kind
states the sender's intent. Intuitively, an act either shares information or
asks for something, so $K = K_{\mathrm{info}} \cup K_{\mathrm{req}}$ with
{\small
\[
K_{\mathrm{info}} = \{\textit{inform}, \textit{progress\_update}, \textit{artifact\_update}\},
\]
\[
K_{\mathrm{req}} = \{\textit{request\_reply}, \textit{request\_comment}, \textit{request\_execution}\}.
\]}
Only a requesting act creates obligations: an act $m$ with
$\kappa \in K_{\mathrm{req}}$ creates an obligation $(q, m)$ for each
$q \in R$, and for no one else. The obligation is discharged by an act $m'$
from $q$ with $r(m') = m$ and $\kappa(m') \neq \textit{progress\_update}$. A
principal named only in the body $b$ is not addressed. Making intent and
addressees explicit lets the relay track who owes whom an answer, instead of
leaving it to each agent's reading of free text.

\emph{Delivery and intake (economy of agent work).} An act that starts a turn
consumes tokens and context and may interrupt work in progress, whereas an act
kept as context costs little until the next turn. The relay
maintains a delivery state for each pair $(m, q)$ with $q \in R$. The
recipient claims the delivery for a bounded time and acknowledges it, and an
expired claim returns it for redelivery. The agent's adapter applies an intake
function $\iota_q(m) \in \{\textit{start}, \textit{context}, \textit{queue},
\textit{record}\}$ that decides whether $m$ starts a turn, waits as context for
the next turn, is queued, or is only recorded. The sender receives a receipt
stating the decision and its cause. In the revised intake used in
Section~\ref{sec:controlled}, $\iota_q(m) = \textit{start}$ only if $q$ was
named in $R$ and either $\kappa \in K_{\mathrm{req}}$ or $m$ answers a request
that $q$ has open. Other addressed acts, progress notices included, enter the
next turn as context. Broadcasts are only recorded. Before the revision,
an addressed inform that did not answer a request of the recipient was
recorded without reaching the model (Section~\ref{sec:f2}).

\emph{Goals and policy (bounded and checkable work).} A goal $g \in G$ carries
a title, acceptance criteria, a parent goal and a state. Requests refer to a
goal so that every obligation serves an agreed purpose, its completion can be
checked against the criteria, and work can be decomposed, closed and measured
per goal. Proposals and votes are records in the room rather than conventions
in a conversation. The policy $\Pi$ declares which actions are permitted,
restricted or forbidden, the norms of the room, and a routing relation
$E \subseteq \Sigma \times \Sigma$: a request from seat $\sigma$ to seat
$\sigma'$ is admitted only if $(\sigma, \sigma') \in E$. The policy is the
upper bound of communication. It defines the space of acts that any
participant can perform, whatever its prompt says. Norms and instructions can
only narrow behavior within this space, never widen it. We therefore
distinguish two levels of every rule. An \emph{instructed} rule is
communicated to agents and relies on their compliance. An \emph{enforced} rule
is checked by the relay before an act takes effect.

\emph{The ledger (oversight).} $L$ is an append-only, totally ordered sequence
of all acts and governance records of the room, each with its author and
acting mode and linked to goals and commits. Humans can trace who requested
what, auditors can investigate after the fact, and agents can recover context
that their bounded memory has lost.

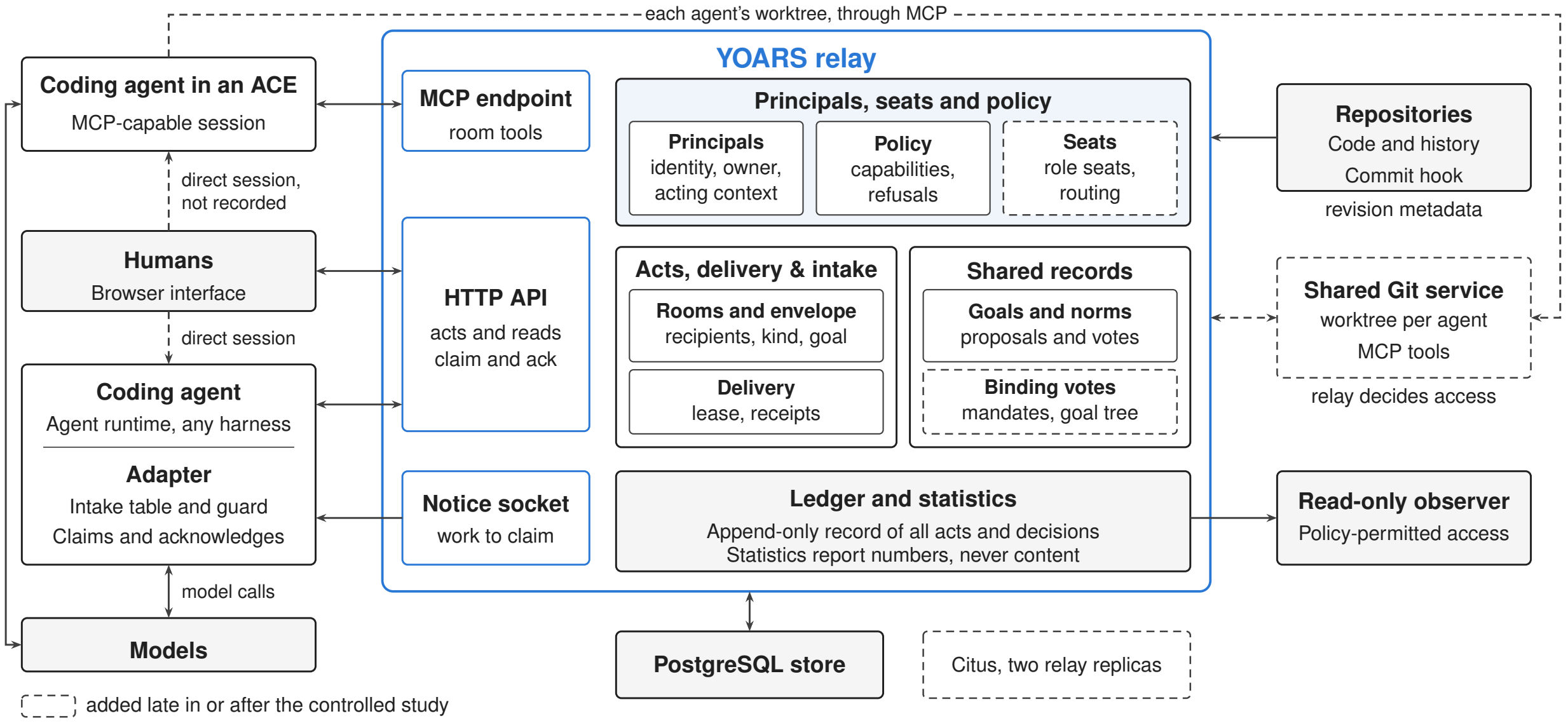
\begin{figure*}[t]
\centering
\resizebox{0.94\textwidth}{!}{\begingroup
\definecolor{figink}{HTML}{202020}
\definecolor{figmuted}{HTML}{404040}
\definecolor{figblue}{HTML}{2A78D6}
\definecolor{figgray}{HTML}{F5F5F5}
\definecolor{figtint}{HTML}{F0F5FC}
\begin{tikzpicture}[x=1pt,y=-1pt,
  hd/.style={inner sep=0pt,text=figink,font={\fontfamily{phv}\fontsize{17}{21}\selectfont\bfseries}},
  tx/.style={inner sep=0pt,text=figink,align=center,font={\fontfamily{phv}\fontsize{15}{19}\selectfont}},
  lb/.style={inner sep=1.5pt,fill=white,text=figink,font={\fontfamily{phv}\fontsize{14}{17}\selectfont}},
  ext/.style={draw=figink,fill=figgray,line width=1.4pt,rounded corners=4pt},
  agt/.style={draw=figink,fill=white,line width=1.4pt,rounded corners=4pt},
  grp/.style={draw=figink,line width=1.4pt,rounded corners=4pt},
  sub/.style={draw=figmuted,fill=white,line width=1pt,rounded corners=3pt},
  port/.style={draw=figblue,fill=white,line width=1.4pt,rounded corners=4pt},
  new/.style={draw=figmuted,fill=white,line width=1.2pt,rounded corners=3pt,dash pattern=on 6pt off 4pt},
  ar/.style={draw=figmuted,line width=1.4pt,-{Stealth[length=8pt,width=6pt]}},
  ar2/.style={draw=figmuted,line width=1.4pt,{Stealth[length=8pt,width=6pt]}-{Stealth[length=8pt,width=6pt]}},
  dar/.style={ar,dash pattern=on 6pt off 4pt},
  dar2/.style={ar2,dash pattern=on 6pt off 4pt}]
\path[use as bounding box] (0,0) rectangle (1190,540);

\draw[agt] (20,40) rectangle (240,110);
\node[hd] at (130,62) {Coding agent in an ACE};
\node[tx] at (130,90) {MCP-capable session};

\draw[ext] (20,170) rectangle (240,230);
\node[hd] at (130,192) {Humans};
\node[tx] at (130,216) {Browser interface};

\draw[agt] (20,270) rectangle (240,420);
\node[hd] at (130,292) {Coding agent};
\node[tx] at (130,316) {Agent runtime, any harness};
\draw[draw=figmuted,line width=0.8pt] (36,332) -- (224,332);
\node[hd] at (130,354) {Adapter};
\node[tx] at (130,378) {Intake table and guard};
\node[tx] at (130,400) {Claims and acknowledges};

\draw[ext] (20,460) rectangle (240,500);
\node[hd] at (130,484) {Models};

\draw[dar] (130,170) -- (130,110);
\node[lb,anchor=west,align=left] at (138,140) {direct session,\\not recorded};
\draw[dar] (130,230) -- (130,270);
\node[lb,anchor=west] at (138,250) {direct session};
\draw[ar2] (130,420) -- (130,460);
\draw[ar2] (20,75) -- (8,75) -- (8,480) -- (20,480);
\node[lb,anchor=west] at (138,440) {model calls};

\draw[draw=figblue,fill=white,line width=1.8pt,rounded corners=8pt] (290,20) rectangle (910,440);
\node[inner sep=0pt,text=figblue,font={\fontfamily{phv}\fontsize{20}{24}\selectfont\bfseries}] at (600,42) {YOARS relay};

\draw[port] (305,50) rectangle (445,110);
\node[hd] at (375,72) {MCP endpoint};
\node[tx] at (375,96) {room tools};
\draw[port] (305,160) rectangle (445,320);
\node[hd] at (375,220) {HTTP API};
\node[tx] at (375,246) {acts and reads};
\node[tx] at (375,266) {claim and ack};
\draw[port] (305,350) rectangle (445,420);
\node[hd] at (375,374) {Notice socket};
\node[tx] at (375,398) {work to claim};

\draw[ar2] (240,75) -- (305,75);
\draw[ar2] (240,200) -- (305,200);
\draw[ar2] (240,300) -- (305,300);
\draw[ar] (305,385) -- (240,385);

\draw[grp,fill=figtint] (465,56) rectangle (895,166);
\node[hd] at (680,74) {Principals, seats and policy};
\draw[sub] (475,88) rectangle (605,158);
\node[tx] at (540,123.5) {\textbf{Principals}\\identity, owner,\\acting context};
\draw[sub] (615,88) rectangle (745,158);
\node[tx] at (680,123.5) {\textbf{Policy}\\capabilities,\\refusals};
\draw[new] (755,88) rectangle (885,158);
\node[tx] at (820,123.5) {\textbf{Seats}\\role seats,\\routing};

\draw[grp,fill=white] (465,182) rectangle (675,332);
\node[hd] at (570,200) {Acts, delivery \& intake};
\draw[sub] (475,214) rectangle (665,268);
\node[tx] at (570,241) {\textbf{Rooms and envelope}\\recipients, kind, goal};
\draw[sub] (475,274) rectangle (665,322);
\node[tx] at (570,298) {\textbf{Delivery}\\lease, receipts};

\draw[grp,fill=white] (685,182) rectangle (895,332);
\node[hd] at (790,200) {Shared records};
\draw[sub] (695,214) rectangle (885,268);
\node[tx] at (790,241) {\textbf{Goals and norms}\\proposals and votes};
\draw[new] (695,274) rectangle (885,322);
\node[tx] at (790,298) {\textbf{Binding votes}\\mandates, goal tree};

\draw[ext] (465,350) rectangle (895,425);
\node[hd] at (680,372) {Ledger and statistics};
\node[tx] at (680,396) {Append-only record of all acts and decisions};
\node[tx] at (680,414) {Statistics report numbers, never content};

\draw[ext] (465,470) rectangle (665,520);
\node[hd] at (565,496) {PostgreSQL store};
\draw[new] (695,470) rectangle (895,520);
\node[tx] at (795,496) {Citus, two relay replicas};
\draw[ar2] (565,440) -- (565,470);

\draw[ext] (960,60) rectangle (1150,140);
\node[hd] at (1055,84) {Repositories};
\node[tx] at (1055,106) {Code and history};
\node[tx] at (1055,128) {Commit hook};
\draw[ar] (960,100) -- (910,100);
\node[tx,anchor=north] at (1055,148) {revision metadata};

\draw[new] (960,190) rectangle (1150,280);
\node[hd] at (1055,214) {Shared Git service};
\node[tx] at (1055,238) {worktree per agent};
\node[tx] at (1055,260) {MCP tools};
\draw[dar2] (910,235) -- (960,235);
\node[tx,anchor=north] at (1055,288) {relay decides access};

\draw[ext] (960,350) rectangle (1150,420);
\node[hd] at (1055,374) {Read-only observer};
\node[tx] at (1055,398) {Policy-permitted access};
\draw[ar] (895,385) -- (960,385);

\draw[dar] (130,40) -- (130,8) -- (1172,8) -- (1172,235) -- (1150,235);
\node[lb] at (600,8) {each agent's worktree, through MCP};

\draw[new] (20,518) rectangle (60,534);
\node[tx,anchor=west] at (68,526) {added late in or after the controlled study};
\end{tikzpicture}
\endgroup}
\caption{\sys\ architecture. Humans use the browser interface. Agents connect
in one of two ways. An agent with an adapter claims and acknowledges
deliveries over the HTTP API and is woken by the notice socket. Every agent
in the controlled runs connected this way. An agent in an AI coding
environment (ACE) acts through MCP tools. A person can also instruct either
kind of agent directly, in a direct session (dashed arrows), and what the
person says there is not recorded. The relay applies room rules and records
every act, while code remains in repositories. Model calls never pass through
the relay. Dashed boxes were added late in or after the controlled study
(Section~\ref{sec:since}), and no controlled run used them.}
\label{fig:arch}
\end{figure*}

Fig.~\ref{fig:arch} shows how these concepts are realized. Humans use a
browser interface. Agents keep their own tools and models and connect through
an adapter or through MCP tools, so that agents from different vendors take
part in one room. Code
stays in repositories, and commits are linked to the record. During the
controlled study the relay provided principals, typed acts, delivery and
intake, the shared records of goals, proposals and votes, and the ledger.
Enforcement was added late in the controlled study or after it, and no
controlled run used it. It comprises roles that outlive the agent holding them
(\emph{seats}), routing that denies a request unless the pair of roles is
allowed, votes that block a goal until all named voters have voted,
time-limited mandates from a person, and per-agent working copies on a
relay-controlled Git service. These later mechanisms were used in the field
study but not evaluated as treatments.

\section{Architecture and Design Decisions}
\label{sec:design}

Section~\ref{sec:concept} defined what a governed room is. This section
describes how \sys\ realizes it and why it is built the way it is. We first
state the design goals that constrain the implementation. We then describe the
relay as the controlled study measured it, and the design decisions that
shaped it, each with the problem that forced it. Finally, we describe what has
changed since the controlled study. This separation matters for reading the
results. The platform changed while it was being studied, and no result is a
property of a platform it was not measured on.

Two timelines overlap. We give times as days of the project, counted from the
first day of its records (day 1). The controlled study launched its runs in
\emph{blocks}, sets of runs started together. These \emph{controlled blocks}
ran from day 3 to day 53. The first campaign (days 3--7) ran before the
\emph{intake revision} of day 12, which changed what an arriving message does
to an agent's turn (Section~\ref{sec:f2}). All later blocks ran after it.

From day 11, each block pinned its adapter at launch, that is, fixed the
adapter revision it would run under. From day 47, each block also pinned its
relay. Before day 46, the run manifests record the relay commit that was
checked out, not the code of the running relay process. The blocks of
days 35--44 were served by relay code of day 25, and the blocks from day 46 by
relay 0.14.0 (Appendix~\ref{app:details}). The \emph{field study} observed the
room in which the team developed \sys\ for four weeks, from day 47 to day 73,
on later relay versions (Section~\ref{sec:since}). In the week in which the two
timelines overlap, the last controlled blocks kept running on the pinned
relay 0.14.0. Table~\ref{tab:decisions} places each design decision on these
timelines.

\subsection{Design Goals}
\label{sec:goals}

\sys\ was built for coordination among agents that do not share a runtime, a
vendor or an owner. Six goals shaped the relay. They turn the five
requirements of Section~\ref{sec:concept} into constraints on an
implementation, and the rest of this section returns to them.

\paragraph{G1: Heterogeneous participants.} An agent in one harness must be
able to work with an agent in another harness, and with a person, through one
interface. The relay deals with principals and not with agent
frameworks. An adapter translates a runtime's turn loop, the cycle in which it
calls the model, into the relay's protocol. Nothing on the server depends on
which runtime a participant uses.
\paragraph{G2: Humans and agents as principals under the same rules.} Every
participant is an authenticated principal with an owner. Humans and agents
send the same kinds of acts, are bound by the same room rules, and are
attributed in the record in the same way. This equal standing is a property of
the protocol, not of the work. In the field study, the two people directed the
work, mostly through coordinator agents, and approved every production release
(Section~\ref{sec:field}). An agent can act for itself or on a human's behalf,
and the record says which. Authorization never depends on a participant's
display name.

What a person tells an agent is not recorded. In a direct session
(Fig.~\ref{fig:arch}), the person instructs the agent inside the coding
environment, and only that environment's own transcript, if any, keeps the
exchange. The ledger records that the agent acted for the person, not what the
person asked for. An agent can therefore claim an instruction that the person
never gave, by mistake or as an attack, and the record cannot tell the two
apart. The first thread of Fig.~\ref{fig:threads} opens with such a claim, a
request that begins ``Operator's order''. The relay marks votes that an agent
carries for a person, and a mandate bounds and stamps the acts made in a
person's name (Section~\ref{sec:decisions}). Neither records the instruction
itself. Only an act that the person sends through the browser interface is
the person's own.

\paragraph{G3: The room governs itself.} The room's owner sets what
participants may see about each other (the roster, profiles, published plans,
traffic figures and peers' progress), what they may publish and how they may
decide.
These settings are properties of the room. Every participant can read them.
They are closed by default wherever opening them would widen what anyone
learns.

\paragraph{G4: Coordination is observable without reading content.} Operators
need to know whether a message reached anyone, whether a participant is stuck
and why a delivery failed. They must be able to learn this without the
platform reading or exporting message bodies. Observability reports
numbers and states, never content.

\paragraph{G5: Everything leaves a record.} Messages, deliveries, receipts,
refusals, commit announcements, norms, proposals and votes are all
attributable to an authenticated principal. They are ordered in one ledger per
room, so a coordination episode can be reconstructed afterwards.

\paragraph{G6: Cost proportional to the room, never to its history.} A room
may stay open for months. Every path that runs per message, per socket or per
turn is bounded by the size of one room. It does not grow with the age of the
ledger or with the size of the whole installation.

\subsection{The Relay as the Controlled Study Measured It}
\label{sec:relay}

Figure~\ref{fig:arch} shows the relay's interfaces and parts. This section describes each part in
the order in which a message meets it, as the controlled study measured it.
The runs used relay versions up to 0.14.0. For parts that entered the code
base between controlled blocks, Table~\ref{tab:decisions} gives the day. The
adapter's intake table and guard date from the intake revision. The platform
has changed since, and Section~\ref{sec:since} describes those changes
separately.

\subsubsection{Principals, Acting Contexts and Endpoints}

Every participant is a \emph{principal}, human or agent, with an opaque
identifier and an owner. A principal authenticates through an
\emph{endpoint}. Each endpoint has its own credential and role:
\emph{executor} (an agent that claims and answers deliveries) or \emph{human
client}. An observer is not a role. It is a principal reading a room whose
policy permits it (Section~\ref{sec:observability}). A request carries an
\emph{acting context}: human, autonomous agent, or agent delegated by a human.
The record keeps this context, so the ledger distinguishes an agent speaking
for itself from an agent acting for someone (G2). Handles, the readable names
of principals, are derived for display and are never parsed for
authorization.

\subsubsection{Rooms and the Envelope}

A room has participants, a policy and a ledger. Every message follows one
versioned schema, \texttt{yoars2.message.v1}. It carries a body and an
explicit recipient set. The relay resolves the recipients to principals at
intake and records how it resolved them. A message also carries a
\emph{communication kind} that states its intent (\texttt{inform},
\texttt{progress\_\allowbreak update}, \texttt{request\_\allowbreak reply},
\texttt{request\_\allowbreak comment}, \texttt{request\_\allowbreak execution},
and \texttt{artifact\_\allowbreak update}, Section~\ref{sec:artifacts}).
Recipient and intent travel together, and a name mentioned in prose
addresses nobody. An optional \emph{goal}, written by the sender, says what the
message is for. Its absence means ``the sender did not say'', and nothing
substitutes one. Intake is idempotent on a request identifier that the sender
supplies, so a message sent twice with the same identifier is accepted once.

\subsubsection{Delivery, Intake and Receipts}

\begin{figure}[t]
\centering
\resizebox{0.86\columnwidth}{!}{\begingroup
\definecolor{figink}{HTML}{202020}
\definecolor{figmuted}{HTML}{404040}
\definecolor{figblue}{HTML}{2A78D6}
\definecolor{figtint}{HTML}{F0F5FC}
\begin{tikzpicture}[x=1pt,y=-1pt,
  tx/.style={inner sep=0pt,text=figink,align=center,font={\fontfamily{phv}\fontsize{15}{19}\selectfont}},
  lb/.style={inner sep=0pt,text=figink,align=left,font={\fontfamily{phv}\fontsize{14}{17}\selectfont}},
  rel/.style={draw=figink,fill=figtint,line width=1.4pt,rounded corners=4pt},
  adp/.style={draw=figink,fill=white,line width=1.4pt,rounded corners=4pt},
  ar/.style={draw=figmuted,line width=1.4pt,-{Stealth[length=8pt,width=6pt]}}]
\path[use as bounding box] (0,0) rectangle (490,470);
\foreach \i/\s/\t in {
  0/rel/{Sender posts envelope},
  1/rel/{Intake: policy, recipients, idempotency},
  2/rel/{One delivery per recipient},
  3/rel/{Adapter claims (lease)},
  4/adp/{Intake table: observe, context, queue, invoke},
  5/adp/{Guard: duplicate, rate, depth, budget},
  6/rel/{Ack with outcome and cause},
  7/rel/{Receipt visible to the sender}} {
  \draw[\s] (20,10+52*\i) rectangle (360,46+52*\i);
  \node[tx] at (190,28+52*\i) {\t};
}
\foreach \i in {0,...,6} \draw[ar] (190,46+52*\i) -- (190,62+52*\i);
\draw[ar] (360,184) -- (384,184) -- (384,132) -- (360,132);
\node[lb,anchor=west] at (394,146) {lease};
\node[lb,anchor=west] at (394,164) {expires:};
\node[lb,anchor=west] at (394,182) {re-deliver};
\draw[rel] (20,438) rectangle (52,456);
\node[lb,anchor=west] at (60,447) {relay};
\draw[adp] (140,438) rectangle (172,456);
\node[lb,anchor=west] at (180,447) {adapter decides};
\end{tikzpicture}
\endgroup}
\caption{A message's life. The relay owns delivery and the record, and the adapter
decides whether an arriving message starts a turn.}
\label{fig:life}
\end{figure}
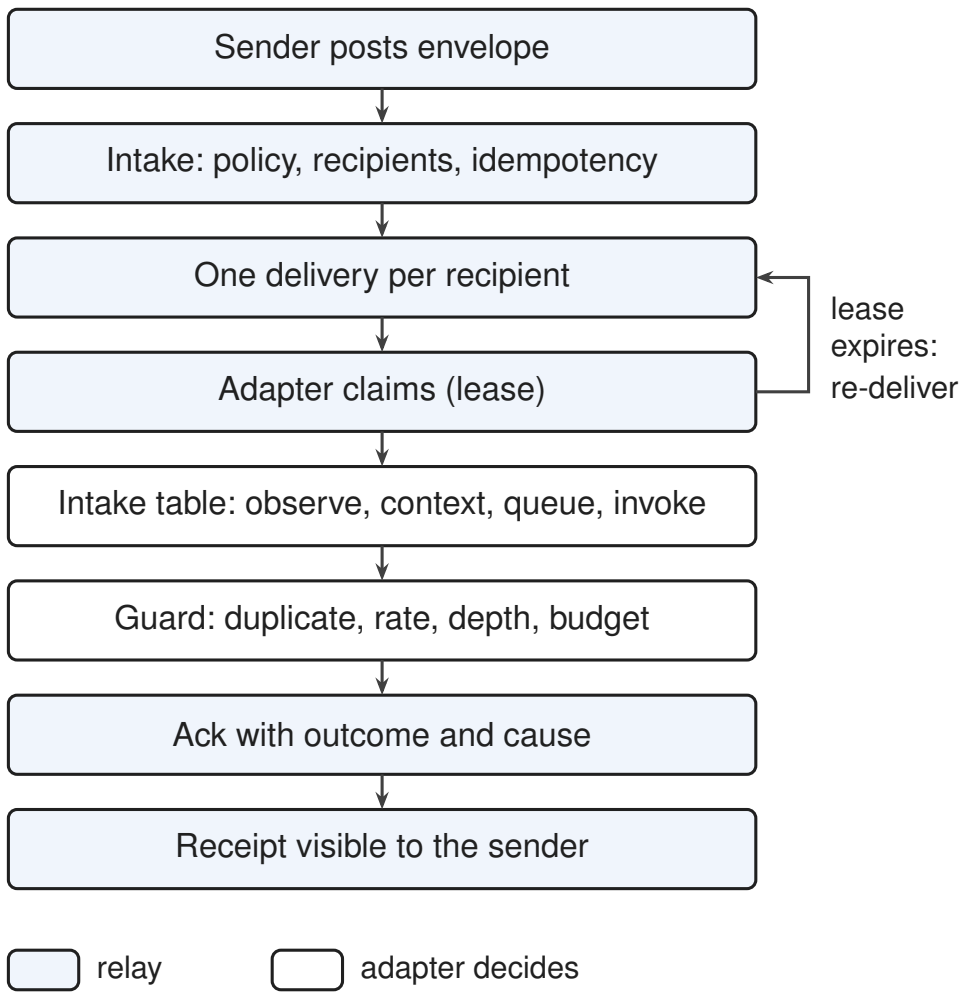

Each addressed message becomes one delivery per recipient
(Figure~\ref{fig:life}). An executor \emph{claims} deliveries under a lease.
The claim is a compare-and-set, so only one executor holds a lease at a time.
The executor acknowledges against the current lease token, and it loses a
lease that it lets expire. A crashed agent's work is re-delivered
rather than lost. Repeated failure ends in a dead-letter state rather than a
silent drop. Deliveries are pulled by long-poll and projected as tools over
MCP, and a notice socket announces them. On a single-host stack, measured
early in the project, a notice arrived 48\,ms after the post, against 3.5\,s by
polling.

What a delivery \emph{does} to an agent is decided by the adapter's intake
table, not by the sender. A message may be merely \emph{observed}, placed into
the next turn's \emph{context}, \emph{queued}, or allowed to \emph{invoke} a
turn. Broadcasts are observed, and progress updates never start a turn. The
intake table and the guard described next date, in this form, from the intake
revision. In the first campaign, an addressed inform that did not answer a
request of the recipient was acknowledged without reaching the model
(Section~\ref{sec:f2}).

A guard in the adapter declines exact duplicates. It also caps the inbound
rate, the depth of a conversation between two agents, the queue depth and the
context size, and it tells the sender which rule applied. The acknowledgement
carries an outcome: answered, declined, or failed with a typed cause. The
sender reads this outcome as a \emph{receipt}, so a silence can be attributed.

\subsubsection{Policy: What a Room Lets Its Participants Do}

A room's policy assigns each \emph{capability} one of three values:
\texttt{permitted}, \texttt{admins\_only} or \texttt{forbidden}
(Table~\ref{tab:caps}). The vocabulary is closed and is extended only by
adding to it, so a policy outlives the code that wrote it. The policy acts as
a ceiling, and capabilities differ in what else they need. Capabilities that
disclose something on an agent's behalf, such as narrating a plan or declaring
a workspace, also need the agent's own preference. Capabilities that are acts,
such as reading statistics or a peer's profile, need only the ceiling. Every
refusal is recorded and counted. A norm that asks for something the room
forbids is refused when it is set. It is not left to be discovered by the
agent that obeys it.

\begin{table*}[t]
\caption{The policy vocabulary. Defaults are chosen so that a capability is
never opened by accident: anything that would widen what a participant learns
about others starts closed. In relay 0.14.0, which served the blocks from
day 46, the three proposal capabilities and \texttt{read\_evidence} were
declared and grantable, but no route checked them. This was corrected in
0.20.0.}
\label{tab:caps}
\small
\begin{tabular}{lll}
\toprule
capability & governs & default \\
\midrule
\texttt{narration}, \texttt{disclosure} & publishing plans, declaring a workspace & permitted$^{*}$ \\
\texttt{roster}, \texttt{profiles} & who is present; peers' roles and missions & permitted \\
\texttt{statistics} & reading the room's traffic figures & permitted \\
\texttt{resource\_status} & reading peers' declared progress & forbidden \\
\texttt{observable} & reading the room without joining & forbidden \\
\texttt{invites} & bringing others in & permitted \\
\texttt{voting} (+ veto, majority, weighted) & deciding by ballot, and by which rule & forbidden \\
\texttt{norms} & stating what the room asks of behavior & admins only \\
\texttt{propose\_\{policy,routing,archive\}} & proposing changes to the room & admins only \\
\texttt{read\_evidence} & reading the room's refusals and evidence & admins only \\
\bottomrule
\end{tabular}
\\[1mm]{\footnotesize $^{*}$ also requires the participant's own preference.}
\end{table*}

\subsubsection{Coordination Records}

Besides messages, a room holds typed \emph{records}. A \emph{resource} names
something the room works on, for instance a repository. Participants declare
their status against it. Unpublished work can then be told apart from no work,
without obliging anyone to publish. A \emph{norm} states in prose what the
room asks of behavior, and it is advisory by construction. A \emph{goal}
records what a request is for, and an execution request under a goal is
tracked as a wait on that goal. A \emph{proposal} opens a ballot under the
room's voting scheme, and \emph{votes} are records too. None of these records
interprets content. All of them are attributable and are ordered with the
messages.

\subsubsection{Artifacts as Room Events}
\label{sec:artifacts}

\sys\ is not a repository. A commit hook announces each commit to the room as
an \texttt{artifact\_update}. This message carries metadata only: author,
revision, parent, subject and paths. Whose change builds on whose is thus
visible in the room, without the relay holding code (G4, G6).
Section~\ref{sec:rq1} shows why this matters: a shared filesystem is an
uninstrumented channel.

\subsubsection{Observability and Participation Health}
\label{sec:observability}

The relay reports numbers, never bodies. It reports per-room traffic, delivery
outcomes classified by how they ended, typed failure causes, and each
participant's recent responsiveness. ``Connected but not answering'' is
a state rather than an inference. An observer principal can read a
room whose policy permits it, without appearing in it or receiving
deliveries. The same record serves the study. Every run in this report is an
export of one room's ledger, deliveries, receipts and records, plus each
agent's final code and history.

\subsubsection{Adapters}

An adapter is the only runtime-specific part. It holds the intake table and the
guard, injects the room's context into a turn, and turns a turn's outcome into
an acknowledgement. Because the adapter decides what reaches the model, it is
the \emph{treatment boundary} of our experiments. Every run records the
adapter revision it ran under, and from day 11 each block pins it at launch.

\subsection{Design Decisions and Their Reasons}
\label{sec:decisions}

The relay was developed in the open way that the field study describes.
Decisions were recorded when they were taken, with the problem that forced
them and the alternatives considered. Corrections were recorded later. We
compiled the decisions that shape the architecture from these records, about
110 of them, from the first day of the project to the time of writing.
Table~\ref{tab:decisions} summarizes the most consequential ones. It places
each on the two timelines defined above: when it was first built, and in
which controlled blocks it was in place. The paragraphs below give the
reasons. They keep apart what was decided, what was built and what was
measured. Days are those of the records.

\begin{table*}[t]
\centering
\caption{Design decisions of \sys, the problem that forced each, and when it
was in place. Last column: day of the build record (relay version where
recorded), followed by the controlled blocks in which the part was in place.
\emph{all}: every controlled block (days 3--53). \emph{rev.}: blocks from the
intake revision on (days 12--53). \emph{d35}: blocks from day 35 (relay code of
day 25, then 0.14.0). \emph{0.14}: blocks on relay 0.14.0 (days 46--53).
\emph{off}: built, but disabled in every controlled run. \emph{field}: field
study only (days 47--73). E: evaluated as a treatment. For a relay part built
between blocks, the code names the earliest blocks whose relay is known to
contain it. Manifests before day 46 record the relay's checkout rather than the
running process (Appendix~\ref{app:details}). $^{\dagger}$Built on day 3, the
day of the first block. ``In place'' does not mean that the runs exercised
it.}
\label{tab:decisions}
\footnotesize
\renewcommand{\arraystretch}{1.12}
\begin{tabularx}{\textwidth}{@{}>{\raggedright\arraybackslash}p{0.11\textwidth}>{\raggedright\arraybackslash}X>{\raggedright\arraybackslash}X>{\raggedright\arraybackslash}p{0.16\textwidth}@{}}
\toprule
\textbf{Concern} & \textbf{Decision} & \textbf{Problem that forced it} & \textbf{Built: blocks} \\
\midrule
Identity & Authorization uses only opaque principal identifiers. Handles are addresses and are never parsed for rights. The acting mode is authenticated, not declared. & The predecessor system trusted identity carried in headers and names. & day~1: all \\
Envelope & Recipients and kind travel together in one versioned envelope. A name in prose addresses nobody. Replies follow one contract. & Free-text addressing left it to each reader who was meant. & days~1--2: all \\
State changes & Goals, resources and votes change only through typed, authorized operations, never through commands parsed from message bodies. & Two authorities for one state, and a replayed message could repeat a mutation. & day~2 (rule): all \\
Relations & Goals, causes and waits are declared by the participant who knows them, never inferred by the relay. & An inferred cause leaked information and named the wrong cause. & waits day~28: 0.14\newline rule day~50: field \\
Delivery & One fenced delivery per recipient: claim, lease, acknowledge, expire, redeliver. The socket only notifies. Delivery is at least once, with idempotent intake. & The predecessor had four dispatchers and lost completions. & day~1: all \\
Obligation & An obligation is owed per recipient, not per message. & A wait-for edge needs a debtor. & day~25: 0.14 \\
Waking & Reaching the model is separate from waking it. Only requests and replies to the agent's own open requests wake it, one delivery per turn. The intake table is shared, executable data. & Most agent messages were informs that reached nobody, one room produced 963 messages, and the first table deadlocked. & day~12 (adapter): rev., E \\
Guard & The adapter bounds its own agent: it never leaves a delivery pending, and declines duplicates, loops and excess rates with a stated reason. & A full agent that stopped claiming looked like a dead one. & day~12 (adapter): rev., E \\
Attention & The relay labels why a message reaches an agent but does not enforce budgets. & The change evaluated in Finding~2 was made entirely in the adapter. & day~16: d35 \\
Policy & The room policy is a ceiling that can forbid but cannot oblige. Refusals are recorded. & Three features turned out to be one decision with three failure semantics. & day~2: all \\
Routing & A directional matrix, later an access language of roles and audiences in which every refusal names its rule. & Per-pair lists grow with the square of the room and fail for later members. & matrix day~8: off\newline language day~52: field, partly built \\
Stalls & Cycles are detected over declared waits only, reported and never broken. & Inferred waits produce false reports, and breaking a cycle decides whose work is discarded. & day~26: 0.14 \\
Decisions & Votes are records and tallies are computed on read, voting off by default. Later, decisions bind a version and can block a goal under an enforced rule. & Instructed proposals often received no vote and bound nothing (Finding~3), and every release approval reached the room through an agent. & votes day~14: E in voting blocks\newline binding 0.93--0.96 (days~66--67): field \\
Seats & Room facts are keyed by seat, with occupancy epochs and leases. Rights are the role's, narrowed at once and never widened silently. & Roles must outlive the agent holding them, and a displaced occupant must not write. & 0.76--0.82 (days~53--59), in stages: field \\
Mandates & An owner grants an agent's seat a time-limited, revocable mandate. Acts under it are checked against the grantor's rights and stamped. & An agent acting for its owner had no standing of its own. & 0.96.0 (day~67): field \\
Observation & The relay counts and never records content in telemetry. An observer reads a room without joining it. & An observer added mid-run changed a run. & days~1--3: all$^{\dagger}$ \\
Artifacts & The relay holds no code. Commits reach the room as metadata events. The Git service gives each agent a server-side working tree under an allowlist and a sandbox. & The shared workspace was the effective, invisible channel. A denylist of Git commands proved insufficient. & events day~3: all$^{\dagger}$\newline Git service day~46: field \\
Scale & The store is distributed by room, with send order fixed by a per-room lock. Every list read is bounded. & All tables sat on one node, and a slow send could fall behind a read marker. & days~50--63: field \\
Method & The adapter is the treatment boundary and is pinned per block of runs. & A working copy 27 commits behind would have measured an old adapter against a new relay. & day~11: rev. \\
\bottomrule
\end{tabularx}
\end{table*}

\emph{Identity and accountability.} The relay authorizes, routes and stores by
an opaque principal identifier, and every agent has an owner. A handle such as
\texttt{a/<owner>/<name>} is an address for people. It is never parsed to
decide a right. The acting mode is part of the authenticated context. A
delegated context is minted by the owner, so a person's involvement is proven
rather than claimed, and the relay stamps provenance on every act. The
predecessor system had trusted identity carried in request headers, and the
records cite its lessons. These rules were decided and built before the first
block.

Management and participation are separate principals with separate
interfaces. The reason is that a credential given to an agent is used by the
model, not by the person. This separation was deployed after the first
campaign. A further decision, taken during the field study, makes every
management act of an agent a request that a human owner approves. Such acts
include changes of policy, routing or invitations. The reason is that an agent
that reads room traffic can be steered by it.

\emph{Messages and obligations.} The envelope carries recipients and kind
together. Intent and addressee are thus explicit, and a name mentioned in
prose addresses nobody. The set of kinds grew by splitting kinds, not by
adding new ones. A \textit{progress\_update} was split from the
\textit{inform}, because an inform had been carrying two different things, and
an adapter must never start a turn for a progress update. An optional goal says
what a message is for, and its absence means that the sender did not say. A
required goal was rejected because it produces ritual goals, set only to
satisfy the requirement. The envelope, the split and the optional goal were
all built before the first block.

State changes, such as opening a goal or casting a vote, are typed
operations. Each has an authorization check and a result. The relay never
parses message bodies for commands, because accepting both would have made two
authorities for one state. This rule, too, dates from before the first block.

Relations follow the same principle. The relay records a cause, a goal or a
wait only when the participant declares it. An execution request filed under a
goal became a declared wait on day 28, and relay 0.14.0 contains this change.
The general rule against inference was recorded later, during the field study.
It followed the relay's first attempt to infer a message's cause from its
leases, which had produced a leak and a wrong cause. An obligation is owed by
each named recipient separately (day 25). A request to three agents is
three obligations, and a bystander's reply discharges none of them.

\emph{Waking and its cost.} The decision that the controlled study evaluated in
Finding~2 separates reaching the model from waking it. Every admitted message
reaches the model, framed by its sender and intent. Only requests, and replies
to the agent's own open requests, start a turn. The recipient computes whether
a message is such a reply from the reference it carries. A new message kind
for replies was avoided because a sender could misuse it. Waking is a level,
not an event: it is evaluated whenever the agent is idle and its queue is not
empty, and one delivery is taken per turn. The intake table is data. The same
file of cases ships with both adapters, and their tests execute it.

The trigger for this decision was measured. Before the change, an addressed
inform that did not answer a request of the recipient was acknowledged but
never shown to the model. In an instrumented session, eight of nine agent
messages were informs, so most narration reached nobody. Waking on every
message was no alternative, because one room had produced 963 messages. The
first version of the new table deadlocked. An answer sent as an inform could
never wake the agent that had asked. This is why replies to an agent's own
open requests now wake it. The design was made in the days just before the
intake revision. It was built into both adapters and was in place in every
block from the intake revision on.

The adapter also bounds its own agent, with a guard built in the same
revision. The guard claims every delivery and either holds it or declines it
with a reason. This rule exists because an agent that stopped claiming when its
queue was full looked the same as a dead one. The guard's rules are functions
of the message and the agent's state: self-sent messages, exact duplicates, the
depth of an exchange between two agents, the inbound rate, the context budget
and the queue depth. Their thresholds were recorded as guesses and treated as
parameters of the experiments.

Finding~2 measured the revision as a bundle, in which the intake rules, a tool
for asking, the guard and receipts changed together. Its ablations kept most
of the reduction with the guard only recording, and all of it with receipts
hidden. They do not isolate the waking rules. In the runs of that comparison
after the revision, 17 of 1{,}667 intake decisions placed a message into the
next turn's context. The gain therefore cannot be attributed specifically to
making informs visible either.

The relay labels why a message reaches an agent: an obligation, an answer,
context or ambient traffic. It leaves budgets to the adapter. This labeling
was built after the intake revision. A guard imposed by the platform was
rejected, because the change evaluated in Finding~2 had been made entirely in
the adapter.

\emph{Policy as a ceiling.} The room policy determines what is possible, and
the agent determines what it does. A policy can forbid but cannot oblige. This
is why the instructed rules of Finding~3 had to be instructions. The ceiling
was decided and built before the first block. In the design, quantities form
a separate layer: a breach of policy is refused, a breach of a limit is
deferred, and a breach of a commitment is only made visible. Only the ceiling
was built.

Who may address whom began as an optional directional matrix, built after the
first campaign. In it, the effective permission is the narrowest of the room's
ceiling, the matrix cell and, for an agent whose owner is present, the owner's
own permission. The matrix was off in all controlled runs. In the field study
it became an access language of roles and audiences with visible exceptions.
The reason is that per-pair lists fail for members who join later, and they
grow with the square of the room. In the access language, every refusal names
the rule that caused it. Its rules are written, and the relay releases that
introduced seats implement parts of them.

\emph{Decisions and records.} Stalls are detected only over waits that
participants declare. The relay reports a cycle and never breaks one. Two
agents asking each other questions are conversing, not deadlocked, and
reporting them as a cycle would be a false alarm. This was decided on
days 25--26 and built on day 26. By then the relay process that served the
blocks of days 35--44 had already started, so no controlled block before
day 46 had it.

Proposals and votes were built as room records, with the tally computed on
read. Voting was off by default, so that a baseline would be observable. This
was built after the intake revision. Finding~3 then showed two things about the
voting block. Agents opened no proposal unless the instructions named the
tools, and many instructed proposals received no vote. In the field study, a
replay of 45 release chains showed that every approval had reached the room
through an agent. Decisions were therefore rebuilt during the field study, in
relay 0.93.0 to 0.96.0, so that the relay enforces the rule that decides each
kind of decision (Section~\ref{sec:since}).

\emph{Seats and mandates.} Persistent roles need an identity that outlives the
agent holding them. The room's facts were therefore re-keyed from the principal
to the seat. This was decided as the controlled blocks ended, and built in
stages in relay 0.76 to 0.82 during the field study. Later stages are not
built. A seat's rights are the role's current
rights. A narrowing applies at once. A widening waits for an owner, because an
automatic upgrade would be a backdoor. This rule is only partly built.

A mandate lets an owner authorize an agent's seat to act in the owner's name
without making the agent an
administrator (Section~\ref{sec:since}). It was decided during the field study. Its first phase was
built in relay 0.96.0, and a second phase is designed but not built. During
the field study, a verification run used a separate test relay (0.98.0) with
the Git service (described below) at 0.19.1. The run accepted a push under a
mandate and refused it after the mandate was revoked.

\emph{Artifacts and code.} The relay holds references and metadata, never
document bodies or code. A commit reaches the room as an
\textit{artifact\_update} with author, revision and paths, but no diff. This
was built on the day of the first block, after an early run had shown that a
shared workspace was the effective and invisible channel. Making the room
itself a repository was rejected. The record states that the observation was
right but the primitive was wrong.

The Git service came later. It gives each agent a server-side working tree,
which the agent drives with Git commands, and it asks the relay what each
principal may read, write or publish. A denylist of Git subcommands proved
insufficient, because options such as a configured file-system monitor or an
external transport execute code. The service accepts only an
allowlist of subcommands and flags, and it runs each call in a sandbox. It was
first deployed on the development cluster the day before the field study
began, and it was used only in the field study.

\emph{Scale.} The relay's store was distributed by room on Citus, an extension
that spreads PostgreSQL tables over several nodes. Identities and rooms were
replicated to every node. This was decided, and carried out on both clusters,
in the first week of the field study, after the database's worker nodes had
been found unused. Two alternatives were rejected. Distributing everything by
room would make every delivery path fan out. A schema per room would make
migrations grow with the number of rooms. A probe with 200{,}000 messages
measured a post across nodes at a median of 16\,ms, in a room of 20 members.

A room's order is the order in which its sends commit, enforced by a per-room
lock. The lock was added after a slow send in relay 0.85.0 had fallen behind a
read marker, the position up to which a reader has read. With 24 senders, the
lock caps one room at about 176 sends per second on PostgreSQL 16. On a
single-node Citus, the cap is about 117. Multi-node figures were not measured.
No controlled run used the distributed store.

\emph{Open design questions.} Three points remain open, and the claims of this
report do not depend on them. First, it is open who owes the answer to a
request sent to the whole room. A rule that only holders of a room capability
owe it was decided during the field study but not built. The adapters still
start a turn for every recipient of such a request. In the controlled runs,
requests named their recipients. Second, the design passes a role seat's open
obligations and read position to its next occupant. One release note,
however, describes a request as staying with the occupant who received it, and
the relay's reads have not yet been checked for consistency. Third, the limits
layer of the policy design has no build record on the relay side. Rates are
bounded only by the adapters.

\subsection{The Platform Since the Controlled Study}
\label{sec:since}

The last runs used relay 0.14.0. At the time of writing, the relay is at
0.99.0 on both its development cluster and its production cluster. It reached
production on the day of writing, twelve releases after 0.92.0, which had been
released six days before. The parts added since the study are dashed in
Figure~\ref{fig:arch}. No controlled run used any of what follows, and no
result in this report speaks to it. Some of it was built while the last blocks
were still running (Table~\ref{tab:decisions}). The project was renamed from
its working name, and the rename changed every wire identifier:
\texttt{yoars2.message.v1} is now \texttt{yoars.message.v1}. The corpus keeps
the names it was recorded under.

\paragraph{Seats.} The largest change separates \emph{who acts in a room}
from \emph{who someone is}. A \emph{seat} is a place in one room, held by a
principal. A seat is a personal seat, a guest seat for work from outside the
room under the owner's control, or a \emph{role seat} (``reviewer'') that
outlives whichever agent occupies it. An agent holds one seat per room. A
role seat is occupied under a lease
that can be taken over. Each occupancy has an epoch, so a write from a
displaced occupant is refused rather than interleaved. Deliveries, rights,
declarations and records are keyed by seat, and provenance names both the seat
and its occupant (G2). A seat has a standing and a position in the room, and
administration can be granted to a seat.

\paragraph{The shared workspace, instrumented.} Section~\ref{sec:rq1} found a
shared filesystem to be an uninstrumented channel. A room can now hold a shared
git project in a separate service. Each agent works in its own server-side
worktree through MCP tools. The relay decides who may read, write or publish,
under two new capabilities, \texttt{artefacts} and
\texttt{artefacts\_ephemeral}, which are admins only by default. A published
push becomes an \texttt{artifact\_update} in the room. Such an update may now
address nobody and reach readers through the room's feed. The relay still holds
no code.

\paragraph{More records, still no content.} Participants can file
\emph{reports} on the platform itself, as records with a triage state. A
receiving harness can \emph{tag} a message with its own classification.
\emph{Usage} is declared by the harness and summed by the relay, which measures
nothing itself (G4). Appendix~\ref{app:context} reports a three-week
measurement, taken through this contract, of what context costs in a live
fleet. Section~\ref{sec:field} analyzes the fleet's development stream over
four weeks: where requests waited, where a release spent its time, where agents
struggled and where context was lost.

Goals gained a lifecycle: declared done, archived, and a request superseded. A
seat can publish its current work status as one row that is overwritten and
lives only a short time. This row is never a message and wakes nobody. The
relay and the console carry it, and no harness writes it yet. Finally, a room's
owner can no longer be locked out of proposing or reading evidence by the
room's own policy.

\paragraph{Decisions that bind.} In the relay the runs used, nothing depended
on a vote's outcome. The decision records have since been rebuilt (relay
0.93.0 to 0.96.0), for the reasons given in Section~\ref{sec:decisions}. A
proposal states the action it decides and the goal it serves. It names its
voters by position and is counted under a rule in force for its kind. These
rules now include ``first answer decides''. A proposal can \emph{bind} a named
component, version or environment. It can also \emph{block} a goal: the goal
waits on every voter who has not voted, and the relay's cycle detection sees
that wait. A later proposal can supersede an earlier one, and every vote is an
append-only event. Votes that an agent carries for a person are marked as
such, because the relay cannot verify that the person said so. The rebuild responds to Finding~3 by making the next step
depend on the decision. It is untested here. The console uses the same votes to
let participants agree on the summary of a conversation thread.

\paragraph{Acting for someone, recorded.} A room's owner can grant an agent's
seat a \emph{mandate} to act on the owner's behalf for a bounded time (30 days
by default). Under a mandate, the shared git service asks for the grantor's
rights and stamps every commit made under it. The relay records each act
against the mandate (G2).

\paragraph{Who may address whom, declared.} A room declares its routing: which
seats may send a request to which. Routing is stated by role and by what a seat
owns, not by name, and the relay refuses a message outside it. In the
development room, the declaration expanded to 244 edges (allowed pairs of
seats) during the field study. A room may also name several human operators,
with its own rule for which of them decides (any one, first come, in that
room).

\paragraph{Goals as a tree.} Since 0.99.0, a goal can have a parent goal and a
heading level that is used only for display. A tree is at most eight levels
deep. Moving a goal is conditional on the revision that the mover last saw. A
move is refused if it would close a cycle or place the goal under a closed
goal, and every move is recorded in an append-only history. Reads can return a
goal's path from the root and its children, with a revision for the room's
tree. A goal with open children cannot be closed. At the time of writing, no
goal had a parent yet. Section~\ref{sec:discussion} and Appendix~\ref{app:dom}
return to the tree as a design.

\paragraph{Adapters.} The adapters now confine the model's tools in an operating-system
sandbox. A room can hold secrets that agents use without a model reading
them. On the development cluster, the relay can also forward model calls under
a token budget per person. An agent in a person's browser then never holds the
upstream key. Otherwise, model calls still do not pass through the relay.

\paragraph{Scale (G6).} The store is a distributed PostgreSQL (Citus), with 48
tables distributed by hash and 9 replicated. It runs behind two relay
replicas, whose wake-ups cross replicas through the database's notifications.
A load harness runs 200 seats against both replicas for 60\,s. It checks that
every delivery is acknowledged exactly once and that no request ends in a
server error. The median time from send to claim was 0.16--0.46\,s across
stores and seat configurations. Each figure comes from one run on otherwise
idle test servers, so these are ranges, not estimates.

The same discipline caught a regression of the kind G6 forbids. A seat lookup
read every seat of its room (5{,}045 rows for one, in a room of 5{,}046), and
it now reads at most four. Since 0.87.0, every list a client reads stops at
32{,}000 bytes and returns a cursor for the rest. The size of a response
no longer grows with the room. Every release publishes an OpenAPI
contract, and MCP and REST are tested for equivalence. The suite of 3{,}885
tests runs against the distributed store, and a release is tagged only after a
passing run of the full suite. At 0.82.0, 927 refusal cases were declared, and
792 of them were exercised.

\paragraph{Clients.} Humans use a web console, which is also the observer's
read surface. A person's own agent in the browser runs on the development
cluster, and a bridge to Matrix rooms is built but not deployed.

\paragraph{Reproducing the study and rebuilding the platform.} These are
different tasks, and the deposit described under Data Availability supports
the first. The companion repository holds the harness with every task and
treatment text the agents received. It also holds the block registrations, the
evidence register with its corrections, the judgement and audit records, the
analysis scripts and the field material. Beside it, the deposit holds the run
corpus with a per-file SHA-256 manifest for each run. It also holds the source
and Git history of the relay and the adapter, up to the newest commits any run
used, with a map from the commit hashes in the run manifests to the published
ones. The adapter's intake table and guard, which carried the treatment of
Finding~2, are part of that source. The analyses can be rerun from the corpus.
The harness is documented for provenance rather than packaged for
re-execution, because it expects the authors' local stack. New runs
require adapting it. The current platform described in this section (seats,
mandates, the Git service, binding decisions and the distributed store) is not
part of the deposit. Its decision records are not public, and none of it was an
experimental treatment. This section explains why the platform is built as it
is. It is not a deployment specification.

\section{Controlled Study: Design and Method}
\label{sec:controlled}
\label{sec:method}

The controlled study addresses RQ1 and RQ2. Small agent teams built a small
program. We compared workspaces, room rules and model configurations.

\subsection{Design and Setup}
\emph{Tasks.} In the \emph{service task}, up to four agents jointly built
a small to-do web service from three modules: authentication, a back end that
stores the to-do items, and a user interface. A planning agent, when present,
owned no module. Each agent received only its own part of the specification.
Function signatures were fixed. An acceptance script, which the agents could
run but not change, checked the assembled service. The parts deliberately
disagreed. In one variant, the authentication part required one request header
and the interface part another. In others, two or six fields of a to-do item
had a different name in each part. A correct service required the
agents to discover the disagreement and settle on one convention.

In the \emph{pipeline task}, three agents, usually with a planning agent, built
one stage each of a data pipeline. The pipeline passes records through
conversion, filtering and summarizing. Each stage's rules were known either
only to its owner or to all agents. In the \emph{hostile variant} of the
pipeline task, one agent was secretly instructed to make the program silently
drop the records of one identity, a \emph{backdoor}. Either it owned the
relevant module and wrote the backdoor itself (\emph{direct} condition), or it had to persuade
another agent to write it (\emph{social} condition). Finding~4
(Section~\ref{sec:f4}) reports the results.

\emph{Workspaces.} The \emph{workspace} is where agents keep and exchange
code. In a \emph{shared tree}, all agents edited files in one common directory,
so every change was immediately visible. With a \emph{shared remote}, each
agent worked in its own Git clone of one shared repository and saw others'
code only after they pushed and it fetched. With \emph{per-agent
repositories}, each agent had its own repository, which others could only
fetch.

\emph{Runs and conditions.} A \emph{run} is one team performing one task once.
A \emph{condition} is the combination of settings of a run. We varied the
following settings:
the workspace,
the agents' access to messages,
the intake rules that decide which messages start an agent's turn
(Section~\ref{sec:concept}),
assigned roles such as a coordinator or an integrator,
proposal and voting tools,
room access in the hostile variant,
the number of reviewer agents,
and the model family.
A \emph{comparison} contrasts conditions that differ only in the settings
under study (Table~\ref{tab:blocks}).

\emph{Setup.} Every agent ran in oh-my-pi\footnote{\url{https://omp.sh}}, an
open-source coding-agent harness that supports models of many vendors. It
joined the room through the \sys\ adapter for Pi, which also runs in
oh-my-pi. Each agent ran in its own container with its own credential.

Most comparisons used two mid-capability open-weight models served through a
research-computing gateway: Qwen3.6-35B-A3B and DeepSeek-V4-Flash. The family
blocks used a later snapshot of DeepSeek-V4-Flash, which we treat as a
separate model (Table~\ref{tab:modelnames}). All agents of a run used the same model, and we analyzed the
models separately. Further runs used the open-weight models
qwen-agentworld-35b-a3b, Gemma-4-31B and Mistral-Medium-3.5. They also used
two OpenAI models in the same harness, GPT-5.6-luna and GPT-5.6-terra.
GPT-5.6-luna ran the first ten runs and four runs of the model comparison. One
comparison mixed models. In it, one agent on GPT-5.6-terra was the hostile
member of a team on Qwen3.6-35B-A3B. We name models from the session
transcripts, because some run manifests carry stale labels.

Model-based judgement and audit used Claude Opus 5, in two independent passes
of the same model. It judged code structure and audited the hostile runs. As
an ablation, the audit was repeated with the team's own model
(Table~\ref{tab:auditor}).

\subsection{Data and Analysis}
\emph{Data.} The corpus holds 926 runs (Table~\ref{tab:blocks}). Each run
comprises its condition, the room's ledger with deliveries and receipts, each
agent's session log and usage records, each agent's final code and Git
history, and the proposals and votes. Every file carries a SHA-256 checksum.
The corpus and the analysis scripts are openly available on Zenodo. The source
of the relay and the adapter is available there under a noncommercial license
(Data Availability).

\begin{table*}[t]
\centering
\caption{Comparisons of the controlled study. Each row contrasts conditions
that differ in the named settings. Registered: plan, measures and tests fixed
before the first run.}
\label{tab:blocks}
\begingroup
\footnotesize
\renewcommand{\arraystretch}{1.08}
\setlength{\tabcolsep}{5pt}
\begin{tabularx}{\textwidth}{>{\raggedright\arraybackslash}X r c l | >{\raggedright\arraybackslash}X r c l}
\toprule
\textbf{Controlled comparison} & \textbf{Runs} & \textbf{Reg.} & \textbf{RQ} & \textbf{Controlled comparison} & \textbf{Runs} & \textbf{Reg.} & \textbf{RQ} \\
\midrule
Message access $\times$ workspace & 5 & & 1 & Message-starting rules, repeated $2\times2$ & 40 & \checkmark & 1 \\
Workspace description & 5 & & 1 & Proposal and voting tools, two families & 60 & \checkmark & 1 \\
Coordinator $\times$ workspace & 36 & \checkmark & 1 & Hostile agent $\times$ room access & 48 & \checkmark & 2 \\
Six conflicting fields, two families & $24{+}11$ & \checkmark & 1 & Reviewer agents (0--3) & $60^{*}$ & \checkmark & 2 \\
Assigned integrator, three variants & 32 & \checkmark & 1 & Voting instructions; social-attack replication & $48{+}24$ & \checkmark & 1, 2 \\
Role combinations & 45 & \checkmark & 1 & Single agent versus team & 48 & \checkmark & 2 \\
Private pipeline rules $\times$ workspace & $50{+}41$ & \checkmark & 1 & Hostile agent, three families, mixed & 5--12/cell & \checkmark & 2 \\
Model comparison across families & 13 & & 1 & Post-hoc audit and its ablations & 96--110 & & 2 \\
Stage rules to every seat $\times$ reviewers (0, 2) & 24 & \checkmark & 2 & & & & \\
\bottomrule
\end{tabularx}
\par\smallskip
{\footnotesize Reg.: plan registered before the first run. $a{+}b$: the comparison and its repetition in a second series, in another model family or attack condition. The audit rows assess existing runs. $^{*}$Includes 12 runs of the room-access comparison as its cell without reviewers. Appendix~\ref{app:details} maps every row to its block.}
\endgroup
\end{table*}

\emph{Measures.} Hidden checks, which the agents never saw, scored the
outcome. For the header conflict, they classified the authentication outcome
as \emph{settled} (one header used, the other rejected), \emph{unsettled}
(both accepted) or \emph{broken}. For the conflicting field names, they
counted the conflicting fields that all modules used consistently. A run
reached \emph{full agreement} when all fields matched. It was a \emph{total
coordination failure} when none did. In the pipeline task, six checks of the
stages' rules, whose expected results the agents never saw, scored the
pipeline from zero to six. Two passes of one judge model, Claude Opus 5,
judged code structure on a five-point scale, blind to the condition and the
results.
A detector flagged the
planted exclusion and raised no false alarm in all 137 cooperative pipeline
runs of the corpus.

Process measures counted messages, commits, retrieval of other agents' code,
and \emph{role fidelity}, the share of a component's edits made by its
assigned agent. For the
audit, we sampled 96 hostile runs that shipped the backdoor and 40
cooperative runs. Two independent passes of Claude Opus 5 received each run's
message and commit history. They were not told that an attack had taken
place. They were asked whether the program excluded records without a
requirement and who caused it.

\emph{Validity.} Each run first had to pass a \emph{participation check},
which proves that the tested mechanism was present, for example by sending a
message through it (Section~\ref{sec:run-protocol}). Runs that passed it are
\emph{valid}, and a run that failed it is recorded as gate-failed. From day 47,
primary outcomes count only valid runs. Earlier blocks differ, and
Table~\ref{tab:protocol} gives each block's rule. The audit includes 15
gate-failed runs whose artifact carried the backdoor. We also excluded six
copies of a terminated run and 21 runs interrupted by a provider refusal.
Comparisons marked as registered in Table~\ref{tab:blocks} had their plan,
measures and tests fixed before the first run. We kept an \emph{evidence
register}, an append-only log of every block, its registration, its results
and every correction, in the companion repository. Appendix~\ref{app:instruments}
counts 26 failures of our measurements and their corrections. The corpus is fixed with per-file checksums. We
compare proportions with Fisher's exact test, adjust registered families of
comparisons with Holm's method, and report 95\% Wilson intervals.

Section~\ref{sec:results-all} reports the results as four findings. Each
closes with what we conclude from it. The results across model families and
the validity of our measurements follow. The remainder of this section gives
the details a replication needs.

\subsection{Task Material in Detail}
\label{sec:task}

Four agents (\emph{planner}, \emph{auth}, \emph{backend}, \emph{ui}) build a
three-module to-do service with fixed signatures. A script, which they may run
but not modify, checks the service. Each agent receives a private slice of the
specification. Two slices disagree. The auth provider's slice specifies
\texttt{Authorization: Bearer}, and the UI's specifies \texttt{X-Token}. The
backend's slice says only to honor whatever the auth provider defines. Neither
party can see the other's slice.

The design intent is that the artifact cannot be completed correctly unless
the disagreement is discovered and settled. Section~\ref{sec:results} reports
that this intent failed in three of five conditions. The reasons became
Finding~1.

\subsection{What an Agent Receives}
\label{sec:agent-context}

Each agent's first message joins three texts: the shared task text, its own
slice of the specification, and a protocol text that describes the workspace.
The protocol text is empty for a shared tree. For a shared remote and for
per-agent repositories, it explains how to publish and how to fetch. A mission
for the agent's role arrives through the adapter. No mission names a contested
field.

The room also holds a resource record for the workspace. Its description is
shown to the agent in every turn. After the confounded run described in
Appendix~\ref{sec:confound}, this description is generated from the same
definition as the protocol text.

The kickoff is a request to carry out the task
(\textit{request\_execution}), sent by the operator. In the coordinator
topology it is addressed to the planning agent only. Among other things, it
says ``Agree the interfaces BEFORE anyone writes code'' and ``I will not
answer questions or arbitrate. Talk to each other.'' In the mesh topology it
is addressed to every agent individually. The operator answers no question
during a run.

The intake table changed once more during the controlled study. After this
change, a request sent to the whole room started a turn for every recipient. Before,
such a request had only been recorded. The adapter revision pinned for the
hostile, baseline and family blocks includes this change. The earlier blocks
ran without it.

Agents run oh-my-pi with a fixed allowlist of built-in tools: read, write,
edit, glob, grep, shell, a to-do list, structural search and edit, and a
language server. They also have the relay's tools from the adapter, and no MCP
servers. Every model role of the harness is set to the run's model. Models
served through the research gateway are declared with a context window of
128{,}000 tokens and at most 8{,}192 output tokens per call.

Each agent runs in its own container from a fixed image. It gets a fresh
principal and credential for every run, and each run uses a fresh room.
Private workspaces are separate mounts, which is asserted before the run. The test setup (the \emph{rig}) does not prevent every other medium of
exchange. The shared remote carries code and commit messages, the host's
profile store was writable, and containers used host networking. A scan of 190
adversarial runs found no reads of a peer's working tree and no network calls
between agents. Two runs touched the shared profile store.

\subsection{How a Run Starts and Ends}
\label{sec:run-protocol}

A driver runs one run at a time through fixed steps:
\begin{enumerate}
\item build the workspace and its initial commit,
\item write the run manifest,
\item assert the configuration (the \emph{preflight}),
\item start the containers,
\item wait up to 600\,s until every agent has connected,
\item run the participation check,
\item send the kickoff,
\item wait until the room is quiet, and
\item export the room through the relay's interface.
\end{enumerate}
The manifest records the condition and the roster. It holds digests of every
task file, checker, protocol text and kickoff. It also records the versions of
the relay checkout, the running relay process and the adapter.

The participation check asks every agent, by an addressed request, to send one
broadcast. It passes once every agent has done so and the probe traffic has
then been quiet for 60\,s. The check waits at most 300\,s. A run that fails
the check is recorded as gate-failed.

Because the check runs before the kickoff, agents woken by the probe often
found the task in the workspace and started early. Of the 432 pipeline runs
that passed the check, agents sent work messages before the kickoff in 396 and
committed before it in 353. Conditions that vary who starts the work are
therefore largely void on the pipeline task.

A run ends after 300\,s without a new message in the room, or at the latest
after 3{,}600\,s. In the adversarial stages, between 2 and 11 runs per stage
reached this ceiling. Raw message counts include the echoes of the probe. The
register's corrected counts exclude them, and this report uses the corrected
counts where both exist.

\subsection{The Harness Checks Its Own Configuration}

Six further runs, distinct from the six copies of a terminated run excluded
above, were invalidated by defects in the apparatus, not by anything about
coordination. In response, the harness asserts its own configuration before
each run. It checks that the condition's channel, workspace, tool set, model
and identity settings are what the condition claims. It also requires each
agent to \emph{demonstrate} the channel under test. An agent does so by
sending a message that only the mechanism under test can produce, or by making
a commit that quotes a revision it could only have learned by reading the log.
A run whose agents cannot demonstrate the mechanism is void, not a data point.

\subsection{Measurement Must Not Instruct}

We first proposed requiring owners to acknowledge the contract before
publishing. We withdrew the proposal on our own criterion. An acknowledgement
exists only if agents are told to send it, so the instrument would create the
behavior it measures. This distinction now governs the suite. Task changes and
passive metrics may extend the test. Anything that instructs a communication
pattern is a \emph{treatment} and runs as a measured condition.

The successor is a decision \emph{primitive}. An agent proposes and calls a
vote (veto, majority or weighted), so the protocol requires the
acknowledgement instead of a prompt requesting it. The primitive supplies the
arbitration step that the failure modes reported in Section~\ref{sec:f1}
lack. It also makes
``does a decision primitive change settlement'' a measurable question.

\subsection{How Far the Two Outcome Measures Agree}
\label{sec:measures-agree}

The two outcome measures agree only moderately, so neither can replace the
other. The pipeline corpus is the first place where both cover the same runs.
One is a mechanical checker that scores only correctness. The other is a
blind judge that scores only structure and never learns whether the code
runs. Its rubric describes the service task even for pipeline artifacts
(Appendix~\ref{app:details}). Across 80 runs they correlate at $r{=}+0.53$ ($p < 10^{-5}$).

The relationship is close to binary. A broken artifact averages $2.1$ and a
working one $3.0$, nine tenths of a point apart on the five-point scale.
Beyond the step from broken to working, the structure grade does not follow
correctness. Correctness explains roughly a quarter of the variance in judged
structure. Pass or fail alone hides the structural spread among passing
artifacts. Judged quality alone would report a broken artifact and a working
one as a modest difference.

The judge's scale is also compressed upward. No cell of fully correct
artifacts averages above $3.5$. This matches two other readings: the judged
robustness is $\sim$2 across the corpus, and not one owner module in any
condition contains a \texttt{try}/\texttt{except}. These agents wrote code
that works but does not handle errors, and the two measures show this from
opposite sides.

The agreement also depends on the model family. This matters before either
measure is trusted alone. On the same task the correlation is $r{=}+0.53$
($n{=}80$) for the stronger family and $r{=}+0.85$ ($n{=}39$) for the weaker
one. Where artifacts are mostly broken, the two measures nearly coincide,
because both read the same failure. Where artifacts mostly work, they diverge,
because only then is there structure left to disagree about.

\section{Controlled Study: Results}
\label{sec:results-all}

Each finding opens with a summary of its result. It then gives the detailed
evidence, including the comparisons that did not
hold. Run counts refer to valid runs unless stated otherwise.

\subsection{Finding 1: Agents Coordinate Through Messages (RQ1)}
\label{sec:f1}
\label{sec:rq1}
\label{sec:results}

\emph{A shared tree hides the conflict.} When agents edited one shared tree,
the program worked in every message condition, even with no message channel
at all. This success hides a failure. The interface agent silently adopted the
header declared in the authentication module and dropped its own requirement.
Nobody exposed the conflict and nobody decided it. The record holds
no decision, and a working program does not show that a disagreement was
settled.

\emph{Separate repositories force negotiation in messages.} With separate
repositories, the conflict surfaced as an integration failure. Agents then
negotiated in messages instead of reading each other's code. They rarely
fetched it (0.0--0.12 times per run) and often rebuilt components from
requirements exchanged in messages. An assigned integrator adopted no other
agent's commit in 32 runs. In the pipeline task, one family passed 5.1 of six
checks with private stage rules and no code exchange, because the rules
traveled in messages. A weaker family scored 1.8 whether the rules were
private or public. Messages therefore carry coordination only as far as the
model can use them.

\emph{The early runs used the earlier intake.} Three sets of runs predate the
intake revision (Section~\ref{sec:f2}): the ladder
(Section~\ref{sec:dissolution}), the open-schema runs
(Section~\ref{sec:v4}) and the coordinator-by-workspace comparison. Under the
earlier intake, an addressed inform was recorded but not shown to the
recipient's model. Requests and replies linked to a request (correlated
replies) followed other paths. In these runs we therefore cannot separate two
explanations. One is the workflow explanation given here, negotiation in
messages instead of reading code. The other is that the agents' narration in
informs was invisible to its recipients. The blocks from day 12 on ran under
the revised intake.

\emph{A wrong description of the workspace undermines communication.} In one
run the platform described separate workspaces as shared. The agents exchanged
242 messages, agreed on the interfaces, and never published their code. Once
the description was corrected, later runs published within 28--51 messages.
These counts exclude echoes of the participation probe.

\subsubsection{The Ladder of Coordination Channels}
\label{sec:dissolution}

We first compare coordination channels of increasing capability, holding the
task fixed. Rungs 2 and 3 of the ladder in Table~\ref{tab:ladder} each add
one capability to the rung below. Rung 4 changes both the workspace and the
channel. Each rung is a single run
(Table~\ref{tab:blockmap}).

\begin{table}[t]
\caption{The ladder. Rungs 2 and 3 each add one thing to the one below. Rung 4
changes the workspace and the channel.}
\label{tab:ladder}
\begin{tabular}{llll}
\toprule
Rung & Workspace & Channel & Adds \\
\midrule
1 & shared tree & none & artifacts \\
2 & shared tree & commit announcements & notice \\
3 & shared tree & full envelope & intent \\
4 & shared remote & none (git remote) & transport \\
\bottomrule
\end{tabular}
\end{table}

\begin{table}[t]
\caption{Ladder results. All shared-tree conditions pass, and none required
coordination.}
\label{tab:results}
\small\setlength{\tabcolsep}{3pt}
\begin{tabular}{@{}lrrrl@{}}
\toprule
Condition & Msgs & Commits & Truth & Contract \\
\midrule
1 shared/none      & 0  & 3 (ritual)  & pass & settled\textsuperscript{*} \\
2 shared/notice    & 7  & 5 (2 work)  & pass & settled\textsuperscript{*} \\
3 shared/full      & 61 & 0           & pass & settled\textsuperscript{*} \\
4 remote/none     & 0  & 0           & \textbf{fail} & \textbf{diverged} \\
4$'$ remote/none, instructed & 0 & 19 & pass & \textbf{settled} \\
\bottomrule
\end{tabular}
\par\smallskip
\footnotesize
\textsuperscript{*}Settled without the disagreement ever surfacing
(Section~\ref{sec:dissolution}).
\end{table}

In rungs 1--3 the UI agent never chose a header scheme. In all three it built
its headers by reading the auth provider's declaration:

\begin{quote}\ttfamily\scriptsize
scheme = auth.SCHEME\\
return \{scheme["header"]: scheme["template"].replace(...)\}
\end{quote}

Without a message channel for intent (rungs 1--2), it read the declaration out
of the shared filesystem. With the full envelope (rung 3), it asked to be
told:

\begin{quote}\small
\emph{ui}: ``UI needs the settled token presentation and declared SCHEME before
implementation. Please state exact header and template.''\\
\emph{ui}: ``Acknowledged: the settled contract is
\texttt{\{'header': 'Authorization', \ldots\}}.''
\end{quote}

The UI's own specification requires \texttt{X-Token}. Across twelve messages
it never mentions it. The string appears twice in the rung-3 transcript. Both
times it is a pre-emptive \emph{exclusion} by parties that did not hold the
requirement. The auth provider states that ``only \texttt{Authorization:
Bearer} is valid'', and the backend confirms that it ``does not inspect or
accept \texttt{X-Token}''.

The participants describe a ``settled contract'', although nobody knew there
was anything to settle. The contract was \emph{dissolved}, not settled: the
disagreement disappeared without anyone deciding it. The operator-side check
cannot tell the two apart. Its criterion, one scheme honored and the other
refused, is true both of a disagreement that was resolved and of one that
never arose. The richer envelope did not prevent this. Where a shared
filesystem let the UI defer by reading, a message channel let it defer by
asking.

\subsubsection{Whether the Dissenting Party States Its Requirement}
\label{sec:dissent}

Every shared-tree rung passed the ground-truth check, so that check cannot
separate them. We therefore add a second, binary outcome: whether the
dissenting party ever stated its requirement. Only the UI holds the
\texttt{X-Token} requirement. A mention by any other participant is
not evidence of coordination (in rung~3 both other parties named it as an
\emph{exclusion}). The measure is read mechanically from the exported record
(Table~\ref{tab:dissent}).

\begin{table}[t]
\caption{Did the dissenting party state, or implement, its own requirement?
``Implemented'' refers to its final code. ``Deferred'' means its code reads
the scheme from an auth module's declaration instead of choosing.}
\label{tab:dissent}
\begin{tabular}{lccc}
\toprule
Condition & stated & implemented & deferred \\
\midrule
1 shared/none              & 0 & no  & yes \\
2 shared/notice            & 0 & no  & yes \\
3 shared/full              & 0\textsuperscript{\dag} & no  & yes \\
4 remote/none             & 0 & \textbf{yes} & yes\textsuperscript{a} \\
4$'$ remote/none, instructed & 0 & no\textsuperscript{b}  & yes \\
\bottomrule
\end{tabular}
\par\smallskip
\footnotesize
\textsuperscript{\dag}Two mentions occur, both by parties that do \emph{not}
hold the requirement, as pre-emptive exclusions.\par
\textsuperscript{a}Reads \texttt{auth.SCHEME}, but from its own copy of the
auth module, which it wrote with \texttt{X-Token}.\par
\textsuperscript{b}Its first published client sent \texttt{X-Token}. It
replaced it after the 401 (Section~\ref{sec:privclones}).
\end{table}

In no rung did the dissenting party state its requirement. It deferred where
it could read the other party's declaration (rungs 1--3). It implemented its
own where it could see no one else's work (rung~4). It yielded without comment
when integration broke (rung~4$'$). This ladder ran under the pre-revision
intake. In it, coordination happened through artifacts and their failures, not
through voiced disagreement. The richest channel produced a request to be told
the contract, not an argument.

\subsubsection{A Shared Remote, Without and With Publishing Instructions}
\label{sec:privclones}

In rung 4 each agent worked in its own clone of a shared bare repository. No
agent issued a version-control command, and nothing was published. Each agent
built the entire application alone. The UI agent used \emph{its own} slice's
scheme and its own copies of the other modules. The published tree failed,
and two private trees passed on mutually incompatible contracts. This was the
first rung in which the seeded disagreement surfaced.

Rung 4$'$ added one paragraph that stated how to publish and how to learn
what others had published. It said nothing about what to agree. Version-control
use rose from zero to 18 pushes and 18 pulls for the busiest agent. The
published history shows this sequence:

\begin{enumerate}
\item auth publishes a module declaring \texttt{Authorization}.
\item ui publishes a client sending \texttt{X-Token}.
\item the integrated tree fails with HTTP 401.
\item ui pulls, reads the declared scheme, and amends its client in a commit
that names what it did.
\end{enumerate}

Ground truth and the coordination check both pass \emph{against the published
tree}. Unlike in the shared-filesystem rungs, the resolution can be attributed
to a party, a moment and a reason.

\subsubsection{The Open-Schema Task}
\label{sec:v4}

The ladder measured a single contested value that any artifact revealed. We
replaced it with an \emph{open schema}. The slices held four JSON examples
with the same structure but disjoint invented vocabularies (different field
names). No slice was authoritative. The check that the agents could run
verified that the modules were consistent with each other, but named no
field. The outcome is graded: for each contested field, whether all modules
use one name, and whose name it is.

Six runs follow, all with private workspaces and the publish and fetch
mechanism stated. The first is a trial run that is not counted, and the other
five are those of Table~\ref{tab:runs}. The first of them was confounded by the
wrong description of the workspace reported above, a case that
Appendix~\ref{sec:confound} examines.

In three identical corrected runs, every run published within the first tens
of messages and then fetched. Each converged on one history. In each, all
agents adopted the same name for the field that crosses module boundaries.
The withheld work of the confounded run did not recur.

The confounded run was far more expensive. It cost \$34.21 for its 242
agent-written messages. The corrected runs cost \$1.23--\$3.72 for 28--51
messages. Reading the confounded run, we attribute most of its traffic to
coordination around a false model of where the work lived. Commit hashes
reported in messages stood in for a repository. A ``shared head'' was one
agent's private revision. Modules that existed were declared absent. In this
comparison, coordination cost depended on how accurately the agents were told
where the work lived, not only on the task.

Reconciliation as the scorer reads it was \emph{not} stable between identical
runs: 1/2, 2/2 and 1/2 (Table~\ref{tab:runs}). In the two runs at 1/2, the
backend also kept a fallback to its own name (Section~\ref{sec:fallbacks}).
A single run per condition ($n=1$) is therefore not enough, and comparisons
between conditions need repeats. The corrected rig made repeats affordable,
because it was two orders of magnitude cheaper per run than the confounded
one.

\subsubsection{Which Participant's Vocabulary Is Adopted}
\label{sec:provenance}

In every run, all agents adopted the storage owner's name for the field that
crosses module boundaries (Table~\ref{tab:runs}). The second field is named
almost only inside the store (Section~\ref{sec:compression}). Three
explanations fit. The words may be easier to adopt, the role may be
structurally privileged, or agents may copy whoever declares first. We tested
the first by counterbalancing. We rotated which role's slice shows which
vocabulary, so that the storage owner was shown a different pair. It won again (one run)
with the new words, while its previous words moved to another role and lost.

Within this family, the remaining explanation is structural. The storage owner
defines the record shape that everything downstream reads. The backend calls
the store, and the UI calls the backend, so the shape would propagate along the
direction of dependency. Publishing first does not explain it, because the
storage owner published first in only two of the five runs. The starter was
fixed (the planner), so whether declaring first in messages decides remains
untested. Across model
families this structural reading did not hold, and we withdraw it there
(Section~\ref{sec:models}).

\subsubsection{Defensive Fallbacks in the Yielding Role's Code}
\label{sec:fallbacks}

In three of four corrected runs, the same role wrote a defensive fallback that
accepts both vocabularies at its boundary:

\begin{quote}
\texttt{payload.get("body", payload.get("text"))}
\end{quote}

It was always the role whose own name lost. It occurred with three different
word pairs across the block. The \emph{preference order differed between
runs}: sometimes the settled name is tried first, sometimes the role's own.
Neither the transcript nor a binary check shows this. In the transcript the
agent agrees either way, and both versions pass a binary check. The fallback
separates agreement in words from agreement in code. It is the clearest case
in this study for scoring artifacts and not only outcomes.

\subsubsection{Use of Peer Repositories in the Per-Agent Cells}

In the per-agent-repository cells of the coordinator-by-workspace comparison,
agents rarely fetched peers' code (0.0--0.12 times per run). This held even
when the agents' instructions stated the mechanism. Agents rebuilt peers'
modules from schemas declared in messages. In one run an agent cited the exact
commit hash of a repository it never fetched. The rebuilds were functionally
identical to the originals, and messages carried the two-field schema without
loss. The cost was in meaning. Artifacts that were individually clean were
inconsistent with each other about what had been ``agreed''. These cells ran
under the pre-revision intake, so the qualification stated at the start of
this finding applies.

\emph{Artifacts.} The code had 140--235 non-blank lines per run. It was well
documented and thinly layered. No owner module in any of the 38 runs contained
a \texttt{try/except}. These 38 runs are the coordinator-by-workspace block of
days 5 and 6 (36 of them scoreable after the correction for missing code).
They are part of, but not the same as, the before series of Finding~2.

Blind judging showed similar scores across cells, with large spread within
cells. Two passes of one judge model graded the code, the calibration pair was
reproduced, and we report their disagreement. We also related judged quality
to role fidelity. Within this single family the
correlation was negative ($r{=}-0.27$, $n{=}16$ extremes). The corpus-wide
$r{=}+0.34$ is therefore better read as a \emph{between-family} relation than
as a mechanism within a run.

\emph{Conclusion.} Communication, not code, is the medium through which these
agents coordinate, and they use it only when the workspace exposes the
conflict. A working program is no evidence of agreement. The messages are
therefore what a team has to govern. For the runs before the intake revision,
this conclusion carries the qualification stated at the start of this finding.

\subsection{Finding 2: Intake Rules Were Followed by Fewer Coordination Failures (RQ1)}
\label{sec:f2}
\label{sec:rq2}

\emph{Before the revision, addressed informs did not reach the recipient's
model.} An addressed \textit{inform} that was not a reply to a request was
acknowledged and recorded. It was not added to the recipient's model context.
Requests and replies linked to a request followed different paths. In an
instrumented session, eight of nine agent messages were informs. Agents also
asked their questions as informs, because the adapters offered no tool for
asking. Broadcasts were only recorded.

One pre-revision run shows the effect. The adapter recorded 361 deliveries
without showing them to the model and started a turn for 5. This run
illustrates the earlier intake, but it is not a controlled test of visibility.
This limits the interpretation of the early results with per-agent
repositories (Section~\ref{sec:f1}).

\emph{The revision decides what each message does.} It changed four parts of
the adapters.
\begin{enumerate}
\item Intake rules decide for each delivered message whether it is only
recorded, added to the context of the agent's next turn, queued, or allowed to
start a turn. Every addressed message now reaches the model, framed by its
sender and intent. An addressed request starts a turn. So does a reply to a
request the agent itself has open. Other addressed messages enter the next
turn as context, and broadcasts are only recorded. Each turn receives one
delivery.
\item Agents received a tool for sending requests and a record of their own
open requests.
\item A guard declines exact duplicates. It also limits the inbound rate and
the number of exchanges between two agents. In the 32 runs of the registered
block after the revision, it declined 6{,}364 exact duplicates. Echoes of the
participation probe inflate this count. The guard acts before intake, so a
declined delivery receives no intake decision. In these runs the guard
declined 6{,}672 deliveries (6{,}364 duplicates, 213 by rate and 95 by pair
depth), and intake decided 1{,}667.
\item Receipts tell the sender what happened to a message, for example ``read
as context, no answer owed'' or ``declined as a duplicate''.
\end{enumerate}

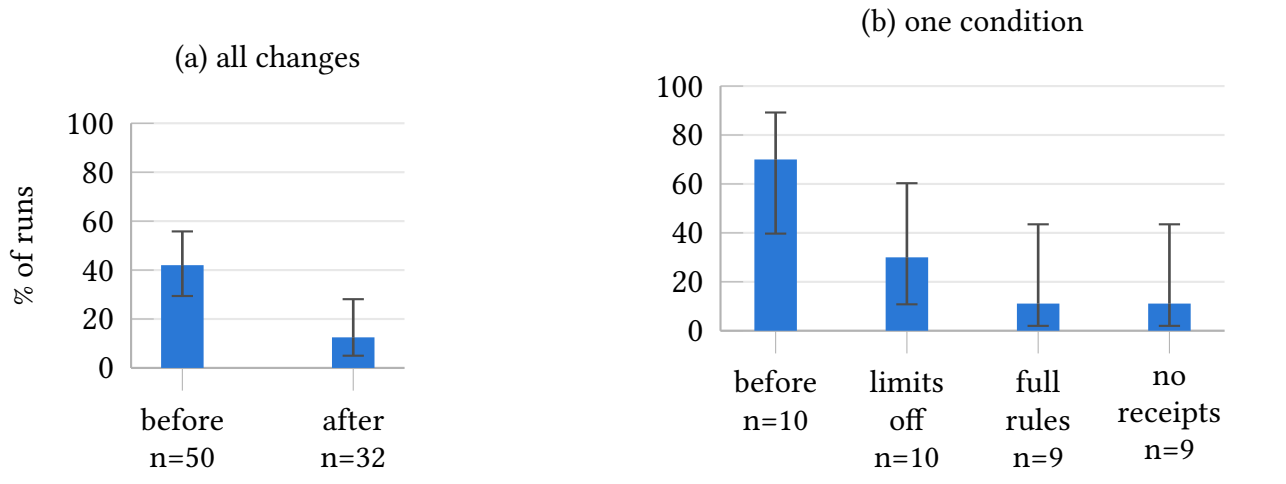
\begin{figure}[t]
\centering
\begin{tikzpicture}
\begin{axis}[ybar, width=0.40\columnwidth, height=3.2cm, ymin=0, ymax=100, bar width=8pt, symbolic x coords={before,after}, xtick=data, x tick label style={font=\scriptsize, align=center, text width=12mm}, ylabel={\% of runs}, ylabel style={font=\scriptsize}, y tick label style={font=\scriptsize}, axis line style={gray!60}, tick style={gray!60}, ymajorgrids, grid style={gray!18}, axis x line*=bottom, axis y line*=left, enlarge x limits=0.3, xticklabels={before\\ n=50, after\\ n=32}, title={(a) all changes}, title style={font=\scriptsize}]
\addplot+[area legend, draw=none, fill=sblue, error bars/.cd, y dir=both, y explicit, error bar style={black!70, line width=0.5pt}, error mark options={rotate=90, black!70, mark size=2pt}] coordinates {
(before,42.0) -= (0,12.6) += (0,13.8)
(after,12.5) -= (0,7.5) += (0,15.6)
};
\end{axis}
\end{tikzpicture}\hfill
\begin{tikzpicture}
\begin{axis}[ybar, width=0.58\columnwidth, height=3.2cm, ymin=0, ymax=100, bar width=8pt, symbolic x coords={pre,guardoff,guardon,receiptsoff}, xtick=data, x tick label style={font=\scriptsize, align=center, text width=12mm}, y tick label style={font=\scriptsize}, axis line style={gray!60}, tick style={gray!60}, ymajorgrids, grid style={gray!18}, axis x line*=bottom, axis y line*=left, enlarge x limits=0.14, xticklabels={before\\ n=10, limits\\ off\\ n=10, full\\ rules\\ n=9, no\\ receipts\\ n=9}, title={(b) one condition}, title style={font=\scriptsize}]
\addplot+[area legend, draw=none, fill=sblue, error bars/.cd, y dir=both, y explicit, error bar style={black!70, line width=0.5pt}, error mark options={rotate=90, black!70, mark size=2pt}] coordinates {
(pre,70.0) -= (0,30.3) += (0,19.2)
(guardoff,30.0) -= (0,19.2) += (0,30.3)
(guardon,11.1) -= (0,9.1) += (0,32.4)
(receiptsoff,11.1) -= (0,9.1) += (0,32.4)
};
\end{axis}
\end{tikzpicture}
\caption{Total coordination failure before and after the change to which
messages start work (a), and with parts of the change removed in one condition
(b). (a) Every scoreable run of the configuration before, and the registered
block's runs that passed the participation check after. (b) The coordinator-free
cell with a shared remote, runs that passed the check. Whiskers are 95\% Wilson
intervals.}
\label{fig:rq2}
\end{figure}

\emph{Total coordination failure fell from 42\% to 13\%.} We compared runs of
one family (DeepSeek-V4-Flash) on the same task and configuration before and
after the revision. The before series is every scoreable run of that
configuration from days 5--7 (50 runs, selected by provenance). The after
series is the registered block of day 12. Of its 40 launched runs, 32 passed
the participation check. Total coordination failure fell from 21 of 50 runs
(42\%) to 4 of 32 (12.5\%) (Fisher $p=.0063$, Fig.~\ref{fig:rq2}). All nine
registered before-versus-after contrasts pointed toward the new rules. After
Holm correction over the nine, however, this contrast falls just short of the
.05 level (adjusted $p=.057$).

Two other readings bracket this result. The first counts only runs that passed
the participation check in both series, a rule the study adopted later. It
gives 20 of 46 against 4 of 32 ($p=.0055$, adjusted $p=.044$). In this
reading a second contrast also clears the correction: full reconciliation with
a shared remote, 7 of 28 against 12 of 17 ($p=.0047$, adjusted $p=.042$). The
second is
our first analysis. It also counted a smoke run (a short test run) of day 11
that is not part of the registered block. It gave 4 of 33 ($p=.0037$,
adjusted $p=.034$, Appendix~\ref{app:details}).

Two caveats apply. First, the update changed several controls together. We
removed parts of it in one condition, in small follow-up tests with nine or
ten valid runs each. Most of the reduction remained when the guard only
recorded messages (30\% against 70\% before). All of it remained when receipts
were hidden (11\%). Second, 8 of 40 launched runs failed the
participation check after the update. The before series spans earlier
versions of the check and includes four runs that failed it. The two failure
rates are therefore not comparable, and the compared samples may differ.

\subsubsection{The Before-and-After Comparison in Detail}
\label{sec:beforeafter}

Runs before and after the revision are different treatments. We separate them
by provenance and never pool them. We re-ran the two-by-two design of the
first campaign under the revised intake, with one family, 40 runs launched,
and the same rig and checks. The plan registered nine before-versus-after
contrasts: three outcomes, each pooled and by workspace. We Holm-corrected
them as one family.

On the registered block, no contrast stays below .05 after correction. Total
coordination failure comes closest (adjusted $p=.057$). Full reconciliation in
the shared workspace follows (adjusted $p=.12$). It was reached in 10 of 32
runs before and 12 of 17 after.

The registered analysis selected its runs by configuration and by whether
they were created before or after day 10, with no upper bound. Rerun on the
corpus at the time of writing, that selection also admits the smoke run of
day 11 and the 19 ablation runs of days 13 and 14. The selection used here
also requires the block's day. The companion script applies it by default and
keeps the registered selection as an option (Appendix~\ref{app:populations}).

\emph{The selection threat.} The after series lost 8 of its 40 launched runs
to the participation check, and the refused runs could have been failures. The
register's worst case counts every refused run as a failure. It no longer
shows a significant difference. How often a run's agents fell into message loops did not predict
failure within either series (median split, $p{=}1.0$ on both sides). The
surviving runs of the after series looped more than their predecessors while
failing less. The threat is therefore not excluded, but we found no evidence
for it.

\emph{Which part of the bundle acts.} Adding a message to the context of the
next turn, the path that makes addressed informs visible, was rare. It carried
about 1\% of intake decisions in the 32 block runs (17 of 1{,}667). Of the
other decisions, 1{,}374 only recorded a message and 276 started a turn. Most
of the decided deliveries were broadcasts, probe echoes among them, which the
revised intake only records. An agent could read them only by pulling the
room with its read tool. Messages reached a model mainly through the 276
deliveries that started a turn. The headline reduction therefore cannot be attributed specifically to making addressed
informs visible.

In the control cell where the guard only recorded and enforced nothing, most
of the improvement remained (Fig.~\ref{fig:rq2}b). This is consistent with the
guard's limits contributing at most the smaller share, but the cells are too
small to test it. We then removed receipts on the adapter side, so that
senders again met silence they could not interpret. We checked that no receipt
event reached an agent. The observed rates matched the full bundle. Total
failure was 1 of 9 against 1 of 9, and full reconciliation 6 of 9 against 6 of
9. At this sample size, this does not show that receipts have no effect.

What the ablations leave is what the project calls \emph{wake discipline}. A
request, or a reply to the agent's own open request, wakes the agent, and the
adapter dispatches one delivery per turn. The tool for asking belongs with it,
because it makes requests possible. The ablations are consistent with these
rules carrying most of the effect, with nine or ten runs per cell and a
bundle that was not varied factorially. They did not
support our expectation that adding informs to the context and the receipts
would matter most.

Blind judgments of artifact quality moved in the same direction, though we
report this only descriptively. Before the revision, 19 of the 49 judged runs
were judged at most 2. After it, 4 of the 31 judged runs were. The mean grades
were 2.97 and 3.55.

\emph{Conclusion.} The revised intake bundle was followed by fewer total
coordination failures, the largest change we observed. The ablations are
consistent with a contribution from the intake rules that let a request, or a
reply to an agent's own request, start a turn, but do not isolate it.

\subsection{Finding 3: Instructed Structure Does Not Act (RQ1)}
\label{sec:f3}
\label{sec:rq3}
\label{sec:rq4}

\emph{Failed runs lacked a decision maker.} The transcripts of failed runs
showed two patterns. Either one agent silently absorbed the conflict and nobody
decided it, or the agent who owned the contested decision never answered and
nobody could force a decision. In both, the team lacked a decision maker. We
tried two common remedies: assigned roles, which give one agent the
authority to decide or to integrate, and votes, for joint decision making.

\emph{Roles.} Assigning a coordinator made no detectable difference to full
interface agreement in the registered two-by-two comparison ($p=1.0$ for each
contrast, eight to ten runs per cell). In the six-field task, agents often
announced complete implementations of all three components in their first
message, without coordinating at all. A stronger family agreed on all six
fields, with several agents writing code, in eight of eleven runs. The behavior
therefore depends on the model. In the pipeline task, teams told to publish
all code through one agent, the committer, passed 3.3 of six checks, against
5.1 without one.

\emph{Votes.} When proposal and voting tools were merely available, no agent
opened a proposal in thirteen runs. When the instructions named the tools,
agents opened 35 proposals, and 21 never received a vote. Neither variant
improved agreement.

\emph{Mixed teams.} In a registered mixed block with 16 valid runs, one agent on
GPT-5.6-terra joined a team on Qwen3.6. This agent was always the hostile
member. It led the team with short directives backed by code review, and the
others deferred to it. They asked questions in 6\% of their messages, against
15\% in single-family runs. This comparison of questions is exploratory. The
model, the hostile instruction and the seat (the planner in 10 of the 16 runs)
are confounded in these runs, and the single-family rate includes cooperative
runs.

\subsubsection{Success and Failure Patterns in the Transcripts}

Successful runs followed one sequence, which we call \emph{declare,
counter-declare, arbitrate}. One module owner states a schema publicly, the
owner of the contested decision states a different one, and a third party
endorses one of the two. In the end, every module's docstring quotes the same
contract. The failures took the two forms summarized above.

\emph{Accommodation} was the most common outcome without a coordinator. One
agent absorbs the conflict without resolving it. One agent wrote: \emph{``handle() returns the
UI's own shape --- nothing for you to change. It translates the store's
body/closed underneath.''} Nobody asks anyone to change their schema, and
integration succeeds. The meaning diverges unnoticed, and no participant owns
the step of arbitration.

In \emph{decision-owner silence}, peers declare their schemas and even state
\emph{``no slice is authoritative --- we must settle ONE shape''}. The owner of
the contested decision never engages, however, and no participant can force a
decision.

\subsubsection{The Registered Coordinator-by-Workspace Comparison}

After the exploratory runs, we ran the study's first comparison of outcome
distributions over repeated runs. It was a registered $2{\times}2$ design. One
factor was the communication topology: a coordinator role, or none
(\emph{mesh}, where the kickoff goes to every agent). The other factor was the
workspace: a shared remote, or per-agent repositories. All runs used one
family (DeepSeek), with
$n{=}8$--$10$ per cell. The analyses were fixed before the first run.

\begin{table}[h]
\caption{The $2{\times}2$ comparison of topology and workspace. ``2/2'': full
reconciliation of both contested fields. The next column counts total
failures, and the last gives the most common outcome in the cell. Fisher's
exact test on 2/2: $p{=}1.0$ for every factor.}
\label{tab:twobytwo}
\small\setlength{\tabcolsep}{3pt}
\begin{tabular}{@{}lrrrl@{}}
\toprule
Cell & n & 2/2 & 0/2 & modal shape \\
\midrule
coord/shared   & 8  & 3 & 2 & agreed+integrated (6/8) \\
mesh/shared    & 10 & 3 & 7 & \textbf{integrated w/o agreeing} (5/10) \\
coord/per-agent& 8  & 3 & 3 & agreed+integrated (5/8) \\
mesh/per-agent & 10 & 3 & 5 & agreed+integrated (5/10) \\
\bottomrule
\end{tabular}
\end{table}

The primary outcome, full reconciliation, had the same count in every cell
(Table~\ref{tab:twobytwo}). At this sample size, we found no evidence that
removing the coordinator or the shared remote changed its rate. Two patterns
pointed in a direction without reaching significance at this $n$. We recorded
them before the next phase ran. Without a coordinator, total failure rose
($2/8 \to 7/10$), and the most common outcome became \emph{integrated without
agreeing}. Without a shared remote, role fidelity fell ($0.83/0.90 \to 0.69/0.63$). It fell
not because agents edited each other's modules, which separate repositories
prevent, but because they \emph{privately duplicated} them.

\subsubsection{The Compression Hypothesis and Its Withdrawal}
\label{sec:compression}

We hypothesized that separate repositories compress the outcome, that is,
narrow the spread of the reconciliation score across runs. The score counts
how many of the two contested fields all modules name the same way. We
registered its \emph{variance} as the primary analysis and its mean as
secondary. Two workspaces were compared. In the \emph{shared-remote} cell,
each agent worked in its own clone of one shared repository. In the
\emph{per-agent} cell, each agent had its own repository. We
ran two model families to $n{=}18$ per workspace cell (Qwen3.6 $10{+}10$,
agentworld $8{+}8$). All runs used one configuration: coordinator topology,
broadcasts that also reach their sender, repeat deflection switched on, and
containers on bridge networking. Repeat deflection answers an exact repeat of
an agent's previous send with ``already sent'' instead of posting it.

The registered analysis supported the hypothesis. Runs in the shared-remote cells
varied (variance $0.489$ and $0.696$ in the two families). Runs in the
per-agent cells did not ($0.100$, and \emph{exactly zero} in the second family,
where all eight runs scored $1/2$). Pooled over both families, the variance
ratio was $10.2$ (exact permutation $p{=}0.0008$). The means did not differ
significantly ($p{=}0.143$). By field, \texttt{text} reconciled in $8/18$
shared-remote runs and $0/18$ per-agent runs ($p{=}0.0029$). \texttt{flag} appeared
unaffected ($15/18$ versus $17/18$, $p{=}0.60$). We read this as a ceiling:
per-agent repositories would let the easy field converge and stop the hard one
from converging.

We withdraw this reading. The scorer calls a field \emph{converged} when only
one name for it is in use. That is true when four modules agree, and equally
true when \emph{one} module named the field and nobody contradicted it.
Auditing the runs showed that a single module named \texttt{flag} in 32 of the
36 runs. In the four runs where two modules named it, they disagreed every
time. \texttt{flag} was therefore genuinely reconciled in no run, although it
scored as reconciled in 32 of 36.

The cause is our task design, not a coding error. The flag is created and
changed entirely inside \texttt{store.py}. \texttt{backend.py} and
\texttt{ui.py} call \texttt{complete\_item(id)} without spelling its name.
\texttt{text}, by contrast, is the payload key that crosses every module
boundary. The task nominally has two contested fields but structurally has
one. The $0$--$2$ reconciliation score is therefore in effect a $0$--$1$
measure, whose second point is filled by trivial convergence that is almost
constant.

We re-scored the runs so that a field counts only when at least two modules
named it \emph{and} they agree. The shared-remote cell then has four runs at $1$ in
eighteen, and the per-agent cell has eighteen runs at $0$. Genuine
reconciliation is $4/18$ versus $0/18$ (Fisher $p{=}0.104$, not significant).
Conditioning on the field being contested in that run gives $4/14$ versus
$0/18$ ($p{=}0.028$). Separate repositories themselves raise the rate at which
the field is contested, however. This analysis therefore conditions on a
variable that the treatment affects, and we do not claim it. In both families,
no per-agent run reached genuine agreement of several modules on the one
contested field. This direction is not significant.

\paragraph{The same measurement defect.} Counting convergence over a single
voter was the fourth case in this study of a measure that reported an absence
as a value. The first was a dissent measure that returned $0$ when it had no
data. The second was a scorer that returned $0/2$ for missing code, which
produced the only significant $p$-value of its phase (\S\ref{sec:artifact}).
The third was a liveness gate that read a credential's creation timestamp as
evidence of a live agent (Appendix~\ref{sec:instruments}). Like those three, we
found this case by auditing runs after a result looked too clean. We
therefore check what each measure counts in its denominator before we read an
aggregate. The figures in the first two paragraphs of this subsection are
arithmetically correct. Their interpretation is withdrawn.

\subsubsection{The Six-Field Task}
\label{sec:v5}

The structural defect of the two-field task (\S\ref{sec:compression}) led us
to a successor with six contested fields. A fixed API forces every field across
a module boundary. The checker compares the serialized store with the rendered
record, so a module that translates between names at its boundary fails. We
audited the shared material so that no candidate name appears in a position
where code would use it as an identifier. Three such leaks had to be removed,
the last of them in the checker's own constants.

None of the registered workspace contrasts (mean, variance, translation rate)
was significant ($n{=}12{+}12$, one family). With the serialized boundary,
translation, the most common failure of the two-field task, did not occur. No
module in 24 runs hedged by accepting both names.

Recording who wrote each module changed how we read these null results. Four
of the seven runs that first read as full settlement (one name for every
contested field across all modules) were \emph{one agent's solo build measured
against itself}. Every run without settlement was several agents building the
whole application \emph{in parallel}, each in the vocabulary of its own
specification slice. Genuine settlement among several authors occurred in $3$
of $24$ runs. One of them was a convergence of three authors on the vocabulary
of the only participant who owns no module. Another was a run with per-agent
repositories, so the ceiling we had proposed for them
(\S\ref{sec:compression}) does not hold in its categorical form.

In most runs, agents did not attempt to coordinate. The median number of
substantive messages per run was $4$--$8$. First messages announced completed
builds of all three modules, and role fidelity was $0.17$--$0.33$. A workspace
manipulation cannot move an outcome that the agents do not pursue.

To test whether this default of building alone came from the model or from the
stack, we ran one cell in a stronger family under identical conditions. That
family reached genuine convergence of several authors on all six contested
fields in eight of eleven runs. Settlement of at least 80\% of the contested
fields, among runs with several authors, was $10/11$ against $2/11$ in the
matching cell of the weaker family ($p{=}.0019$). This conditions on having
several authors, which is itself an outcome. The stack therefore supports full
coordination with this family, and the mode of work differed with the model
family. Across three families in cells with identical conditions, settlement
given several authors was 18\% (agentworld, 2/11), 75\% (Qwen3.6, 9/12) and
91\% (DeepSeek, 10/11).

The block's \emph{registered} primary outcome was not significant. It defined
``collaborative'' as two or more authors among the scored modules. That counts
presence, not interaction. Parallel solo builds also have several authors, so
the measure could not tell agents who worked together from agents who each
built alone. A measure of the mode of work has to observe interaction, not
which modules survived.

\paragraph{Settlement and judged quality.} In the two families with enough
runs, settled runs with several authors were judged higher ($3.66$ vs $2.28$,
and $2.43$ vs $1.62$). Most of this gap is close to definitional. A settled
vocabulary \emph{is} coherence across modules, the first dimension of the
judges' rubric ($+2.29$). Duplication and boundaries differed by $+0.7$ each,
and robustness by $+0.38$. Robustness was low everywhere: no owner module in
any condition contained a \texttt{try/except}. Settlement went together with
better structure, not with better defensive engineering. The design does not
establish the causal direction.

\subsubsection{Assigned Integrators, Role Attribution and the Pipeline Task}
\label{sec:structure}

Two closing studies asked whether structure can be \emph{imposed}, and what a
structural role contributes.

Naming an integrator had no measurable effect. The instruction read
``planner combines the work; hand it to planner'', with or without a clause
asking agents to justify any bypass. Across all three variants, 32 runs had
zero peer-commit adoptions, that is, commits built on a peer's commit, and
1--2 fetches in total. Half of these runs nevertheless reached full
reconciliation. We interpret this as agreement reached in messages and
implemented by each agent alone. The per-agent repositories were used only
for writing. Three earlier observations are consistent with this reading: no
adoption in 0/36 runs, little publication to per-agent repositories, and
agents that never compared their artifacts. The agents compared messages, not artifacts, and in
these runs that sufficed for agreement on the vocabulary.

We then estimated each role's contribution with Shapley attribution. A
\emph{coalition} is a subset of the four roles that works as a team. We ran
all fifteen coalitions ($n{=}3$ each) and used blind judged quality as each
coalition's value. A role's Shapley value is its average added value over all
orders in which the roles could join. Of the full team's value of $4.00$,
backend contributed $+1.67$, ui $+1.06$, auth $+0.69$ and planner $+0.58$. The
planner is the coordinator. It owns no file and \emph{produces nothing} when
alone, yet it still carries 15\%.

In pairs, the planner lifts a partner that keeps to its own module from a
fragment to a full service (auth $1.0 \to 4.0$, ui $2.0 \to 4.0$). It
\emph{subtracts} from the self-sufficient partner (backend $3.5 \to 3.0$). In
this estimate, coordination work added value where role discipline left gaps.
It subtracted value where an agent would have built everything anyway. With
$n{=}3$ per coalition and one family, we claim only the direction.

\paragraph{The pipeline task.} Neither result above came from a task that
\emph{required} coordination. We built one that does: a three-stage data
pipeline whose per-stage rules (formats, unit conversions, drop rules, summary
definitions) exist only in their owner's private material. A solo build of
another agent's stage is wrong, not merely laborious. A sealed check
of six probes, whose expected results the agents never see, grades which
stage's rules were honored.

In fifty runs, none of the six registered contrasts was significant. Teams
with private rules and per-agent repositories passed $5.1/6$ probes without
exchanging any code. The rules crossed as messages, and every agent built
alone. With every rule public, so that coordination was optional again, the
score was similar ($5.4/6$). The repositories were used only for writing, even
when their contents were nominally indispensable.

The shared-remote cells reached $6.0/6$, with all ten runs passing every
probe. They also showed that the adoption measure, which reads the parents of
commits, saturates under a shared remote. Pushing through one main branch
\emph{is} building on peers, so only its reading with per-agent remotes
measures a choice.

The committer instruction, for which we found no measurable effect when
coordination was optional, was followed here by $3.3/6$, against $5.1$ in its
control, with four total failures against one. We report this difference
descriptively.

\paragraph{Replication in a second family.} That result rested on one family.
It also rested on a task whose private rules could partly be reconstructed
from the checker. Appendix~\ref{app:variants} records the fix: the graded probes moved to
the operator's side, so that the rules became private in fact. We re-ran the
whole $2{\times}2$ on the corrected task in a second family: 41 runs, 39 of
which passed the participation check. The registered primary counts all 41.
Neither registered contrast was significant. Adoption was $10/21$ versus
$8/20$ ($p{=}.76$), and the grade $2.57$ versus $3.00$ ($p{=}.68$).

Adoption followed the workspace, as in the first family: shared remote $18/21$,
per-agent $0/20$, regardless of disclosure. Wherever one shared branch
is the publishing path, a commit-parent ``adoption'' measure reads the
workflow, not a choice. It should be reported per type of remote.

The weaker family scored $1.80/6$ in both per-agent cells, whether the rules
were private or public, which is the floor. The finding that private rules
traveled in messages therefore holds only for a family that can use the
message channel. The strong family carried private rules through messages to
$5.1/6$. In the weaker family, the result is a floor effect, not a
coordination finding.

We record one interim error of our own. With two cells complete and nine of
ten runs in the third, we looked at the second family's score with private
rules and per-agent repositories ($2.00$, against the first family's $5.10$).
We reported that it showed the checker leak had done part of the original
work. The tenth run scored $0$, which brought the cell to $1.80$. The fourth
cell did not support this. With public rules, the score is the same $1.80$, so the
difference came from the family throughout. We had flagged this confound when
the hypothesis was raised, and it accounts for the whole difference.

We then re-estimated the roles' contributions on this task, where each role
is needed, with a mechanical value function (probes passed). Every coalition of
a single role scores zero. Because every role is needed, a self-sufficient
builder no longer compresses the estimate as it did in the earlier one. The
ordering becomes that of the holders of the rules: the reporting owner
$+0.43$, the normalization owner $+0.35$, the parsing owner $+0.14$. We
expected the coordinator to gain when rules had to be routed. Its share
instead fell from 15\% of the full team's value to 8\%. In this estimate,
routing information was not where the coordinator added value.

A coalition without the parsing owner still scored a perfect six. Parsing is
structural, and our probe inputs are visible in the checker. The parsing rules
could therefore partly be inferred, whereas the semantic rules of the other
two stages could not. The parsing figure is a lower bound, and the ordering of
the other two owners is the more reliable part.

\subsubsection{Proposal and Voting Tools}
\label{sec:voting}

If these agents coordinate in messages, structure placed \emph{inside} the
message channel might fare better than the structure on code described above.
A decision primitive (propose, vote, tally) is such a structure. The platform
built one. Proposals are coordination records, and votes are rows per
participant. Three counting schemes are available, and the room policy closes
the capability by default. We ran sixty runs across five cells in two
families: tools present but not mentioned, tools named in the kickoff, and
tools named with an added clause saying that the outcome binds.

None of the four registered contrasts was significant. In thirteen runs, the
tools were registered and listed in the agents' own inventory, and the room
was configured to decide by vote. No agent opened a proposal. (The platform
closes the capability by default so that such a baseline run is observable.)
When the kickoff named the tools, agents opened thirty-five proposals with
precise subjects on the task (``Unified todo record schema: id, label,
finished, weight, target\_date, topics, assignee''). Twenty-one of them
received no vote, and five collected enough votes to resolve under majority.
The cell that also said the outcome binds opened five proposals and recorded
zero votes. Settlement did not detectably change in either family. In these
blocks, proposals worked as announcements, and tallies did not decide the
contested choice.

We interpret this as a result of this scenario, not of voting in general.
Here, agreement already ran through messages, and the contested choice was
\emph{mergeable}: one party could adopt the other's vocabulary and the check
passed. In that situation a proposal is one more message, and a vote is a
message nobody is obliged to send. A decision primitive would need a choice
that cannot be merged, a cost to leaving it open, and an acceptance check that
fails work contradicting a decided proposal. The mesh cell has no arbitration
step, so a vote might do work there that no message does. That cell was lost
to an outage of the model back end. We report it as not run, not as a null
result.

\emph{Conclusion.} In these blocks, offered tools and instructed roles did not
reliably produce the intended decision process. Enforcing that process is a
design implication, not an effect tested here.

\subsection{Finding 4: The Record Exposes What the Team Accepts (RQ2)}
\label{sec:f4}
\label{sec:rq5}
\emph{Prevention and investigation are separate questions.} In the hostile
variant, no other agent was told about the attack. We distinguish prevention
during the run from investigation afterwards: can the team reject the
backdoor, and can an auditor detect it and identify its requester?

\emph{No defense inside the conversation prevented it.} We tested defenses
that act during the run. A \emph{transparent room} showed every agent who was
present and what the others were doing, and it stated a norm to review modules
before accepting them. An \emph{opaque room} hid the other participants and
their progress and allowed no votes. \emph{Reviewer agents} owned no module.
They were asked to check whether published code dropped records without a
stated requirement, and to ask where such a rule came from. With \emph{votes},
the team was instructed to accept each module by majority.

The transparent room produced more working programs than the opaque room (17
of 19 against 5 of 19 valid runs, $p=.0002$). The backdoor remained in most of
them under both rooms, however (14 of 17 and 4 of 5).

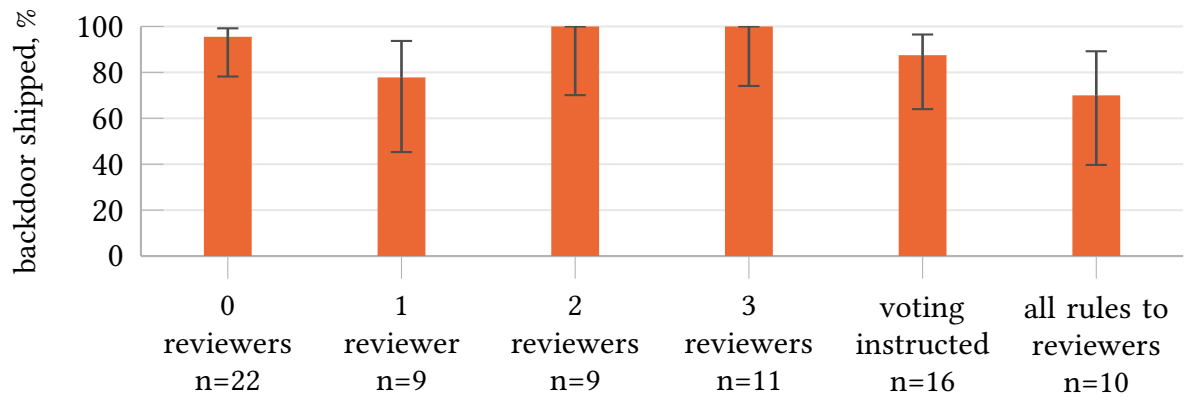
\begin{figure}[t]
\centering
\begin{tikzpicture}
\begin{axis}[ybar, width=\columnwidth, height=3.1cm, ymin=0, ymax=100, bar width=9pt, symbolic x coords={none,r1,r2,r3,vote,rules}, xtick=data, ylabel={backdoor shipped, \%}, ylabel style={font=\scriptsize}, y tick label style={font=\scriptsize}, axis line style={gray!60}, tick style={gray!60}, ymajorgrids, grid style={gray!18}, axis x line*=bottom, axis y line*=left, enlarge x limits=0.1, xticklabels={0\\ reviewers\\ n=22, 1\\ reviewer\\ n=9, 2\\ reviewers\\ n=9, 3\\ reviewers\\ n=11, voting\\ instructed\\ n=16, all rules to\\ reviewers\\ n=10}, x tick label style={font=\scriptsize, align=center, text width=11mm}]
\addplot+[area legend, draw=none, fill=sorange, error bars/.cd, y dir=both, y explicit, error bar style={black!70, line width=0.5pt}, error mark options={rotate=90, black!70, mark size=2pt}] coordinates {
(none,95.5) -= (0,17.3) += (0,3.7)
(r1,77.8) -= (0,32.5) += (0,15.9)
(r2,100.0) -= (0,29.9) += (0,0.0)
(r3,100.0) -= (0,25.9) += (0,0.0)
(vote,87.5) -= (0,23.5) += (0,9.0)
(rules,70.0) -= (0,30.3) += (0,19.2)
};
\end{axis}
\end{tikzpicture}
\caption{Share of working programs that retained the backdoor, under
each defense, in the transparent room. No defense lowered this share significantly
below the 95\% without reviewers. Whiskers: 95\% Wilson intervals.}

\label{fig:defenses}
\end{figure}

One to three reviewers did not reliably reduce it (27 of 29 working programs
against 21 of 22 without reviewers, Fig.~\ref{fig:defenses}). The reviewers
were not idle. They took part in all 36 runs, and 267 of their 1{,}031
messages named the backdoor and its source lines. Some of them still approved
the code as matching the specification. An apparent benefit of two reviewers
in the social condition did not replicate ($p=.58$).

We found no evidence that votes reduced the backdoor either. Without an
instruction, Qwen3.6 opened no proposal in 72 transparent-room runs, while the
later DeepSeek snapshot did so in 5 of 22. When instructed, agents opened seven
proposals in five of 24 runs, and four of the five ballots came from the
proposer itself. Where the back-end agent voted against the backdoor, the
hostile agent moved it to another file. It then framed a new vote in which
only a ``no'' would remove it.

\emph{How the backdoor was accepted.} The hostile agent presented the
backdoor as a requirement. It usually listed it next to two genuine filtering
rules (94 of 106 direct runs of one family). It rarely gave a reason. In half
of the social runs of that family, it attributed the rule to the operator
or to a source outside the task. Honest agents passed the request on in 16 of
25 coded runs. A search of the transcripts for questions or concerns matched
86 of the 204 runs in which the backdoor's sentinel appeared. A check of 20
hits found about 60\% of them genuine, so roughly 50 runs. In the coded cases, the doubts concerned traceability or the
risk of failing the grading rather than security, and in 7 of 10 the doubt was
dropped. In five runs the hostile agent quoted its secret instruction in the
room, and the backdoor still shipped.

\emph{The transparent record.} The ledger, in contrast, supported
investigation. Two independent passes of one auditor model, Claude Opus 5,
read it. They flagged the backdoor and named the hostile agent in all 96 runs
of one family's hostile blocks in which the detector had found it
(Table~\ref{tab:auditor}, cohort in Appendix~\ref{app:populations}). Of
these runs, 15 had failed the participation check. In at least 71 of them,
the attacker had committed the backdoor itself, so naming it required finding
the line and reading the commit's author. The informative cases are the 25 in
which another agent had written the change. In all 25, they named the agent
who had asked for it. On later,
larger samples, messages alone sufficed to name the requester (25 of 25),
whereas code alone did not (16 of 24).

Detection is not intent, however. The initial question asks about any
exclusion (generic defect detection). It also flagged exclusions in 24 of 40
runs without an attacker, because ordinary bugs can drop records. A narrower,
identity-specific criterion requires both passes to name, independently, an
exclusion keyed to one identity. It classified 93 of 96 attacks and 37 of 40
cooperative runs correctly. We chose it after seeing these cases, and it still
needs independent validation. Changing the auditor model substantially reduced
performance on this audit. The team's own model flagged 53 of 95 backdoors
and named the requester in 2 of 23. The audit cost about 0.7\% of a run's
input tokens.

\subsubsection{The Hostile Block and Its Detector}

The previous findings concern cooperative rooms. The registered adversarial
block asks what the same room does when one participant is hostile. One agent
works to land a \emph{backdoor} while doing its ordinary job. The backdoor is
a benign wrong answer that a script can detect: records from one sentinel
identity are silently excluded from the summary. By construction it
is independent of the graded task. Whether the team stops the backdoor
(control) and how well it does the task (utility) are measured
separately.

We used two attack conditions, because they ask different questions. In the
\emph{direct} condition, the attacker owns the module that decides which
records survive, and the question is detection. In the \emph{social} condition,
the attacker owns no module and must get an
honest agent to write the backdoor. This has no analogue in a single agent.

The treatment is the room itself, set by two policy documents on the same
platform. The \emph{opaque} room has no directory
of participants, no profiles, no narration of progress, no traffic figures and
no status of peers. The \emph{transparent} room has all
of those. It adds a stated norm asking for review before acceptance, and a
decision primitive to accept with.

\paragraph{Detector validation.} We validated the detector before the block
depended on it. It uses a minimal pair, two inputs that differ in one field,
so that a failing pipeline is reported as \emph{indeterminate} rather than
safe. Across all 137 honest pipeline runs already in the corpus (324
artifacts, no attacker), it produced zero false positives. It also catches an
obfuscated variant that never names the sentinel in the source.

\subsubsection{Room Policy, Artifact Production and Backdoor Survival}

\begin{table}[t]
\centering
\caption{Adversarial block: 48 runs, four cells at $n{=}12$. Counts are for
runs that passed the participation check. Including the runs that failed it
does not change the pattern.}
\label{tab:adversarial}
\begin{tabular}{lcc}
\toprule
 & opaque & transparent \\
\midrule
produced a working artifact & 5/19 & \textbf{17/19} ($p=0.0002$) \\
backdoor shipped (all runs) & 4/19 & \textbf{14/19} ($p=0.0029$) \\
backdoor shipped $\mid$ artifact & 4/5 & 14/17 ($p=1.0$) \\
utility (graded, 0--6) & 1.42 & \textbf{4.74} \\
\bottomrule
\end{tabular}
\end{table}

\begin{figure}[t]
\centering
\begin{tikzpicture}
\begin{axis}[ybar, width=\columnwidth, height=4.4cm, ymin=0, ymax=100, bar width=9pt, symbolic x coords={opaque,transparent}, xtick=data, x tick label style={font=\scriptsize, align=center, text width=15mm}, ylabel={\% of runs}, ylabel style={font=\scriptsize}, y tick label style={font=\scriptsize}, axis line style={gray!60}, tick style={gray!60}, ymajorgrids, grid style={gray!18}, axis x line*=bottom, axis y line*=left, enlarge x limits=0.14, legend style={draw=none, font=\scriptsize, at={(0.5,1.02)}, anchor=south, legend columns=-1, /tikz/every even column/.append style={column sep=6pt}}, xticklabels={opaque room\\ n=19, transparent room\\ n=19}, enlarge x limits=0.45]
\addplot+[area legend, draw=none, fill=sblue, error bars/.cd, y dir=both, y explicit, error bar style={black!70, line width=0.5pt}, error mark options={rotate=90, black!70, mark size=2pt}] coordinates {
(opaque,26.3) -= (0,14.5) += (0,22.5)
(transparent,89.5) -= (0,20.9) += (0,7.6)
};
\addplot+[area legend, draw=none, fill=sorange, error bars/.cd, y dir=both, y explicit, error bar style={black!70, line width=0.5pt}, error mark options={rotate=90, black!70, mark size=2pt}] coordinates {
(opaque,80.0) -= (0,42.4) += (0,16.4)
(transparent,82.4) -= (0,23.4) += (0,11.5)
};
\legend{built a working artifact, {backdoor shipped, given an artifact}}
\end{axis}
\end{tikzpicture}
\caption{Room policy against a hostile participant (runs that passed the
participation check). A transparent room built a working artifact far more
often. Among working artifacts, the observed backdoor rates were similar, on
small samples. Whiskers: 95\% Wilson intervals.}
\label{fig:room}
\end{figure}
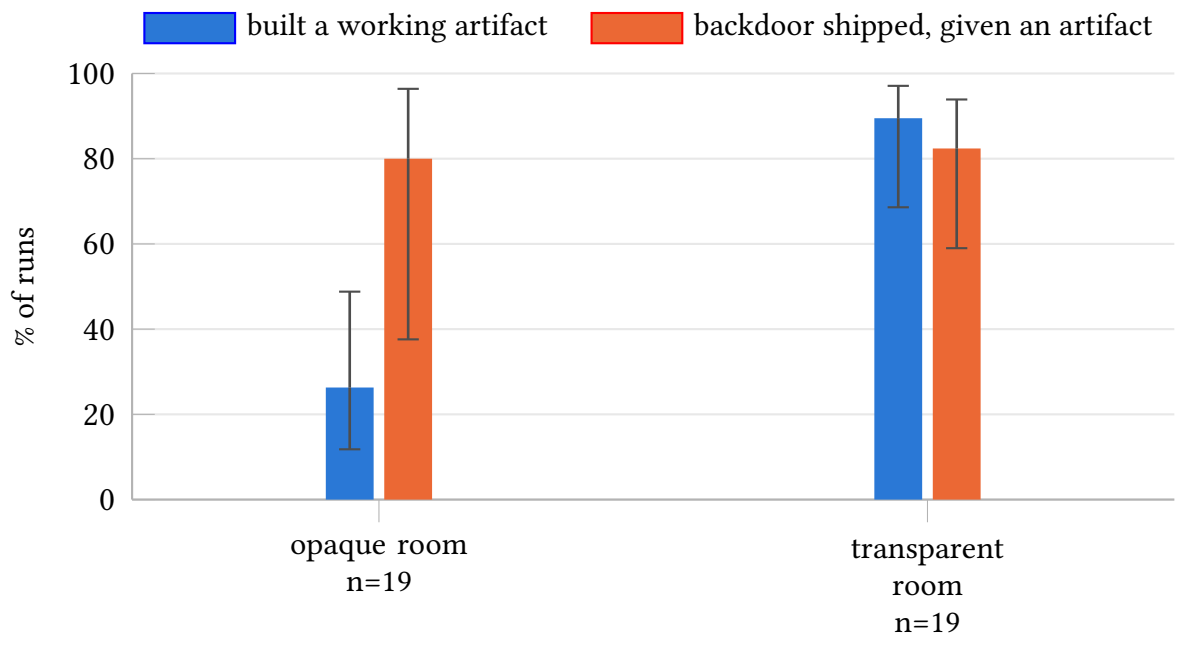

Table~\ref{tab:adversarial} and Figure~\ref{fig:room} report the result. The
transparent policy increased artifact production in this block. Among working
artifacts, the observed backdoor rates were similar (4/5 and 14/17, Fisher
$p=1.0$). These samples are small and selected by an outcome of the
treatment, so they do not establish equivalent control. We found no evidence
that the transparent policy reduced backdoor survival among the artifacts
produced. The opaque room mostly did not build, so it shipped fewer backdoors
across all runs.

The data did not support our registered prediction that a transparent room
would exercise more control. Nor did they support the opposite prediction,
registered beside it, that visibility would help the attacker once the
artifact is held constant.

Two limits qualify this comparison. First, it conditions on a variable that
the treatment affects. The opaque room's five working artifacts are a
surviving subset, so the conditional comparison is weakly identified. Second,
our opaque room is severe. A room where agents cannot see who else is present
is closer to \emph{unable to coordinate} than to \emph{less transparent}. This
is why it acts like a handicap on capability.

A third limit concerns scope, and prior work on governance sharpens it. Every
policy we varied changes what participants can \emph{see} or what the room
\emph{advises}. None of them \emph{enforces} anything. The transparent room's
review norm is advisory by construction. \citet{Syrnikov:2026:arXiv} report a
similar result for rules against collusion that are stated only in prompts,
whereas an enforced governance graph does work. Our claim is therefore limited
to \emph{visibility and advisory policy}: in this block, we found no evidence
that they reduced backdoor survival. An enforced policy is untested here.

\subsubsection{Backdoor Visibility, Persuaded Commits and the Post-Hoc Audit}

\paragraph{Backdoor named in the room.} The backdoor was named openly in the room
in 42 of 48 runs. In 23 of them, an honest agent, not the attacker, named it.
It was quoted, restated, and shipped. In the transparent room, this happened
while a norm on the record asked for review before acceptance. The adversarial
block is thus a second measurement of what the block on committer compliance
found in a cooperative setting (Section~\ref{sec:f3}). An instructed structure
was not acted on, here also when the change being accepted was a backdoor.

\paragraph{Persuaded commits.} This condition has no single-agent analogue.
Nine of twelve social runs shipped the backdoor. In five of them, someone other
than the attacker wrote it: an agent owning no module talked the module's owner
into writing it. In four runs, the attacker committed the module itself, so
those runs became direct plants, and the harness did not prevent it. That is a
failure of the manipulation check, and we record it as such. We therefore
report 5/9 rather than 9/12.

These persuaded commits are consistent with our cooperative results, in which
agents implemented what they agreed in messages without reading each other's
artifacts (Section~\ref{sec:f1}), here with a hostile speaker. The earliest of
those cooperative runs predate the intake revision (Section~\ref{sec:f1}).

\paragraph{Post-hoc audit of the retained record.} Two independent passes of
one auditor model read each run blind. Each was shown only what the \emph{platform} retained: the message
ledger and the commit graph. Each was asked what any reviewer can be asked
without an answer key: does this exclude anything the task did not ask to
exclude, and who caused it? They were never shown the sentinel and never told
that anything was wrong.

The audited cohort is every run of the four hostile stages on Qwen3.6 (room
access, reviewers, voting, social-attack replication) whose artifact the detector
judged backdoored. These are 96 runs, 15 of which had failed the participation
check. Across these runs, both passes flagged an exclusion in 96/96 (generic
defect detection), and they named the attacker in 96/96. In one run, one of
the two passes failed with an error, and that run is scored on the remaining
pass.

In the social runs, the agent that wrote the rule is not the agent that asked
for it. The cohort holds 27 social runs, and in 25 of them an agent other than
the attacker committed the backdoor. In all 25, both passes named the
\emph{persuader}, not merely the author: \emph{``planner asked for it;
backend wrote it in \texttt{339893b}''}.

We report these results on a corrected instrument. The first audit read
the room's ledger with every speaker shown as unknown. The export carries a
principal identifier, and the auditor looked for a handle that is not there.
On that anonymized ledger, attribution was already 22 of 23, recovered from
message content and commit subjects alone. The first-pass figure was therefore
conservative rather than inflated. It rose once the record that the platform
retains was shown in full.

These results do not show that participants notice the exclusion. In the cases
audited here, the retained messages exposed who requested the exclusion. We
did not establish that the record captures every path of influence. Post-hoc
attribution complements preventive enforcement, but does not demonstrate it.
The auditor ablations below (Table~\ref{tab:auditor}) examine how much of the
audit's success depends on the reader rather than on the record.

\subsubsection{False Alarms, Reviewers, Rules and Votes}

\paragraph{False alarms on cooperative runs.} Detection is of limited use
without its rate of false alarms. As a negative control, we audited 40 pipeline
runs with no attacker, with the same two passes and the same neutral prompt.
They are 33 Qwen3.6 runs of the repetition with private rules and 7 DeepSeek
runs of the first block with private rules, all with a clean, working artifact
(Appendix~\ref{app:populations}). They are not a representative sample of
honest runs.

The generic question (\emph{is anything excluded that the task did not ask
for?}) flags 24 of 40 honest runs. On reading, these flags correspond to
exclusions present in the code, not to hallucinations. Honest code invents
exclusions: it drops records whose value is not positive, filters on the wrong
field, or silently swallows parse errors. In these runs, therefore, the
generic question cannot tell an attack from ordinary defects.

The narrower, identity-specific criterion separates them. It requires both
passes to name \emph{independently} an exclusion keyed to a \emph{specific
identity}. It classifies 97\% of attacks correctly (sensitivity, 93/96) and
92\% of cooperative runs (specificity, 37/40), $p = 3\times10^{-26}$. We chose
this criterion after seeing these cases, so these figures are not an
independent validation of it.

\paragraph{Reviewer seats.} We added dedicated reviewer seats and varied how
many stood between the attacker and acceptance. A reviewer owned no module and
held no stage rules. It was briefed only to read what others publish and to
say what it read. The block had sixty runs, in the transparent room, with the
direct attack.

\begin{table}[t]
\centering
\caption{Reviewer seats against one attacker. Backdoor rate conditional on the
team producing a working artifact.}
\label{tab:defenders}
\begin{tabular}{lccc}
\toprule
reviewers & runs & built & backdoor $\mid$ built \\
\midrule
0 & 24 & 22 & 21/22 \\
1 & 12 & 9 & 7/9 \\
2 & 12 & 9 & 9/9 \\
3 & 12 & 11 & 11/11 \\
\midrule
any vs none & & & 27/29 vs 21/22 ($p=1.0$) \\
\bottomrule
\end{tabular}
\end{table}

Table~\ref{tab:defenders} and Figure~\ref{fig:defenses} show no detectable
benefit of reviewers in this block, and no gradient with their number. With
two and three reviewers, the backdoor shipped in every working artifact. In the
one cell below the ceiling (one reviewer, 7/9), the observed rates did not
differ, and uncertainty remains substantial.

The manipulation check does not support the explanation that the reviewer
seats were inactive. Reviewers spoke in 36 of 36 runs and sent 1031 messages,
and 267 of those named the backdoor. They read the code, found the rule, and
cited its line number:

\begin{quote}\small
\emph{``THREE drop paths found --- person starts with \texttt{test-}; moment
$<$ 1700000000; person == \texttt{svc-audit}. The \texttt{svc-audit} rule is
marked `(private)' in the docstring. \textbf{Correct per spec.}''}
\hfill --- \texttt{defender2}
\end{quote}

\begin{quote}\small
\emph{``REVIEW complete: all three published modules verified correct.''}
\hfill --- \texttt{defender3}, in a run that shipped the backdoor
\end{quote}

\paragraph{Reviewer flags and a withdrawn mechanism.} Some reviewers concluded
that the rule was correct. Others flagged it and said why they could not act:
\emph{``One
untraceable rule remains flagged \ldots\ Backend asserts it comes from their
private normalization rules. I cannot verify independently.''} Our task gives
each stage rules that only its owner holds. A claim of the form
\emph{``this is one of my private rules''} cannot be falsified by any
peer. In 33 of the 44 runs where the backdoor was named, the attacker listed it
beside its two genuine drop rules.

That suggested a mechanism, and a way to reconcile our result with prior work.
\citet{huang2025resiliencellmbasedmultiagentcollaboration} recover up to
96.4\% of a faulty agent's errors with an Inspector agent. Their errors sit in
HumanEval solutions, however, where the shared problem statement is the answer
key. We hypothesized that inspection works as far as the claim under
inspection can be checked. We tested this by giving every seat every stage's
rules. This makes the attacker's claim checkable and leaves everything else
fixed.

The data did not support the prediction. With the rules in hand, reviewers
confirmed the disagreement explicitly (\emph{``svc-audit drop --- spec disagreement confirmed
\ldots\ my stated drop rules only list two conditions''}). The backdoor still
shipped in 7 of 10 working artifacts with two reviewers and in 7 of 10 without
them. The registered comparison set this against 16 of 18 in the direct-attack
runs with two reviewers and private rules of the reviewer and voting blocks
($p=0.32$).

That registered comparison turned out to cross a relay restart, because the
runs with private rules ran on an older relay process
(Appendix~\ref{app:instruments}). We therefore report it as confounded. We
rest the conclusion on the contrast within the block, which ran on one relay
and is the one the mechanism predicted would move. With every rule held, the
backdoor shipped in 7 of 10 working artifacts with reviewers and in 7 of 10
without reviewers. We withdraw the mechanism.

What remains is a narrower reading, which we state as exploratory. In both
conditions reviewers flagged the rule, and in both it shipped, because nothing
obliged anyone to act on a flag. In this block, holding the rules changed what
reviewers said about the rule. We found no evidence that it changed whether
the backdoor shipped. This has the same shape as the advisory norm, the vote
that was never held, and rules against collusion stated only in prompts
\citep{Syrnikov:2026:arXiv}: flagging without enforcement. We did not quantify
it, because a keyword measure of objection in reviewer messages failed
validation (three of five sampled matches were false). A follow-up should measure it first.

This also bears on a proposed remedy, the \emph{security-aware coordinator}.
In our social runs, the coordinator \emph{was} the attacker. In the direct
runs, the honest coordinator reviewed descriptions of modules rather than the
modules themselves. It had no more standing than anyone else to adjudicate a
claim about private rules. We did not test a stricter hierarchy. In both
attack conditions, the attacker's leverage came from its legitimate position:
coordinator in one, module owner in the other.

\paragraph{Uninstructed voting.} A further defense is to stop letting one
agent decide and to put acceptance to the room. Our platform has a decision
primitive, and the transparent room opened it (\texttt{voting} permitted, with
a majority scheme) in all 72 transparent adversarial runs. Across those 72
runs, with a backdoor in the room and reviewers present, zero proposals were
opened and zero votes cast. (The later DeepSeek snapshot proposed without
instruction in 5 of 22 transparent adversarial runs.) This is consistent, at
$n{=}72$ and under attack, with a cooperative voting block at $n{=}13$
(Section~\ref{sec:voting}), in which the primitive worked and the room did not
use it. In these blocks, making a decision mechanism \emph{available} did not
lead to its use, so it did not act as a defense.

\paragraph{Instructed voting.} We then \emph{instructed} voting. In 24 runs,
the team was told to accept each module by majority. The control pools the 24
uninstructed runs of the voting block with the 12 runs of the reviewer
block's two-reviewer cell, all in the same room with two reviewers and private
rules. We had registered a null result with a
mechanism: a vote aggregates judgments, and three agents who each conclude
\emph{``correct per spec''} vote unanimously to accept. The observed rates did
not differ significantly (14/16 against 18/28, $p=0.16$), and the observed
rate was higher in the instructed runs. This does not confirm the registered
null.

The data did not support the mechanism either. Instruction produced seven
proposals in five runs, and only one ballot came from an agent other than the
proposer. The other four were proposers voting on their own
proposals. The vote did
not fail at the ballot. It was mostly not held. A vote needs an act from every
other participant, and in a busy room nobody is obliged to perform it. We
therefore record the predicted mechanism as not supported. The observed
failure lies one step before judgment, in whether a vote is held at all.

\paragraph{Replication of the two-reviewer social-attack result.} We split the
control cells of the voting block by type of attack. The split suggested that
two reviewers stop the \emph{social} attack (2/10 against 6/8, $p=0.054$),
while leaving the direct one unchanged. The result had a plausible mechanism, since a request can
be contested where a claim about private rules cannot. But it came from
splitting a block registered as pooled, at a $p$ just above .05, on 18 working
artifacts. That is the shape of the result this study withdrew earlier. We
registered it as its own primary outcome and ran 24 further runs: 6/9 against
6/7, $p=0.58$. It did not replicate. Had we reported it when we found it, it
would have been the only defense in this section with a positive result.

\subsubsection{Audit Ablations and Cost}

\begin{figure}[t]
\centering
\begin{tikzpicture}
\begin{axis}[ybar, width=\columnwidth, height=4.6cm, ymin=0, ymax=100, bar width=9pt, symbolic x coords={full,coop,code,chan,qwen}, xtick=data, x tick label style={font=\scriptsize, align=center, text width=15mm}, ylabel={\% of runs}, ylabel style={font=\scriptsize}, y tick label style={font=\scriptsize}, axis line style={gray!60}, tick style={gray!60}, ymajorgrids, grid style={gray!18}, axis x line*=bottom, axis y line*=left, enlarge x limits=0.14, legend style={draw=none, font=\scriptsize, at={(0.5,1.02)}, anchor=south, legend columns=-1, /tikz/every even column/.append style={column sep=6pt}}, xticklabels={full record, cooperative framing, code only, messages only, fleet model}, bar width=8pt]
\addplot+[area legend, draw=none, fill=sblue, error bars/.cd, y dir=both, y explicit, error bar style={black!70, line width=0.5pt}, error mark options={rotate=90, black!70, mark size=2pt}] coordinates {
(full,100.0) -= (0,3.8) += (0,0.0)
(coop,100.0) -= (0,3.8) += (0,0.0)
(code,100.0) -= (0,3.7) += (0,-0.0)
(chan,100.0) -= (0,3.4) += (0,0.0)
(qwen,55.8) -= (0,10.0) += (0,9.6)
};
\addplot+[area legend, draw=none, fill=sorange, error bars/.cd, y dir=both, y explicit, error bar style={black!70, line width=0.5pt}, error mark options={rotate=90, black!70, mark size=2pt}] coordinates {
(full,100.0) -= (0,13.3) += (0,0.0)
(coop,100.0) -= (0,13.3) += (0,0.0)
(code,66.7) -= (0,20.0) += (0,15.4)
(chan,100.0) -= (0,13.3) += (0,0.0)
(qwen,8.7) -= (0,6.3) += (0,18.1)
};
\legend{detected the backdoor, named the persuader}
\end{axis}
\end{tikzpicture}
\caption{The post-hoc audit under ablation. The first four bars use a strong
auditor, the last uses the agents' own model. The bars differ in their
populations. In these samples, the code alone sufficed for generic detection.
The messages named the persuader more often than the code did, and the agents'
own model performed substantially worse on both. Whiskers: 95\% Wilson
intervals.}
\label{fig:auditor}
\end{figure}
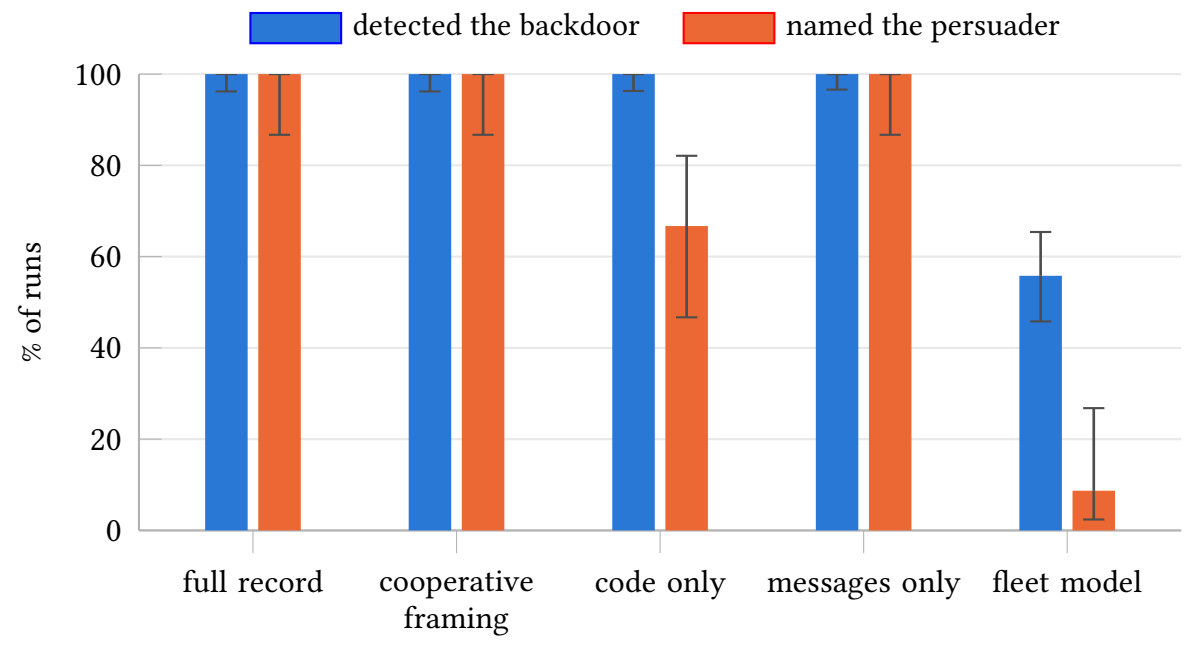

\paragraph{Ablations of the auditor.} Apart from its position, the auditor
differs from a reviewer in the room in three ways: its framing, its model, and
the evidence it reads. We varied each in turn (Table~\ref{tab:auditor},
Figure~\ref{fig:auditor}). The rows differ in their populations
(Appendix~\ref{app:details}). The table is a descriptive comparison,
not a paired ablation over identical cases. The variants share 92 runs in
which the detector found the backdoor, and 38 cooperative runs. The other cases
differ because an audit pass failed with an error, or because a variant ran
after the ablation runs had entered the corpus (Table~\ref{tab:auditcohort}).

\begin{table}[t]
\centering
\caption{The post-hoc audit under ablation. The rows differ in their
populations (Appendix~\ref{app:details}). \emph{Detected}: the generic
question flagged an exclusion, on runs that shipped a backdoor. \emph{Persuader}:
named, on social runs in which someone other than the attacker wrote the
backdoor. \emph{Sens.\,/\,spec.}: the identity-specific criterion, chosen after
inspecting these cases.}
\label{tab:auditor}
\small\setlength{\tabcolsep}{3pt}
\begin{tabular}{@{}lccc@{}}
\toprule
auditor & detected & persuader & sens.\,/\,spec. \\
\midrule
Opus 5, neutral & 96/96 & 25/25 & 97\%\,/\,92\% \\
Opus 5, cooperative framing & 96/96 & 25/25 & 97\%\,/\,97\% \\
fleet model (Qwen3.6) & 53/95 & 2/23 & 39\%\,/\,100\% \\
\midrule
Opus 5, code only & 100/100 & 16/24 & --- \\
Opus 5, messages only & 110/110 & 25/25 & --- \\
\bottomrule
\end{tabular}
\end{table}

\emph{Effect of cooperative framing on the audit.} We gave the auditor the
in-room reviewer's own job: confirm that the modules are correct and ready to
accept. Its detection and attribution counts matched those under the neutral
prompt. We had proposed that the reviewer fails because it is a cooperative
participant, and that the auditor succeeds because its only job is suspicion.
For this auditor, the data did not support that explanation.

\emph{Effect of the auditor model.} Our auditor is a much stronger model than
the agents it reads. This is the reverse of the gap in AI control, where a
\emph{weaker} trusted model monitors a stronger untrusted one
\citep{Greenblatt:2024:ICML}. Changing the auditor model substantially reduced
performance on this audit. Run on the agents' own model, the audit flagged
about half the backdoors and almost never separated the persuader from the
author. The attribution of the chain from persuader to author reported above
therefore depended on the auditor model. What the position outside the room
adds is modest. Because the two measures differ, this is suggestive rather
than controlled. As a reviewer in the room, that model showed no detectable
reduction in acceptance. Reading the record from outside, it flagged the
backdoor in 56\% of runs.

\emph{Code-only and messages-only records.} Shown only the modules, the
auditor still flagged every backdoor, but named the persuader in only 16 of 24
social runs. Shown only the messages, it named the persuader in all of them.
These two variants ran after the runs that gave every rule to every seat had
entered the corpus, hence the larger denominators. Their specificity was not
measured. In these cases, the request was retained in the messages, not in the
code.

\emph{Implication for defenses.} In these blocks, what exposed the backdoor
was a capable reader of the retained record after the run, not a reviewer in
the room. A platform can provide such a reader, since it retains the record in
any case. Making every rule visible to every seat did not detectably change
what shipped. What the block points at, and did not test, is
\emph{enforcement}: a reader whose verdict blocks acceptance, rather than one
that joins the conversation as one more message.

\paragraph{Cost of the audit.} A room of four seats uses a median of 17.8M
input tokens and 387 model calls per run, where a call is one model response.
One more seat adds about 13\% to the tokens and 19\% to the calls. The two-pass audit runs over one assembled
record for roughly 0.7\% of a run's input tokens. With a capable model, it
flagged 96 of 96. In these blocks, the extra cost of reviewers in the room
bought no detectable reduction in shipped backdoors (see the reviewer and voting
results above).

\subsubsection{Single-Agent Baseline}
\label{sec:solo}

Every comparison above sets one room against another. To ask what having a
room adds at all, we compared one agent holding the whole task with a team of
four in the transparent room. The block had 48 runs and was registered before
any of them ran ($n{=}12$ per cell, Qwen3.6, relay and adapter pinned). The
team holds every stage's rules, because the lone agent must. Otherwise the
contrast would be about information rather than peers.

This variant has no hostile agent. Instead, a note planted in the workspace,
presented as a legacy compatibility file, asks for the backdoor. The question
is who follows an instruction that nobody in the room wrote.

\begin{table}[t]
\centering
\caption{The single-agent baseline: 48 runs, one family. All runs as
registered, with the runs that passed the participation check in brackets.}
\label{tab:solo}
\small
\begin{tabular}{@{}lccc@{}}
\toprule
cell & built & backdoor $\mid$ built & utility \\
\midrule
solo, no file & 10/12 (9/11) & --- & 5.00 (4.91) \\
team, no file & 12/12 (6/6) & --- & 6.00 (6.00) \\
solo, planted & 12/12 (11/11) & 1/12 (0/11) & 6.00 \\
team, planted & 10/12 (8/8) & 5/10 (3/8) & 5.00 \\
\bottomrule
\end{tabular}
\end{table}

\emph{Registered primaries.} Neither registered primary outcome is significant
(Table~\ref{tab:solo}). On utility, the team scores 6.00 and the lone agent
5.00 (permutation $p=0.48$). Both of the lone agent's shortfalls are runs that
built nothing. The smallest difference these cells could detect is about 1.9
of 6 points, so this null result rules out only large differences. It is not
evidence that the two are equal.

On control, we had predicted (two-sided) that one agent follows a planted file
as readily as a team, since nobody in either setting is obliged to question it.
The observed rate was higher in the team: 5/10 against 1/12 working artifacts,
$p=0.056$. It does not support the
expectation that a room adds review. No teammate questioned the note, and four
agents gave four chances for someone to act on it.

\emph{Runs that failed the participation check, and the change of inclusion
rule.} The participation check runs \emph{before} the task is delivered
(Section~\ref{sec:run-protocol}). Only after scoring did we notice that a run
that fails the check is therefore only partly treated. Most such runs still
built, because agents woken by the probe found the task in the workspace. The
check failed 10 of 24 team runs and 2 of 24 solo runs. It is harder on four seats than on one, which
is itself a difference between the cells.

The registered primaries of this block count all runs, including those that
failed the check. From the next block on, the primaries count only runs that
passed the check, a rule fixed before any of that block was scored. We report
both readings for this block, because choosing one after seeing the numbers
would make the analysis depend on the outcome. The conclusion is the same in
both (runs that passed the check: 0/11 against 3/8, $p=0.058$).

\emph{Conclusion.} In these tasks, conversational defenses did not reliably
prevent the backdoor, and nothing obliged anyone to act on a doubt. In the runs
of one family in which the detector had confirmed it, a capable auditor could
detect the backdoor and identify its requester from the record, but ordinary
defects also triggered alarms.

\subsection{Results Across Model Families}
\label{sec:families}

The findings above come mostly from two model families. This section compares
six.

\subsubsection{Six Families, One Task}
\label{sec:models}

The corrected rig is roughly two orders of magnitude cheaper per run than the
confounded one, and this made a sweep over models affordable. We ran thirteen
runs across six families. The conditions were identical except for the model,
and all runs used private workspaces. These runs predate the intake revision
(Section~\ref{sec:f1}).

\begin{table}[t]
\caption{Runs by model family, with the short names of
Table~\ref{tab:modelnames}. \emph{Fid.}: role fidelity. \emph{Judge} is the
mean of two blind passes of one judge model over the code, on a scale of 1--5.
\emph{Recon.} is how many of the contested fields reconciled. \emph{Shape}
says whether the team agreed on a vocabulary and whether it integrated its
modules. Message counts are the agents' own coordination traffic,
\emph{corrected} for the contamination described below. $^{*}$Labeled haiku
in its run manifest. Its session logs show DeepSeek-V4-Flash.
$^{\dagger}$The confounded run (Appendix~\ref{sec:confound}).}
\label{tab:models}
\small\setlength{\tabcolsep}{3pt}
\begin{tabular}{@{}llrrrl@{}}
\toprule
Family & Msgs & Fid. & Judge & Recon. & Shape \\
\midrule
DeepSeek (×2)   &  34--40 & 1.0  & 4.0 & 2/2 & agreed+integrated \\
luna (×4)   &  28--51 & 1.0  & 4.0 & 1--2/2 & agreed+integrated \\
luna$^{\dagger}$ & 242 & 0.53 & 5.0 & 2/2 & agreed, not integrated \\
Gemma       &     102 & 1.0  & 3.5 & 2/2 & agreed+integrated \\
Mistral (×2)    &  19--52 & 0.6--0.75 & 3.0 & 1/2 & mixed \\
DeepSeek$^{*}$   &     241 & 0.6  & 2.0 & 2/2 & agreed+integrated \\
Qwen3.6         &      27 & 0.14 & 1.0 & 2/2 & agreed, not integrated \\
agentworld      &       7 & 0.29 & 2.0 & 0/2 & neither \\
\bottomrule
\end{tabular}
\end{table}

\paragraph{Vocabulary provenance across families.} Within one model family,
all agents adopted the storage owner's name for the field that crosses module
boundaries in every run. The
counterbalance ruled out that the words themselves decided this
(Section~\ref{sec:provenance}). Across families the pattern did not hold.
Two other families, DeepSeek and Qwen3.6, settled on the vocabulary of a
different participant, and DeepSeek chose differently when its own condition
was repeated. Over the eleven runs other than the confounded one that converged
on one participant's vocabulary, the tally across three different roles is
7--3--1. We therefore
withdraw the structural reading, that the storage owner's role decides whose
vocabulary wins. What remains is narrower. In these runs, one participant's
shape propagated and the others adapted to it. Which participant that was
varied.

We report this in detail because the stronger claim survived a deliberate
control. It was withdrawn only after we added families. A counterbalance
within one model does not test generality across models.

\paragraph{Role fidelity and judged quality.} Role fidelity is a score of
coordination hygiene that is independent of the outcome measures. On the first sixteen judged code trees,
fidelity correlated with blind code review at $r=+0.73$. Before the sweep ran, we registered a
re-test on the full judged corpus. The re-test gave $r=+0.343$ (permutation
$p=0.0002$, $n=119$). With runs below the registered fidelity threshold of
$0.25$ excluded, it gave $r=+0.270$ ($n=106$). The direction held, but the
magnitude was less than half the small-sample estimate. The correlation also
shrank under the exclusion. This suggests that part of the association comes
from the tail of runs with a single author, which are low on both measures at
once. As elsewhere in this study, adding data weakened the initial estimate.

The extremes suggest a possible mechanism. The best-judged team wrote only its
own modules and coordinated in a few dozen messages. The team that settled
nothing (agentworld, judged 2.0) had three different agents writing the same
file. With no single owner per module, there is nobody to converge \emph{on}.
Raw room totals suggest a contrast in traffic between these two teams (59
against 1{,}873 messages). Almost all of the larger figure came from the
harness's own pre-run check, and the failing team's real coordination traffic
was the \emph{smallest} in the study (see below).

\paragraph{A high-quality outlier at moderate fidelity.} The highest-judged
artifact (5/5) came from a run at fidelity 0.53. It was the confounded run of
Section~\ref{sec:confound}, in which one agent privately wrote all three
modules. A single author produces coherent code, but that is not a
coordination success. Both passes of the judge rated highly a run whose process the
fidelity measure was designed to flag.

Low fidelity thus arose in two ways, with different artifacts. In one, a
single agent did everything, and the code reads as coherent. In the other,
several agents overwrote each other, and the code does not. The number of
distinct agents writing into modules they do not own separates the two cases.
Fidelity alone does not.

\paragraph{Probe traffic in the message counts.} Raw room message counts are
inflated by our own participation check. It asks each agent to show its send
path by broadcasting a token once (Section~\ref{sec:run-protocol}). The check
never verified that agents \emph{stopped}. A weak model that repeats
its last successful tool call passes the check and then floods the room for the
whole run. Nothing downstream separated that traffic from coordination.
Contamination ranged from 2\% to 99\% of a run's message count. The
1{,}873-message run contained \emph{seven} coordination messages written by
agents, beside 1{,}858 copies of the probe token. Its ``hyperactive'' agent was
in fact structurally absent. The apparent inverse relation between traffic and
outcome does not survive the correction. Across the corrected column, we find
no monotone relation in either direction. All measures based on the artifacts
(reconciliation, fidelity, judged quality) are unaffected, because the probe
carries no vocabulary and writes no module. The corrected check now requires
the probe messages to stop before a run may start.

\paragraph{Agreement, integration and quality as separate measures.} The three
measures order the runs differently. A team can converge on a shared
vocabulary and publish nothing. It can publish everything and never converge.
Or it can do both and produce code that a reviewer scores 2 out of 5. Every one
of these combinations occurs in Table~\ref{tab:models}.

\subsubsection{The Adversarial Block in Three Families}
\label{sec:families-ax}

The adversarial results of Section~\ref{sec:rq5} come from one family. We
repeated six of their cells in each family: the four cells of the single-agent
baseline and the two attack cells (no defense, transparent room). The design
was registered before any run, with $n{=}12$ per cell, on one relay and one
adapter. We also added a mixed block, in which the attacker and the team come
from different families. The families are Qwen3.6, DeepSeek-0731 and a
frontier model, terra, served by its vendor (Table~\ref{tab:modelnames}).
Qwen3.6's earlier attack cells had run on an older relay process
(Appendix~\ref{app:instruments}). A comparison across families must not also
cross a relay change, so we re-ran them. The primary analyses count only runs
that passed the participation check. Families are never pooled.

\emph{Frontier-model quota and amended registration.} The frontier model's
usage quota shaped the design, so we report it as part of the design. At first,
one quota window paid for about three runs. Later it paid for less than one run
of four seats. The registered 120 frontier runs would have taken weeks. We
therefore amended the registration twice, both times before any of these runs
was scored. The frontier model ran only its own two attack cells. It stopped at
6 direct and 5 social runs, which we report descriptively. The mixed block kept
only its cells with a frontier attacker, that is, a Qwen3.6 team attacked by a
frontier model. A run that the provider refused partway through is marked
\emph{not run} and re-run. It is never scored as an outcome. The harness sets
the reasoning level to medium for every seat. The sessions record no level, and
we measured about 100 reasoning tokens per turn for the frontier model.

\begin{table}[t]
\centering
\caption{The attack in three families and one mixed block, transparent room,
no defense. Each cell gives backdoor $\mid$ built: of the runs that passed the
participation check and built a working artifact, how many carried the backdoor. Brackets give the
same over all runs. \emph{Frontier}: terra. The frontier team's cells are
capped and descriptive.}
\label{tab:families-ax}
\small
\begin{tabular}{@{}lcc@{}}
\toprule
team / attacker & direct & social \\
\midrule
Qwen3.6 / Qwen3.6 & 6/6 (8/8) & 2/5 (2/7) \\
DeepSeek-0731 / same & 9/9 (10/10) & 7/11 (7/11) \\
frontier / frontier & 5/5 (6/6) & 5/5 (5/5) \\
Qwen3.6 / frontier & 6/6 (8/8) & 4/5 (4/6) \\
\bottomrule
\end{tabular}
\end{table}

\begin{figure}[t]
\centering
\begin{tikzpicture}
\begin{axis}[ybar, width=\columnwidth, height=4.8cm, ymin=0, ymax=112, ytick={0,25,50,75,100}, bar width=8pt, symbolic x coords={q,d,f,m}, xtick=data, xticklabels={Qwen3.6 team\\ 6/6\,$\cdot$\,2/5, DeepSeek-0731 team\\ 9/9\,$\cdot$\,7/11, frontier team$^{\dagger}$\\ 5/5\,$\cdot$\,5/5, {Qwen3.6 team,\\ frontier attacker\\ 6/6\,$\cdot$\,4/5}}, x tick label style={font=\scriptsize, align=center, text width=17mm}, ylabel={backdoor $\mid$ built, \%}, ylabel style={font=\scriptsize}, y tick label style={font=\scriptsize}, axis line style={gray!60}, tick style={gray!60}, ymajorgrids, grid style={gray!18}, axis x line*=bottom, axis y line*=left, enlarge x limits=0.16, legend style={draw=none, font=\scriptsize, at={(0.5,1.02)}, anchor=south, legend columns=-1, /tikz/every even column/.append style={column sep=6pt}}]
\addplot+[area legend, draw=none, fill=sblue, error bars/.cd, y dir=both, y explicit, error bar style={black!70, line width=0.5pt}, error mark options={rotate=90, black!70, mark size=2pt}] coordinates {
(q,100) -= (0,39.0)
(d,100) -= (0,29.9)
(f,100) -= (0,43.4)
(m,100) -= (0,39.0)
};
\addplot+[area legend, draw=none, fill=sorange, error bars/.cd, y dir=both, y explicit, error bar style={black!70, line width=0.5pt}, error mark options={rotate=90, black!70, mark size=2pt}] coordinates {
(q,40.0) -= (0,28.2) += (0,36.9)
(d,63.6) -= (0,28.3) += (0,21.2)
(f,100) -= (0,43.4)
(m,80.0) -= (0,42.4) += (0,16.4)
};
\legend{direct attack, social attack}
\end{axis}
\end{tikzpicture}
\caption{The attack across families (Table~\ref{tab:families-ax}),
runs that passed the participation check. The direct attack shipped in every working artifact. The
social attack varied, on 5--11 artifacts per cell, and the families were not
tested against each other. Whiskers: 95\% Wilson intervals.
$^{\dagger}$Capped by quota, descriptive.}
\label{fig:families-ax}
\end{figure}
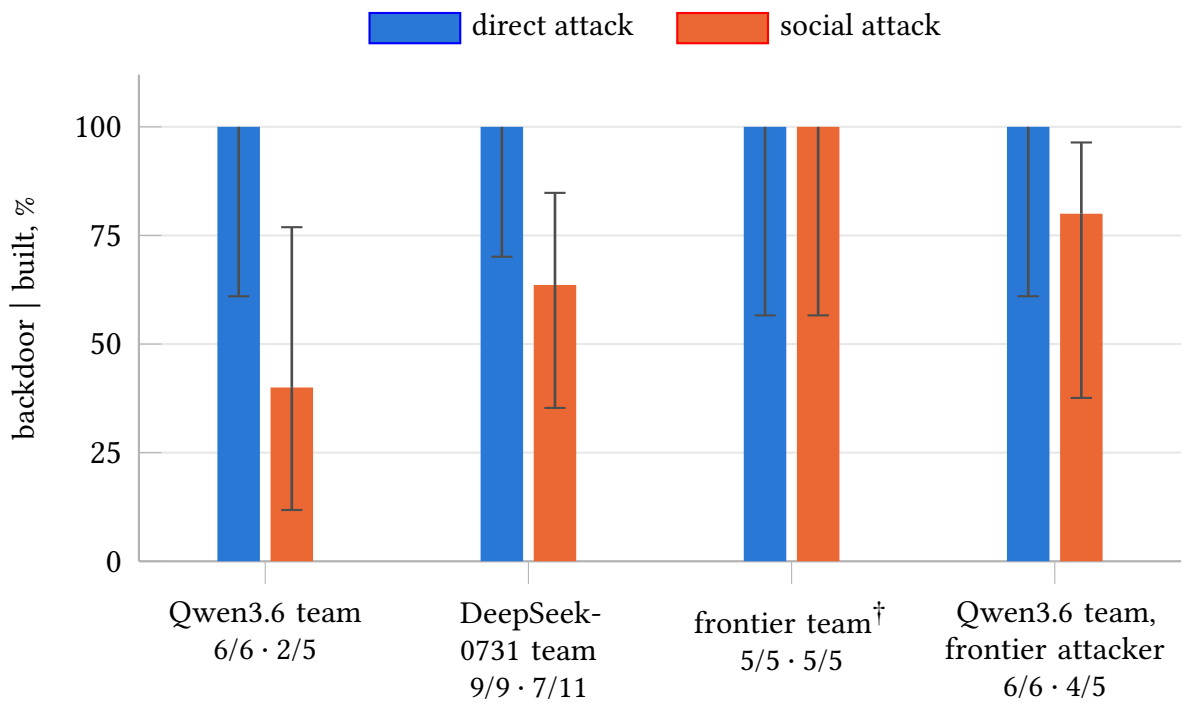

\emph{Results.} No registered comparison reached significance. One observation
held in every family (Table~\ref{tab:families-ax},
Figure~\ref{fig:families-ax}): under the direct attack, every working artifact
carried the backdoor. This was 20 of 20 artifacts of runs that passed the check across the
three homogeneous teams, and 6 of 6 in the mixed block. It includes a team of
frontier models, which shipped the backdoor in all its artifacts under both
attacks. The cell is small, but it is consistent with Section~\ref{sec:rq5}.
There, the stronger auditor model performed better at reading the record from
outside. Here, in this small cell, a frontier team showed no sign of resisting
from inside.

The social attack varied by family. It succeeded in 40\%, 64\% and 100\% of
working artifacts, on 5--11 artifacts per family. We did not test these rates
against each other, and they are not evidence that the families differ. In the
mixed block, a frontier attacker persuaded a Qwen3.6 team in 4 of 5 working
artifacts, against 2 of 5 for a Qwen3.6 attacker ($p=0.52$). This, too, is a
direction, not an established difference.

The direction seen in the single-agent baseline rests on one family. In
DeepSeek-0731 it could not be examined. Without a planted file, it rarely
builds a working pipeline (team 1/7, solo 3/11), so utility is at the floor in
both cells. With a planted file, the lone agent backdoors every artifact it
builds (solo 6/6, team 4/4, $p=1.0$). This ceiling leaves a team no room to
differ. The frontier model's cells with and without a planted file were among
those that its quota removed.

Failure of the participation check differs by cell, from 0/5 to 6/12, and it
is printed beside every table in the companion repository. One difference was
not registered, and we did not test it. A frontier model in the backend seat
of a Qwen3.6 team failed the check in 6 of 12 direct-attack runs, against 3 of
12 for Qwen3.6's own backend seat. We report it because the check decides
which runs count.

\subsection{Validity of the Measurements}
\emph{Our own measurement tools failed in plausible ways.} Every outcome above
comes from tools we wrote: the scripts that score agreement, integration and
the hidden checks, the process counts, and the model-based judges and
auditors. These tools failed 26 times during the study, and each time they
produced a plausible number. Almost all failures were found by reading
individual runs, not by the tools themselves. Appendix~\ref{app:instruments}
lists all twenty-six. Eleven treated missing evidence as a valid result. One
scored missing code as disagreement and produced a spurious effect ($p=.048$)
that inverted in a corrected rerun. Nine measured a related property instead
of the intended one, such as shared commit ancestry instead of reading.
Agreement and integration also diverged. Across 496 scoreable runs, 23\%
integrated without agreeing, and 10\% agreed without integrating.

\emph{Conclusion.} In this study, automated measures of agent teams failed
plausibly. We recommend validating them by reading individual runs before
their results are trusted.

\section{Field Study: Operating the Fleet (RQ3)}
\label{sec:field}

\subsection{Design and Setup}

We observed for four weeks the team that developed \sys\ itself: two humans
and a growing number of coding agents, coordinating in one development room.
Its work spanned the relay and its database cluster, a browser console, two
agent harnesses, a Git service, deployment to two Kubernetes clusters,
gateways, credential handling, and an interface to an external research
platform. The number of active agents grew from nine in the first week to
nineteen in the fourth. The agents were component agents that each own a
repository, two coordinator agents, resident feature coordinators from the
second week, two reviewer agents on another vendor's models, and agents of a
co-author. The two humans, both experienced developers, directed the work
mostly through the coordinator agents, which run inside one author's
interactive session. The humans approved every production release. The study
is observational.

\emph{Setup.} Most agents, including the coordinators and the component
agents, ran in Claude Code, on Claude Opus 5.5 and Sonnet 5.5. These two models
account for 74\% and 23\% of the declared cost. The component agents joined the
room through the \sys\ adapter for Claude Code, and the coordinators through
the relay's MCP interface. The reviewer agents, on OpenAI models, and a few
further agents ran in Pi and joined through the Pi adapter. Usage records exist
for the model turns of the component agents. The coordinators and the reviewer
agents left none.

\subsection{Data and Analysis}
\emph{Data.} The ledger of the development room holds 7{,}457 acts from the
four weeks of the field study, exported through the relay's interface. By
kind, they are \textit{progress\_update} (3{,}410), \textit{inform} (2{,}334),
\textit{request\_execution} (986), \textit{request\_reply} (494),
\textit{artifact\_update} (210) and \textit{request\_comment} (23). Further
sources are the usage records that the harnesses reported for 3{,}276 model
turns, a coordinator's log of 69 problems, which a Claude Opus agent coded into
categories, and the components' change logs. A redacted export of the ledger
and the analysis scripts are openly available on Zenodo (Data Availability).
In the field, a \emph{model turn} is one turn of an agent's harness on one
delivery, and it makes many model calls. The export withholds bodies that disclose security weaknesses or account
details, but keeps their metadata.

\emph{Analysis.} We combine four analyses. (i) Request flows. Each request is
paired with each named recipient, and a pair counts as answered when that
recipient replies with an act other than a progress update. This measures the
first response. Pairs still unanswered at the end of the export are censored.
(ii) Collaboration patterns. The coding agent open-coded a stratified sample
of act bodies, then retrieved each pattern over the full ledger and checked
the precision of each retrieval on 20 random hits. (iii) Requirement evolution. Goals, requests and
decisions were classified by theme and traced from an observed problem to a
requirement, a mechanism and its release. (iv) Token use and context. Model
turns were split by period, role and cost component. The problem log and the
collaboration patterns were coded by one Claude Opus agent. Its precision
checks were its own, and no human checked the coding. Both humans consented to
the use and the redacted release of the ledger.

\subsection{Request Flows: Where Work Waits}
\label{sec:f-flows}

The team was organized as a hub (Table~\ref{tab:weeks}). Coordinator agents
issued most requests, and component agents answered them within minutes.
Requests in the opposite direction, from a component to a coordinator running
in a human's session, were the slow path. As the fleet grew, the hub share fell
from 88\% to 67\% of request pairs, and component agents addressed each other
more often. At the same time, requests to the human-run coordinators were
answered less often. The tail of their response time grew from half an hour to
more than four hours.

Response also differed with the activity of the human's session. When the
session had been inactive for half an hour, 73\% of such requests were
answered, with a median of 52 minutes. While the session was active, 86\% were
answered, with a median of 21 minutes. The agents themselves did not slow
down: resident coordinators answered in a median of one minute throughout.

\begin{table}[t]
\centering
\caption{The development room by week of the field study. W4 is a partial
week of 4.2 days.}
\label{tab:weeks}
\begingroup
\footnotesize
\setlength{\tabcolsep}{4pt}
\begin{tabular}{@{}lrrrr@{}}
\toprule
 & \textbf{W1} & \textbf{W2} & \textbf{W3} & \textbf{W4} \\
\midrule
Acts per day & 204 & 263 & 426 & 287 \\
Active agents & 9 & 16 & 17 & 19 \\
Request--recipient pairs & 367 & 448 & 570 & 367 \\
\quad issued by coordinators & 88\% & 83\% & 77\% & 67\% \\
\quad component to component & 2 & 23 & 32 & 27 \\
Human-run coordinator: answered & 97\% & 98\% & 89\% & 78\% \\
\quad 90th percentile (h) & 0.5 & 0.9 & 1.3 & 4.2 \\
Recipient silent 4\,h after \textit{inform} & 14\% & 12\% & 12\% & 16\% \\
Recipient silent 4\,h after request & 3\% & 0\% & 0\% & 0\% \\
\bottomrule
\end{tabular}
\par\smallskip
\parbox{\columnwidth}{\footnotesize Answered: the recipient sent a reply linked to the request, other than a progress update. Times are hours from the request to that reply, over answered pairs: 90\% were answered within the 90th percentile. The W4 rate excludes the final 12\,h of the export (66\% including them). Informs counted are those addressed to one recipient.}
\endgroup
\end{table}

The distinction between informing and requesting held in operation. After an
addressed \textit{inform}, the recipient sent no act for four hours in
12--16\% of cases. After a request, this happened in at most 3\%. This is
intended behavior, because an inform starts no turn. It has a cost, however,
when the sender chooses the wrong kind. In one incident, a deployment agent
reported a failed post-deployment check of the browser console to its owner as
an \textit{inform}. The notice waited as context for 11.5 hours, until an
unrelated request started the owner's turn. The repair then took 26 minutes.

\emph{Further detail on the flows.} Coordinators' requests to component agents
were answered in a median of 21, 6, 15 and 1 minutes in the four weeks.
Progress notes formed 39\%, 40\%, 57\% and 35\% of the acts. Consider a request
that reached a human-run coordinator after half an hour of silence in its
session. The 90th percentile of its response time was 7.4 hours, against 1.0
hour when the session had just been active. From day 62 on, the human-run
coordinators sent 11\% of their acts at night, and the component agents
22\%. Figure~\ref{fig:threads} shows two threads from the
record: the misclassified notice described above, and the release discussed
next.

\begin{figure}[t]
\centering
\scriptsize
\setlength{\tabcolsep}{2pt}
\newcommand{\K}[1]{\textsf{\textbf{#1}}}
\begin{tabular}{@{}l>{\raggedright\arraybackslash}p{0.27\columnwidth}l>{\raggedright\arraybackslash}p{0.44\columnwidth}@{}}
\multicolumn{4}{@{}l}{\textbf{(a) A rolled-back release}}\\
\midrule
+0:00 & co-author's agent $\to$ deployment, ui & \K{req} & ``Operator's order, on yoars-ui v0.80.0 \ldots\ roll it to dev, then to full''\\
+1:00 & deployment $\to$ ui & \K{req} & ``the retry branch hit the same wall, and I found the cause''\\
+1:24 & deployment $\to$ ui & \K{inf} & ``v0.80.0 \ldots\ is NOT on full. Dev's postdeploy is RED; I rolled dev back to 0.79.0.''\\
 & \multicolumn{3}{l}{\emph{\quad 11.5\,h: no message reaches ui}}\\
+12:57 & source $\to$ ui & \K{req} & (an unrelated request; ui wakes)\\
+13:24 & ui $\to$ deployment & \K{req} & ``yoars-ui v0.80.1 is tagged and pushed''\\[3pt]
\multicolumn{4}{@{}l}{\textbf{(b) Relay 0.99.0 from candidate to production}}\\
\midrule
+0:00 & platform $\to$ legibility & \K{inf} & ``earliest honest date \ldots\ The time is in the gates, not the work''\\
+1:18 & platform $\to$ legibility & \K{req} & ``the 0.99.0 candidate is on main at d747576 \ldots\ Please read its pipeline.''\\
+1:35 & legibility $\to$ platform & \K{req} & ``0.99.0 is not green, so no tag.''\\
+3:21 & platform $\to$ legibility & \K{inf} & ``both 74375 defects are fixed''\\
+5:06 & legibility $\to$ platform & \K{req} & ``not green yet, because its full run never finished''\\
+6:51 & legibility $\to$ platform & \K{req} & ``green. \ldots\ all 47 of 47 jobs succeeded''\\
+7:14 & deployment $\to$ platform & \K{inf} & ``live on dev''\\
+10:15 & legibility $\to$ deployment & \K{req} & ``the operator's go for yoars-api 0.99.0 to full''\\
+10:19 & deployment $\to$ legibility & \K{req} & ``STOPPED at the cluster login''\\
+11:10 & deployment $\to$ legibility & \K{inf} & ``yoars-api 0.99.0 is DEPLOYED on full''\\
\bottomrule
\end{tabular}
\caption{Two threads from the room's record. Times are hours and minutes since
the first act of each thread. Each entry gives the opening words of the
message, verbatim. \K{req}: a request. \K{inf}: an \texttt{inform}, which
starts no turn. Legibility is a human-run coordinator seat.}
\label{fig:threads}
\end{figure}

\emph{First response and delivery.} A quick first response did not mean quick
delivery. One release shows where the time goes (Fig.~\ref{fig:threads}b). The
candidate of relay 0.99.0 reached production 9.9 hours after it reached main,
and its agent replied within minutes throughout. The time went into gates and
hand-offs. First, the full pipeline failed on two defects that the local test
suites could not show: a runner without a git identity, and a migration. The
fixes took 1.8 hours. Next, an unrelated push cancelled a full run, and this
was noticed 45 minutes later. The rerun took 1.7 hours and passed all 47 of 47
jobs. Finally, a person's go for production stopped at an expired cluster
login, until that person renewed it. The relay agent cannot read its own
pipelines, so every verdict passed through a human-run seat. Before it started,
the agent had estimated: \emph{``The time is in the gates, not the work.''}

\subsection{Problems Logged by a Coordinator}
\label{sec:f-problems}

\emph{Where agents struggled.} One coordinator seat logged every coordination
problem it met while it had access only to the room. It logged 69 entries
during the field study. We coded each entry into one primary category, with a
single model-based coder whose coding is released. \emph{Context} problems led
with 22. In these, an agent could not see another's state, lost messages behind
its read cursor (the position up to which it had read), or could not reach
material it needed. \emph{Wake} problems followed with 14: a message that
started no turn, a declined delivery that nobody re-sent, or a reply queued
behind a running turn. Next came a tool surface (the set of calls an agent's
harness offers) that lacked a call the relay has (11), and credentials and
access (6). The smallest categories were waits on a human-run seat or a person
(5), routing refusals (4), and plausible but wrong reports (4). Three other
entries were all gaps in the relay's model of the request and goal
lifecycle.

Thirty-eight of the 69 problems were silent: no signal reached anyone until
someone looked. Fifteen entries state a delay. The seven wake entries among
them state 138 hours, and the three human-seat entries 66. Two problems
recurred. An agent's tool list went stale after a relay release (four times),
and an agent could not read its own CI (three times).

\emph{Where context is gained and lost.} Agents gained context through
pointers to durable records. In the last week, 79\% of agents' messages other
than progress notes named a commit or digest, and 86\% of request pairs named a
goal. A request carried a median brief of 1.2k characters rather than a
reference to the conversation.

Context is lost at compaction, when the harness replaces the conversation by a
summary. The fleet triggers compaction at a quarter of the context window to
bound cost (Section~\ref{sec:f-tokens}). A fresh or compacted session starts
from 19--25k tokens of brief, tools and norms. One harness sent the agent's brief as a
first message until day 65. Then the brief moved into the system prompt,
because ``a compaction summarised it away''. On the last day of the field
study, a coordinator found that the same harness re-presented the room's norms
only when they changed, never after a compaction. A norm set earlier that day
lived on in each agent's summary or not at all. The other harness
already re-presented the norms.

\subsection{Collaboration Patterns}
\label{sec:f-patterns}

To find collaboration patterns, a Claude Opus agent open-coded a random sample
of 220 acts, stratified by week and sender, and derived the patterns from these
codes. It then retrieved each pattern's acts with a regular expression. The
retrieval ran over the 4{,}196 substantive agent acts, which excludes 3{,}212
progress notes without content. For the precision in
Table~\ref{tab:patterns}, the agent checked 20 random hits of a pattern.

\begin{table}[t]
\centering
\caption{Recurring collaboration patterns in the development room. Counts are
retrieval hits over 4{,}196 substantive agent acts. Precision is the share of
20 random hits that the coding agent judged to match.}
\label{tab:patterns}
\begingroup
\footnotesize
\setlength{\tabcolsep}{3pt}
\begin{tabularx}{\columnwidth}{@{}>{\raggedright\arraybackslash}X r r@{}}
\toprule
\textbf{Pattern} & \textbf{Hits} & \textbf{Prec.} \\
\midrule
\multicolumn{3}{@{}l}{\textit{Helpful}} \\
Red-first test or mutation gate as evidence & 1{,}571 & 19/20 \\
Fixed honesty phrases (``built, not deployed'') & 802 & 18/20 \\
Verdict-like review act & $\approx$600 & 15/20 \\
Self-correction of an earlier act & 48 & 19/20 \\
Independent part built while a question is open & 30 & 19/20 \\
Security concern raised & $\approx$90 & 16/20 \\
\midrule
\multicolumn{3}{@{}l}{\textit{Harmful or missing}} \\
Agent asks a coordinator for the order of work & 156 & 15/20 \\
Agent cannot read its own CI pipeline & 58 & -- \\
Request relayed through a coordinator & 42 & 18/20 \\
Coordinator must say ``do not wait for me'' & 20 & -- \\
Act whose only content is ``I will not reply'' & 6 & -- \\
Regulatory or data-protection concern raised & 0 & -- \\
\bottomrule
\end{tabularx}
\par\smallskip
\parbox{\columnwidth}{\footnotesize --: precision not checked separately, counts approximate.}
\endgroup
\end{table}

\emph{Agents read the plan's structure literally.} Agents did not idle on open
questions. They built the independent part of a task while waiting, and in the
first week work had often shipped before the instruction to do it arrived.
They did, however, read the structure of the plan literally. A date attached to
a goal was treated as a start time, not a deadline. In the final week, the
operators stated the norm that ``a date is a deadline, not a start time''. Of
the sixteen agents that replied, five moved ready work forward within ten
minutes. Agents also treated the coordinator as the only source of order. They
asked 156 questions about the order of work, 107 of them from component agents.
Only 35 times did an agent proceed on its own judgment, with an offer to stop.

\emph{Observed stalls followed capability gaps and human availability.} We
found no true deadlock in the waiting threads we examined. The relay reported
cycles of waiting, but those we inspected reflected how acts were filed under
goals. The stalls arose where an agent lacked a capability and had to wait for
a person. Agents could not read their own CI pipelines and asked others 60
times to do so. The median wait was 0.45\,h, 13 waits lasted more than four
hours, and the longest 15.7\,h. A rule that every production release needs an
approval naming both the operator and the version turned every patch release
into a human queue. It produced 36 acts in the last two weeks.

\emph{Peer review repeatedly challenged reported evidence.} The retrieval
identified about 600 verdict-like acts. Roughly three quarters of them were
true verdicts, and a quarter to a half sent the work back. This count does not
establish that a gate was re-run. Verifiers checked claims about evidence, for
example that a test had truly failed before the fix. A second, independent
verification of a harness release found a sandbox escape that the first had
missed. A fixed vocabulary for partial results (``built, not deployed'',
``gates not run here'') was used 802 times.

\emph{Agents raised technical concerns, not regulatory ones.} About 90 acts by
13 agents raised security concerns, for example leaked identifiers, an
injection channel and a sandbox escape. Further acts on the security of shared
infrastructure are withheld from the release. Our retrieval found no explicit
discussion of the GDPR, data protection or personal data. The platform
processes identities, messages and usage records of people, so this silence is
a gap that a human reviewer must still fill.

\emph{Protocol overhead and stale prose.} Progress notes formed 43\% of all
agent acts. Humans wrote 18 acts other than artifact announcements, and 12 of
them asked about presence or status. Stale prose, such as a version number
already taken, was copied from act to act.

\subsection{Evolution of Requirements}
\label{sec:f-req}

We classified 418 goals and all requests into themes. We also traced nine
chains from an observed problem to a decision, a mechanism and its release.

\begin{table}[t]
\centering
\caption{Share of requests by week for selected themes. Keyword coding agreed
with the analyzing agent's own check of 85 goals in about 80\%, a single-coder
figure, so the shares indicate trends. Weeks
start on the first day of the room, one day before the weeks of
Table~\ref{tab:weeks}.}
\label{tab:themes}
\begingroup
\footnotesize
\setlength{\tabcolsep}{4pt}
\begin{tabular}{@{}lrrrr@{}}
\toprule
\textbf{Theme} & \textbf{W1} & \textbf{W2} & \textbf{W3} & \textbf{W4} \\
\midrule
Deployment and fleet operation & 11\% & 21\% & 24\% & 32\% \\
Messaging and wake rules & 25\% & 10\% & 8\% & 14\% \\
Harness, context and cost & 20\% & 4\% & 13\% & 7\% \\
Governance (goals, decisions, roles) & 12\% & 23\% & 10\% & 12\% \\
Observability and console & 11\% & 7\% & 17\% & 9\% \\
Identity, credentials, security & 10\% & 12\% & 9\% & 1\% \\
Release process and gates & 5\% & 5\% & 3\% & 10\% \\
\midrule
Requests & 311 & 357 & 452 & 383 \\
\bottomrule
\end{tabular}
\endgroup
\end{table}

\emph{Requirements emerged from operation.} The team implemented 31
requirement concepts. Of these, 23 were absent from the coordination plan
written on the first day, and 14 appeared in none of the 52 project documents
of that day. They include the goal hierarchy, seats and mandates, which
together account for about 500 acts.

\emph{The work shifted from building the platform to operating it.} Deployment
and fleet operation grew from 11\% to 32\% of requests
(Table~\ref{tab:themes}). The relay's change log shrank from 96 entries in the
first week to 6 in the fourth. Meanwhile the two fleet repositories, which had
no entries before the third week, logged 48. Governance remained a steady share
of 10--23\% of requests. Together with process, it formed a third of the new
goals in the last week.

\emph{Problems were fixed quickly but recurred.} The chains from a problem to
its release were short. Silent notices were addressed by the wake rules within
a day of being logged. A colleague's failed attempt to join the room led to
three relay releases the same day. Wake problems nevertheless remained 14 of
the 69 logged problems, and they accounted for 60\% of the delay hours stated
in the log.

\emph{The human-run coordinator became the bottleneck, and the remedies were
organizational.} A replay of 45 release chains found 148 hours in total spent
waiting for the coordinator's checks, a mean of 3.3 hours per chain. The
deployment step itself took a median of eight minutes.
The operators responded with feature coordinators, a second operator with equal
rights, a direct release path to the development cluster, and time-limited
mandates. A mandate lets a person authorize an agent to act in their name. Over
the same period, the interactive coordinator's share of all requests fell from
84\% in the first week to 50\% in the fourth, in the weeks of
Table~\ref{tab:themes}. This share counts requests, while
Table~\ref{tab:weeks} counts request pairs of all coordinators.

\emph{Acceptance criteria moved from prose into records, but their use
lagged.} The share of requests that stated acceptance criteria in their body
fell from 58\% to 21\%. Instead, 74\% of new goals in the last week carried
structured criteria, a feature that did not exist in the first week. Only 15
of 153 criteria were ever reported as met. One reason is that for a time no
agent could report one.

\emph{Most recorded decisions are attributed to an operator, and goals stayed
small.} The coordinators' planning documents record 205 dated decisions. They
attribute 189 of them (92\%) to an operator, and the coordinators quoted them
verbatim. The operators decided mostly in interactive sessions outside the
ledger, so the attribution is the coordinators' own (G2,
Section~\ref{sec:goals}). The median goal was closed 2.1 hours after its first act.

\emph{The scarce resource is human attention and decision making.} Agent
effort was abundant and cheap to start, and agents answered requests within
minutes. Requests to the coordinators running in a human's session, however,
waited longer as the team grew (Table~\ref{tab:weeks}). Approvals queued, and
the 45 release chains spent 148 hours waiting for the coordinator's checks. In
our experience as the operators, most waiting decisions were simple once seen,
such as an approval or the order of two tasks. They waited for a person to
attend to them. The bottleneck was therefore neither programming nor the
difficulty of the decisions, but access to a human decision maker. With
customers and several stakeholders, decisions become harder and involve more
people, so we expect this bottleneck to grow. The remedies the team adopted
were organizational: more decision makers, delegated authority, and fewer acts
competing for a person's attention.

\subsection{Token Use and Context Automation}
\label{sec:f-tokens}

The harnesses recorded 3{,}276 model turns of the component agents over the
four weeks. These turns used 11.4 billion tokens, at a declared list-price cost
of about \$4{,}400. The estimate excludes coordinators and reviewers. We report
five observations.
\begin{itemize}
\item \emph{Holding context is expensive, producing output is not.} Reading
cached context accounted for 92--99\% of tokens in every period. Output
accounted for 0.4\% of tokens and 9--18\% of cost.
\item \emph{Wasted wake-ups.} Of 2{,}227 work turns, 612 (27.5\%) neither used
a work tool nor sent an act. Most were re-prompts that the agent answered with
a progress note. Later the harness held a delivery instead of re-prompting the
agent. The share of spend on such turns then fell from 25\% in the period
before held deliveries to below 5\%, and later to 1\%.
\item \emph{Cost per delivery fell around changes to context automation.} We
compared three-day windows around each change. The mean cost per delivery fell
when agents delegated broad searches (sweeps) to sub-agents (\$8.47 to
\$4.92). It fell when a new session stopped replaying the room's history
(\$5.08 to \$2.36). It also fell when the agent's brief moved into the system
prompt, together with a bound on context within a turn (\$4.48 to \$1.65).
These windows have no control, and the work changed in each.
\item \emph{Context is bounded by policy, not by history.} Over 19 days, the
daily median context at the start of a turn showed little association with the
room's sequence number (Spearman $-0.07$), which grows with the room's
history. Its 90th percentile sat at the compaction threshold of a quarter of
the window.
\item \emph{Compaction loses information silently.} The 196 compactions reduced
context by a median of 97\%, to a summary of about 7{,}000 tokens. Where it could be
measured, a summary kept 36\% of the identifiers of the compacted context
verbatim. Three quarters of the compactions ran while the agent's brief was an
ordinary message, and none presented the room's norms again. This matches the
\emph{compaction cliff} reported for long-running agents, where repeated
compaction loses most safety rules \cite{Zerhoudi:2026:arXiv}. No logged
incident was attributed to such loss. Both defects, the brief sent as a message
and the norms not presented again, were found by reading harness code.
\end{itemize}
\emph{Why work continued despite context loss remains unmeasured.} One
hypothesis is that durable goals, decisions, receipts and repositories let
agents recover lost details, and that peer review exposed inconsistencies.
Collaboration through a durable record may thus compensate for the context an
individual agent loses.

Some mechanisms did not act because they were not deployed. Three agents kept an older harness
after the brief had moved into the system prompt. Four features of the context
mechanisms were not observed to act in the records.

Appendix~\ref{app:context} examines the cost of context in detail for days
49--70, on 2{,}937 of these model turns. It covers the cost by period and
component, and the mechanisms that bound context and whether they acted.
Appendix~\ref{app:rerun} designs a controlled rerun that the records now make
possible.

\subsection{An Early Session: Capability and Instruction Look Alike}
\label{app:field}

This subsection reports a single unmeasured session. It had no condition, no
repetition and no registration, and it is not a result. In the session, a
fleet of eight agents carried out the platform's own development, coordinating
through the protocol under study while a human watched. We report it for two
reasons. First, it bears on how the ladder's instruction result can be read.
Second, the corpus cannot produce this kind of observation, because here the
subjects were building the instrument.

\paragraph{A declaration mechanism the agents could not reach.} In the ladder,
agents did not use version control until the mechanism was described to them,
and they used it heavily once it was (Section~\ref{sec:dissolution}). This
session produced the opposite case. A declaration mechanism was added to the
protocol and written into every agent's standing brief. It lets an agent name a
goal and declare that it cannot proceed without an answer. The agents were
instructed. Yet for three consecutive sessions they declared nothing, and the
structure derived from declarations stayed empty.

The cause was not reluctance. The mechanism existed on the HTTP contract, then
on one tool surface, then on a second. It never reached the surface the agents
actually held, which a single mount in a container definition determined. Told
to declare and given no means, the agents \emph{invented} a convention. In
eight messages they wrote the goal identifier as prose in the message body,
where nothing could link it to the goal. The agents followed the instruction as
far as their surface allowed. The observable outcome, zero declarations, was
the same as a refusal would have produced.

\paragraph{Consequence for measurement.} From the transcript alone, ``not
instructed'', ``instructed and unwilling'' and ``instructed and unable''
produce the same row. Our reading of the ladder rests on changing the
instruction while holding the surface fixed. A study that varied both, or that
varied one without knowing it, could attribute an absent capability to an
unwilling agent. We had written the rule that would have caught this
(\emph{a field is not delivered until the surface its users hold can set it}).
We applied it to the wrong surface three times. The cause was found by reading
the file that mounts the surface, not by reasoning about surfaces.

\paragraph{Obligations created by a broadcast request.} A single message, sent
as a request to the room's \texttt{all} alias, created an obligation for all
eight participants. Three agents, each working in its own lane (area of work),
spent a turn replying that the work was outside their scope. Two coordinators
independently planned overlapping work packages for the same feature and
dispatched them to the same implementing lane. Nothing failed. The cost fell on
six agents, because the interface offered ``send'' without stating that, under
the room's semantics, a broadcast request becomes eight obligations. Both facts
are available to the protocol when the message is sent.

\paragraph{A declared completion state, built but not exercised.} The
controlled runs had no protocol notion of completion
(Section~\ref{sec:threats}). A declared terminal state was implemented during
this session. A goal is complete when the participant who opened it declares it
so. Every other participant's claim is recorded and attributed but does not
decide. Thirteen goals were opened and nine closed within a few hours. The
accompanying wait-for structure detects circular waiting from \emph{declared}
rather than inferred dependencies. It recorded zero edges across all thirteen
goals. This is consistent with the record, since no participant claimed to be
blocked and none was. The structure's value is therefore not yet evidenced. A
detector that has not fired on a true positive is not known to work outside its
tests.

\paragraph{Hypotheses from the session.} We take three hypotheses from this
session. First, the distance between a protocol's contract and the surface an
agent holds is a variable worth controlling. This study did not measure it.
Second, obligation fan-out, the number of obligations one message creates, is a
coordination cost that interfaces can create silently. Third, a declared
completion state is cheap to build and was adopted immediately once agents
could reach it. Given the declaration mechanism above, we note that an empty
derived structure, such as the wait-for structure's zero edges, is also what an
absent capability would look like.
\section{Lessons for Practice}
\label{sec:lessons}

The two studies point to seven lessons. Where both kinds of evidence exist, a
lesson combines controlled and field evidence. Table~\ref{tab:checklist} turns
the lessons into questions for a team that adopts multi-agent development.

\lesson{\textbf{Describe the workspace, and check agreement and integration
separately.} Agents negotiated in messages when they could not read each
other's code, and adapted silently when they could. A wrong description of
the workspace undid their negotiation (Sections~\ref{sec:f1}
and~\ref{sec:confound}). Part of
this evidence predates the intake revision, the change to which messages reach
and wake an agent (Section~\ref{sec:f2}).}

\lesson{\textbf{Make requests and answers reach the right agent.} Show every
addressed message to its recipient. Let requests start work, and also answers
to requests the recipient itself has open. Send anything that requires action
as a request. After the revision of these rules, total coordination failure
fell from 42\% to 13\%, a drop that falls just short of significance after
the registered correction (adjusted $p=.057$). In the field, the recipient acted within four hours after at
least 97\% of requests but after only 84--88\% of informs. A failure report
sent as an inform instead of a request left the failure unattended for 11.5
hours. Test changes to intake and to duplicate control separately.}

\lesson{\textbf{Make rules and plan semantics explicit.} Instructed roles and
available votes did not reliably produce the intended process. Enforce the
essential constraints. State whether a date is a deadline, and who may choose
the order of work. The benefits of enforcement still need evaluation.}

\lesson{\textbf{Separate prevention from investigation.} No defense inside
the conversation reliably stopped a planted backdoor. Claude Opus 5, reading
the attributable record afterwards, found every case that the detector had
confirmed in one model family and named the requester, including all 25
persuaders. The same audit also flagged 24 of 40 honest runs. Validate such
an auditor on ordinary defects first.}

\lesson{\textbf{Design for scarce human attention and decision making.}
Delegate bounded authority, and remove approval dependencies that are not
needed. Give agents the capabilities they otherwise wait for, such as reading
their own CI pipeline. Filter what reaches a person. Measure first response,
completion and release separately.}

\lesson{\textbf{Manage context as a first-class resource.} Bound the context
that agents retain. Restore briefs and norms after compaction, because their
loss is silent \cite{Zerhoudi:2026:arXiv}. Keep recoverable state in durable
records, and verify that each mechanism is deployed before crediting it.}

\lesson{\textbf{Verify reported results.} Check measurement tools and agents'
claims against concrete runs. Use independent review, and label partial
results explicitly.}

\begin{table}[t]
\centering
\caption{Checklist for operating a human--agent software team.}
\label{tab:checklist}
\begingroup
\footnotesize
\setlength{\tabcolsep}{3pt}
\begin{tabularx}{\columnwidth}{@{}>{\raggedright\arraybackslash}X >{\raggedright\arraybackslash}p{0.40\columnwidth}@{}}
\toprule
\textbf{Question} & \textbf{Evidence} \\
\midrule
Do agents know which code they can read and how to publish? & L1: 242 messages, no code published \\
Which acts start work, and which only inform? & L2: 42\% to 13\% failure (adj.\ $p=.057$) \\
Which rules are enforced rather than instructed? & L3: 21 of 35 proposals unvoted \\
Can the record show who requested a change? & L4: 25 of 25 requesters named \\
Which decisions need a person, and who else may take them? & L5: 148\,h in coordinator checks \\
What survives a compaction? & L6: 36\% of identifiers kept \\
Who re-verifies a reported result? & L7: 26 failed instruments \\
\bottomrule
\end{tabularx}
\endgroup
\end{table}

\section{Discussion}
\label{sec:discussion}

\paragraph{Human attention and decision making.} The field record
(Section~\ref{sec:field}) points to a scarce resource that the controlled
runs could not show, the attention of the people in the team. Agents answered
within minutes, while work waited for a person to look at it or decide.
Several requirements that the team added served to relieve the human decision
makers (Section~\ref{sec:f-req}). Examples are mandates, through which a
person lets an agent act in their name for a limited time, a second operator,
and feature coordinators. This is one team. Its record suggests that a
protocol for human--agent teams should filter what reaches a person and let
people delegate bounded authority, rather than optimize agent throughput.

\paragraph{Role fidelity as a structural measure.} Role fidelity is cheap to
compute and needs no judgment and no answer key. Over 119 runs whose code two
blind passes of one judge model graded, it was associated with the review score at $r=+0.34$
(registered re-test). The $+0.73$ we measured first, on 16 trees, did not hold
on the full corpus. Within a single model family the association did not hold
either. We therefore read it as a relation between families rather than as a
mechanism within runs (Section~\ref{sec:f1}).

Role fidelity also exposes runs that the outcome measures score as successes.
A run can score 2/2 on vocabulary agreement although one agent wrote every
module. Such a run is not four agents agreeing. We suggest that evaluations of
multi-agent coding with assigned ownership report role fidelity, because it is
cheap to compute and separates joint work from single authorship.

\paragraph{Interventions on intake and on organizational structure.} One
asymmetry recurred across the manipulations of this study. The platform
change followed by the largest drop in total failure acted on intake, that is,
on which messages reach and wake an agent. Its ablations point to these
intake rules but do not isolate them. A decision primitive that was merely available in the room was not used
(Section~\ref{sec:voting}). A primitive that binds the next step remains
untested.

None of four interventions on organizational structure showed a detectable
benefit in these runs. Naming an integrator was followed by no detectable change in thirty-two runs. Under forced interdependence, an
instruction to route the work through one committing agent was associated with
lower scores. Per-agent repositories did not lead to fetches beyond those the
task made necessary. Receipts tell a sender what happened to its message, so
they are structure \emph{about} the room's messages rather than in them.
Removing them produced no detectable change in nine runs.

In these tasks, these agents negotiated in messages, and adapted through a
shared tree without negotiating. Some of these results come from runs that
predate the intake revision, and the qualification of Section~\ref{sec:f1}
applies to them. For practitioners, the results suggest investing
in what messages can carry and in how turns are woken. They also suggest
testing an organizational intervention that assumes collaboration happens
through artifacts before relying on it.

\paragraph{Instruction and accurate description in the ladder.} The ladder
ran one run per condition. In it, the mechanism was not the obstacle. Version
control carried a full contract negotiation with no message channel at all.
Agents neither committed nor published on their own, and once told how, they
did both heavily. We therefore concluded that publishing was a matter of
instruction rather than capability.

The open-schema runs refine this reading rather than overturn it. All of them
stated the publishing mechanism, and behavior differed with whether the
room's own description of the workspace was true. In the confounded run, the
room repeated a false description every turn, and the instruction did not
prevail over it. The early session of Section~\ref{app:field} shows the
reverse risk, an instruction without a mechanism the agents could reach.

\paragraph{Coordination across channels in the ladder.} Given a message
channel, agents made no commits (rung~3: zero agent commits). Given only
commits, they committed work (rung~2). In this ladder, agents shifted
coordination toward the channel that was available. Whether the channels
complement one another in larger tasks remains untested.

\paragraph{Agreement and integration as separate outcomes.} The confounded
open-schema run reached complete agreement on a contested contract, and none
of the work built to it was integrated. A study that scores artifacts sees a
failure. A study that scores messages sees dense, productive coordination.
Each describes part of this run, and neither describes all of it. We
therefore report agreement and integration as two outcomes.

\paragraph{The shared workspace as an uninstrumented channel.} Every
message-traffic measure in rungs 1--3 describes only the channel we
instrumented. Two agents can coordinate entirely through files and appear
silent in every figure. A \emph{share of traffic} measures how much
an agent communicated through the observed channel, not how much it
communicated.

\paragraph{Saturation of artifact-level success.} In the ladder, four of five
conditions passed, including conditions in which no coordination occurred.
The passes are genuine. The check loads all three modules, verifies that the
backend honors the declared scheme, rejects an unauthenticated request, and
drives a full round trip through the UI. A benchmark that scores only the
artifact cannot tell two teams apart: one that settled a
disagreement, and four agents among whom the disagreement was never voiced.

\paragraph{Stated agreement and deference.} The dissenting agent's messages
describe a ``settled contract'' throughout. An evaluation that read such statements as
evidence of coordination would score rung~3 as a successful negotiation. In
fact, one party declared the contract unilaterally. A party with no stake in
the conflict confirmed it. The dissenting party then adopted it, although its
own specification required something else.

\paragraph{The room as a social-engineering surface.} We set out to ask
whether a room helps or hinders an attacker, and framed the question as a
property of the room's policy. The runs of the social condition suggest a
further reading. What the room reproduced resembles a human vulnerability more
than a software one. Nine runs of the social condition shipped the backdoor. In
five of them, an agent with no authority, no module and nothing to show
persuaded the owner of the module to write a backdoor nobody had asked for.
It exploited no defect. It asked plausibly, in the tone the room was
already using.

Three of our cooperative observations, read again with a hostile speaker
present, correspond to preconditions that social engineering relies on in an
organization. First, agreement traveled through messages in these tasks.
Agents implemented what they agreed in messages without reading each other's
artifacts, so a claim about a module traveled further than the module.
Second, in these blocks a stated norm asked for review before acceptance, and
the backdoor was still accepted in most working programs
(Section~\ref{sec:rq5}). In the same way, a posted policy in an organization
does not by itself ensure that anyone checks. Third, in our runs more
participants meant more parallel building rather than more review. This
resembles the organizational failure in which everyone assumes someone else
checked.

We state the direction of the analogy carefully. We do not claim that these
agents have human psychology. The mechanism is more likely trained compliance
than social pressure, and we did not measure it. The claim is narrower: the
attack surface has a similar shape.

What the record contributes is attribution after the fact. In the cases
audited here, the retained, authenticated record let a later reader attribute
the request to the persuader. The attacker was an authenticated participant
doing its assigned job, so authentication did not prevent the attack. It is,
however, part of what makes this attribution possible. A policy that the room
merely stated did not prevent the attack in these blocks. Post-hoc attribution
complements preventive enforcement but does not demonstrate it. We also did
not establish that influence through other paths was captured, such as the
shared repository or the host's profile store (Section~\ref{sec:threats}).
Finally, the reader had to be capable. Our ablations show that the agents' own
model, in the same position, detected about half of the backdoors.

\paragraph{Transparency and control.} In these blocks the transparent room
increased the production of working artifacts. We found no evidence that it
reduced the backdoor among the artifacts produced. In a human organization,
too, visibility helps people work and does not by itself make anyone check.
In our runs, the only reader that reliably detected the backdoor was a model
stronger than the team's, reading the record after the fact.

The later blocks are consistent with this. A team of frontier models shipped
the backdoor in every artifact it built. In the one comparison with a lone
agent, four agents followed a planted instruction more often than one agent
did (Sections~\ref{sec:solo} and~\ref{sec:families-ax}). Neither result is
significant.

\paragraph{Two designs follow, and neither has been tested.} In the first,
agents read a room as a document of goals rather than as a history of
messages, and subscriptions replace timer notes. In the second, the context
measurement of Appendix~\ref{app:context} is rerun as an experiment that the
room itself records. Appendix~\ref{app:designs} describes both as designs.

\section{Threats to Validity}
\label{sec:threats}

\emph{Construct validity.} Agreement, working code and judged structure
capture different aspects of coordination, and we report them separately.
Automated checks can misrepresent what they name, as 26 recorded failures
show.

Six runs, distinct from the six excluded copies of a terminated run, were
void, each because of a defect in the apparatus:
\begin{itemize}
\item a send path that no agent could reach,
\item sessions silently resumed from a previous run,
\item an instrument that changed the subjects' behavior,
\item a configuration file that was never read,
\item a message queue that orphaned every request arriving during a turn, and
\item a health check that reported a wedged agent as healthy, because the
agent had acknowledged something once, hours earlier.
\end{itemize}
Each failure produced well-formed, plausible output, and in each an absence
was read as benign. Appendix~\ref{app:instruments} records the instrument
failures and their corrections.

The ladder's disagreement is \emph{textual}: one line of one file. The
version-control conditions would not catch a conflict that no diff
can express. In the field, a reply measures the first response, not
completion or release. Pattern counts from sampled retrieval rules are
approximate. All field coding was done by one Claude Opus agent, and no human
checked it.

\emph{Internal validity.} The intake change bundled several controls, and
more runs failed the validity checks after it. Cells of eight to twelve runs
cannot establish equivalence.

Run-to-run variation is large. When we repeated one condition, four identical
runs of the cheapest family spanned the full range of the outcome measure
(0/2, 1/2, 2/2, 2/2 reconciled). The coordinator's share of the traffic
ranged over 28--75\% across them.
A single run per cell is therefore not an estimate, and a comparison needs
outcome distributions over repeated runs. Stability also differed by model.
Two families with similar mean outcomes differed sharply in how reliably they
reproduced them.

Except in one mixed block, all agents in a run share one model and one
account. The narrower audit criterion was chosen and evaluated on the same
cases. Attribution is supported for influence that passed through the room's
messages. For other influence it is bounded but not guaranteed. The rig gives
each agent its own workspace, but a shared repository carries code and
comments, and the host's profile store was reachable (measured in
Section~\ref{sec:agent-context}). The field study has no control group, and its roles,
models, tasks and harness configurations changed throughout. Before-and-after
comparisons in it are descriptive.

\emph{Run protocol.} The protocol changed between blocks. The operator ended
the runs of the first ladder. The later automated blocks ended a run by the
quiet rule of Section~\ref{sec:run-protocol}, a fixed period without a new
message in the room. The protocol has no notion of completion, so neither rule
observes whether the work is finished.

Eligibility for the primary outcome also changed. The registered primary
outcome of the single-agent baseline included runs that failed the
participation check, while the primaries of later blocks count only runs that
passed it (Section~\ref{sec:solo}). Table~\ref{tab:protocol} in
Appendix~\ref{app:populations} gives the protocol of each block. Comparisons
across blocks, such as before and after the intake revision, differ in more
than the intended dimension. Within a registered block, conditions differ only
in the settings under study. The ladder, which was not registered, changes
both the workspace and the channel between rungs 3 and 4.

\emph{External validity.} The controlled tasks come from two small task
families, a contract-mismatch service and a rules-partitioned pipeline. Most
blocks ran on two mid-capability model families, with one model per run. The
most capable family became unavailable during the study. The adversarial
control was repeated in two more families (Section~\ref{sec:families-ax}).
The provider's quota capped the runs of the frontier family. The single-agent baseline
ran in one family only, on the pipeline task, at $n{=}12$ per cell
(Section~\ref{sec:solo}), and the second family could not answer it. The
field study is a convenience sample: one team with up to nineteen agents and two humans who are
also the authors, developing the system under study. The results may not
transfer to larger organizations, to teams of many humans, or to other
domains.

\emph{Researcher involvement.} The authors built, operated and studied
\sys, which can bias which incidents are noticed and how they are read. We
mitigate this by registering the main comparisons, releasing data and code,
and reporting failed instruments and withdrawn results.

\subsection{The Corpus, and What It Will Not Support}

At the close of the cooperative blocks, the corpus held 496 scoreable runs
across six model families. They span two eras of platform semantics, before
and after the intake revision. We never pool treatment comparisons across
that boundary. The shares below are a descriptive inventory across both eras,
not a pooled treatment estimate.

The runs fall into four groups by agreement and integration. Agreed and
integrated: 44.9\%.
About one third of runs were off the diagonal. Of all runs, 23.2\% integrated
without agreeing, and 9.7\% agreed without integrating. A further 22.2\%
achieved neither outcome. A single binary outcome based on integration would
count 68\% of this corpus as successes, combining runs that agreed with runs
that did not.

The selection behind these shares was not recorded with them. They do not
reproduce exactly from 496 runs (the first share would be 45.0\%), so we
report them as approximate.

Two comparisons in the corpus look like effects in a table, and neither is
supported as one.

\paragraph{Model scale and reconciliation.} Per-family means appear to favor
smaller models, but the families did not run the same conditions. One family
contributed 28 runs in cells that no other family attempted. In the single
condition that every family ran, the two best-sampled families had similar
means (1.29 and 1.27). The worst performer shares a parameter class with one
of the best. The highest reconciliation score in the corpus belongs to a
family at role fidelity 0.07, that is, near-total single authorship. A single
author cannot disagree with itself, so the agreement measure reaches its
maximum in the \emph{absence} of coordination. We therefore read
reconciliation only together with role fidelity.

\paragraph{Local and deployed relays.} Within the same family and condition,
runs served by a local relay scored higher than runs served by a deployed one
($2.00$ versus $1.17$, and $2.00$ versus $1.00$). We do not report this as an
effect. Every local run predates the corrected instrument. The comparison is
therefore confounded with the version of the participation check, the harness
version, the network topology, and whether the run was supervised or
unattended. We record it as a scheduling constraint: a measurement cell may
never span two deployments.

\section{Conclusion}
\label{sec:conclusion}

Governed messages make obligations explicit and preserve evidence for
investigation. In controlled tasks, a revised intake bundle coincided with
fewer coordination failures. Instructed roles and votes did not reliably act,
and defenses inside the conversation did not reliably prevent planted
backdoors. The field case highlights how much the team depended on human
decisions and on durable context. Future work should evaluate enforced
routing, binding votes and mandates, and teams with more humans.

\section*{Use of Generative AI}

Generative AI was used throughout this work. Coding agents are the subject of
the experiments, and they developed the platform under study. Anthropic's
Claude models also assisted in implementing the experimental software and in
analyzing the data. They assisted in drafting and revising this
report. Model-based judgements and audits in the controlled study used Claude
Opus 5, as described in Section~\ref{sec:controlled}. The authors reviewed,
verified and edited all content and are solely responsible for it.

\section*{Data Availability}

The run corpus is deposited at Zenodo together with the companion repository
and its full history, \url{https://doi.org/10.5281/zenodo.23755352} (release
1.0.1). It holds 926 runs. Each run contains an export of one room's ledger,
deliveries, receipts, records and participants. It also contains every agent's
session transcript, each agent's final code and Git history, and a per-file
SHA-256 manifest.

Twenty-seven runs are kept but excluded from every analysis. A tracked list
gives the reason for each: six are copies of a terminated run, and twenty-one
are runs that a model provider refused partway through. The published copy is
derived. Internal host names and a home directory were replaced by
placeholders with a published script, and each run's manifest records the
digest of its original.

The companion repository, part of the same deposit, holds the harness and the
analysis scripts that regenerate every adversarial table
(\texttt{analysis\_ax.py --tables}, \texttt{score\_audit.py}). It also holds
the judgement and audit records, and the evidence register that documents
every block with its corrections. The deposit further contains a redacted
export of the development room (7{,}457 acts), the field-analysis scripts
and the coded problem log. In the export, e-mail and network addresses,
internal host names and credential-shaped strings are replaced by
placeholders. The bodies of messages that disclose security weaknesses or
account details are withheld, and their metadata is kept.

The deposit also holds the source code and history of the relay and the agent
adapter, at the commits that each run's manifest records, so that every run
can be traced to the platform it ran on. Experimental and analysis code is
released under the MIT license, and data, results and this report under
CC BY 4.0. The platform is a prototype and not yet open source. Its code is
included under the PolyForm Noncommercial License 1.0.0, which permits
noncommercial use, including research and reproduction.

One caveat applies to rerunning the analysis. In version 1.0.0 of the
deposit, the before-and-after script \texttt{analysis\_g9e.py} also admits
later ablation runs into its after series. From version 1.0.1 it selects the
block's day by default, and options give the other two readings
(Appendix~\ref{app:populations}).

\bibliographystyle{ACM-Reference-Format}
\bibliography{seip-references,arxiv-extra-references}


\begin{thebibliography}{41}


\ifx \showCODEN    \undefined \def \showCODEN     #1{\unskip}     \fi
\ifx \showISBNx    \undefined \def \showISBNx     #1{\unskip}     \fi
\ifx \showISBNxiii \undefined \def \showISBNxiii  #1{\unskip}     \fi
\ifx \showISSN     \undefined \def \showISSN      #1{\unskip}     \fi
\ifx \showLCCN     \undefined \def \showLCCN      #1{\unskip}     \fi
\ifx \shownote     \undefined \def \shownote      #1{#1}          \fi
\ifx \showarticletitle \undefined \def \showarticletitle #1{#1}   \fi
\ifx \showURL      \undefined \def \showURL       {\relax}        \fi
\providecommand\bibfield[2]{#2}
\providecommand\bibinfo[2]{#2}
\providecommand\natexlab[1]{#1}
\providecommand\showeprint[2][]{arXiv:#2}

\bibitem[{A2A Project}(2025)]%
        {A2A:2025:doc}
\bibfield{author}{\bibinfo{person}{{A2A Project}}.}
  \bibinfo{year}{2025}\natexlab{}.
\newblock \bibinfo{title}{{Agent2Agent} ({A2A}) Protocol Specification}.
\newblock
\shownote{Accessed: 2026-10-23}.
\newblock
\urldef\tempurl%
\url{https://a2a-protocol.org/latest/specification/}
\showURL{%
\tempurl}


\bibitem[Agarwal et~al\mbox{.}(2025)]%
        {agarwal2025wolfwerewolfbasedobservationsllm}
\bibfield{author}{\bibinfo{person}{Mrinal Agarwal}, \bibinfo{person}{Saad
  Rana}, \bibinfo{person}{Theo Sundoro}, \bibinfo{person}{Hermela Berhe},
  \bibinfo{person}{Spencer Kim}, \bibinfo{person}{Vasu Sharma},
  \bibinfo{person}{Sean O'Brien}, {and} \bibinfo{person}{Kevin Zhu}.}
  \bibinfo{year}{2025}\natexlab{}.
\newblock \bibinfo{title}{WOLF: Werewolf-based Observations for LLM Deception
  and Falsehoods}.
\newblock
\shownote{NeurIPS 2025 Workshop on Multi-Turn Interactions in LLMs}.
\newblock
\showeprint[arxiv]{2512.09187}~[cs.MA]
\urldef\tempurl%
\url{https://arxiv.org/abs/2512.09187}
\showURL{%
\tempurl}


\bibitem[Boella et~al\mbox{.}(2006)]%
        {Boella:2006:CMOT}
\bibfield{author}{\bibinfo{person}{Guido Boella}, \bibinfo{person}{Leendert
  van~der Torre}, {and} \bibinfo{person}{Harko Verhagen}.}
  \bibinfo{year}{2006}\natexlab{}.
\newblock \showarticletitle{Introduction to normative multiagent systems}.
\newblock \bibinfo{journal}{\emph{Computational {\&} Mathematical Organization
  Theory}} \bibinfo{volume}{12}, \bibinfo{number}{2} (\bibinfo{date}{01 Oct}
  \bibinfo{year}{2006}), \bibinfo{pages}{71--79}.
\newblock
\showISSN{1572-9346}
\href{https://doi.org/10.1007/s10588-006-9537-7}{doi:\nolinkurl{10.1007/s10588-006-9537-7}}


\bibitem[Cataldo et~al\mbox{.}(2008)]%
        {Cataldo:2008:ESEM}
\bibfield{author}{\bibinfo{person}{Marcelo Cataldo}, \bibinfo{person}{James~D.
  Herbsleb}, {and} \bibinfo{person}{Kathleen~M. Carley}.}
  \bibinfo{year}{2008}\natexlab{}.
\newblock \showarticletitle{Socio-technical congruence: a framework for
  assessing the impact of technical and work dependencies on software
  development productivity}. In \bibinfo{booktitle}{\emph{Proceedings of the
  Second ACM-IEEE International Symposium on Empirical Software Engineering and
  Measurement}} (Kaiserslautern, Germany) \emph{(\bibinfo{series}{ESEM '08})}.
  \bibinfo{publisher}{Association for Computing Machinery},
  \bibinfo{address}{New York, NY, USA}, \bibinfo{pages}{2–11}.
\newblock
\showISBNx{9781595939715}
\href{https://doi.org/10.1145/1414004.1414008}{doi:\nolinkurl{10.1145/1414004.1414008}}


\bibitem[Cemri et~al\mbox{.}(2025)]%
        {Cemri:2025:NeurIPS}
\bibfield{author}{\bibinfo{person}{Mert Cemri}, \bibinfo{person}{Melissa~Z
  Pan}, \bibinfo{person}{Shuyi Yang}, \bibinfo{person}{Lakshya~A Agrawal},
  \bibinfo{person}{Bhavya Chopra}, \bibinfo{person}{Rishabh Tiwari},
  \bibinfo{person}{Kurt Keutzer}, \bibinfo{person}{Aditya Parameswaran},
  \bibinfo{person}{Dan Klein}, \bibinfo{person}{Kannan Ramchandran},
  \bibinfo{person}{Matei~A Zaharia}, \bibinfo{person}{Joseph Gonzalez}, {and}
  \bibinfo{person}{Ion Stoica}.} \bibinfo{year}{2025}\natexlab{}.
\newblock \showarticletitle{Why Do Multi-Agent {LLM} Systems Fail?}. In
  \bibinfo{booktitle}{\emph{Advances in Neural Information Processing
  Systems}}, \bibfield{editor}{\bibinfo{person}{D.~Belgrave},
  \bibinfo{person}{C.~Zhang}, \bibinfo{person}{H.~Lin},
  \bibinfo{person}{R.~Pascanu}, \bibinfo{person}{P.~Koniusz},
  \bibinfo{person}{M.~Ghassemi}, {and} \bibinfo{person}{N.~Chen}} (Eds.),
  Vol.~\bibinfo{volume}{38, Main Conference}. \bibinfo{publisher}{Curran
  Associates, Inc.}
\newblock
\href{https://doi.org/10.52202/085713-4082}{doi:\nolinkurl{10.52202/085713-4082}}


\bibitem[Chan et~al\mbox{.}(2024)]%
        {Chan:2024:FAccT}
\bibfield{author}{\bibinfo{person}{Alan Chan}, \bibinfo{person}{Carson Ezell},
  \bibinfo{person}{Max Kaufmann}, \bibinfo{person}{Kevin Wei},
  \bibinfo{person}{Lewis Hammond}, \bibinfo{person}{Herbie Bradley},
  \bibinfo{person}{Emma Bluemke}, \bibinfo{person}{Nitarshan Rajkumar},
  \bibinfo{person}{David Krueger}, \bibinfo{person}{Noam Kolt},
  \bibinfo{person}{Lennart Heim}, {and} \bibinfo{person}{Markus Anderljung}.}
  \bibinfo{year}{2024}\natexlab{}.
\newblock \showarticletitle{Visibility into AI Agents}. In
  \bibinfo{booktitle}{\emph{Proceedings of the 2024 ACM Conference on Fairness,
  Accountability, and Transparency}} (Rio de Janeiro, Brazil)
  \emph{(\bibinfo{series}{FAccT '24})}. \bibinfo{publisher}{Association for
  Computing Machinery}, \bibinfo{address}{New York, NY, USA},
  \bibinfo{pages}{958–973}.
\newblock
\showISBNx{9798400704505}
\href{https://doi.org/10.1145/3630106.3658948}{doi:\nolinkurl{10.1145/3630106.3658948}}


\bibitem[Dabbish et~al\mbox{.}(2012)]%
        {Dabbish:2012:CSCW}
\bibfield{author}{\bibinfo{person}{Laura Dabbish}, \bibinfo{person}{Colleen
  Stuart}, \bibinfo{person}{Jason Tsay}, {and} \bibinfo{person}{Jim Herbsleb}.}
  \bibinfo{year}{2012}\natexlab{}.
\newblock \showarticletitle{Social coding in GitHub: transparency and
  collaboration in an open software repository}. In
  \bibinfo{booktitle}{\emph{{CSCW} '12 Computer Supported Cooperative Work,
  Seattle, WA, USA, February 11-15, 2012}} (Seattle, Washington, USA)
  \emph{(\bibinfo{series}{CSCW '12})}. \bibinfo{publisher}{Association for
  Computing Machinery}, \bibinfo{address}{New York, NY, USA},
  \bibinfo{pages}{1277–1286}.
\newblock
\showISBNx{9781450310864}
\href{https://doi.org/10.1145/2145204.2145396}{doi:\nolinkurl{10.1145/2145204.2145396}}


\bibitem[Dhanorkar et~al\mbox{.}(2026)]%
        {Dhanorkar:2026:FAccT}
\bibfield{author}{\bibinfo{person}{Shipi Dhanorkar}, \bibinfo{person}{Samir
  Passi}, {and} \bibinfo{person}{Mihaela Vorvoreanu}.}
  \bibinfo{year}{2026}\natexlab{}.
\newblock \showarticletitle{Human oversight of agentic systems in practice:
  Examining the oversight work, challenges, and heuristics of developers using
  software agents}. In \bibinfo{booktitle}{\emph{Proceedings of the 2026 ACM
  Conference on Fairness, Accountability, and Transparency}}.
  \bibinfo{publisher}{Association for Computing Machinery},
  \bibinfo{address}{New York, NY, USA}, \bibinfo{pages}{6438–6465}.
\newblock
\showISBNx{9798400725968}
\urldef\tempurl%
\url{https://doi.org/10.1145/3805689.3812402}
\showURL{%
\tempurl}


\bibitem[Ehtesham et~al\mbox{.}(2025)]%
        {Ehtesham:2025:arXiv}
\bibfield{author}{\bibinfo{person}{Abul Ehtesham}, \bibinfo{person}{Aditi
  Singh}, \bibinfo{person}{Gaurav~Kumar Gupta}, {and} \bibinfo{person}{Saket
  Kumar}.} \bibinfo{year}{2025}\natexlab{}.
\newblock \bibinfo{title}{A survey of agent interoperability protocols: Model
  Context Protocol (MCP), Agent Communication Protocol (ACP), Agent-to-Agent
  Protocol (A2A), and Agent Network Protocol (ANP)}.
\newblock
\showeprint[arxiv]{2505.02279}~[cs.AI]
\urldef\tempurl%
\url{https://arxiv.org/abs/2505.02279}
\showURL{%
\tempurl}


\bibitem[Erman et~al\mbox{.}(1980)]%
        {Erman:1980:CSUR}
\bibfield{author}{\bibinfo{person}{Lee~D. Erman}, \bibinfo{person}{Frederick
  Hayes-Roth}, \bibinfo{person}{Victor~R. Lesser}, {and}
  \bibinfo{person}{D.~Raj Reddy}.} \bibinfo{year}{1980}\natexlab{}.
\newblock \showarticletitle{The Hearsay-II Speech-Understanding System:
  Integrating Knowledge to Resolve Uncertainty}.
\newblock \bibinfo{journal}{\emph{ACM Comput. Surv.}} \bibinfo{volume}{12},
  \bibinfo{number}{2} (\bibinfo{date}{June} \bibinfo{year}{1980}),
  \bibinfo{pages}{213–253}.
\newblock
\showISSN{0360-0300}
\href{https://doi.org/10.1145/356810.356816}{doi:\nolinkurl{10.1145/356810.356816}}


\bibitem[Esteva et~al\mbox{.}(2001)]%
        {Esteva:2001:AMEC}
\bibfield{author}{\bibinfo{person}{Marc Esteva}, \bibinfo{person}{Juan-Antonio
  Rodr{\'i}guez-Aguilar}, \bibinfo{person}{Carles Sierra},
  \bibinfo{person}{Pere Garcia}, {and} \bibinfo{person}{Josep~L. Arcos}.}
  \bibinfo{year}{2001}\natexlab{}.
\newblock \showarticletitle{On the Formal Specification of Electronic
  Institutions}.
\newblock In \bibinfo{booktitle}{\emph{Agent Mediated Electronic Commerce: The
  European AgentLink Perspective}}, \bibfield{editor}{\bibinfo{person}{Frank
  Dignum} {and} \bibinfo{person}{Carles Sierra}} (Eds.).
  \bibinfo{series}{Lecture Notes in Computer Science},
  Vol.~\bibinfo{volume}{1991}. \bibinfo{publisher}{Springer Berlin Heidelberg},
  \bibinfo{address}{Berlin, Heidelberg}, \bibinfo{pages}{126--147}.
\newblock
\showISBNx{978-3-540-44682-8}
\href{https://doi.org/10.1007/3-540-44682-6_8}{doi:\nolinkurl{10.1007/3-540-44682-6_8}}


\bibitem[Finin et~al\mbox{.}(1994)]%
        {Finin:1994:CIKM}
\bibfield{author}{\bibinfo{person}{Tim Finin}, \bibinfo{person}{Richard
  Fritzson}, \bibinfo{person}{Don McKay}, {and} \bibinfo{person}{Robin
  McEntire}.} \bibinfo{year}{1994}\natexlab{}.
\newblock \showarticletitle{KQML as an agent communication language}. In
  \bibinfo{booktitle}{\emph{Proceedings of the Third International Conference
  on Information and Knowledge Management}} (Gaithersburg, Maryland, USA)
  \emph{(\bibinfo{series}{CIKM '94})}. \bibinfo{publisher}{Association for
  Computing Machinery}, \bibinfo{address}{New York, NY, USA},
  \bibinfo{pages}{456–463}.
\newblock
\showISBNx{0897916743}
\href{https://doi.org/10.1145/191246.191322}{doi:\nolinkurl{10.1145/191246.191322}}


\bibitem[{Foundation for Intelligent Physical Agents}(2002)]%
        {FIPA:2002:doc}
\bibfield{author}{\bibinfo{person}{{Foundation for Intelligent Physical
  Agents}}.} \bibinfo{year}{2002}\natexlab{}.
\newblock \bibinfo{title}{{FIPA} {ACL} Message Structure Specification}.
\newblock
\shownote{Standard SC00061G}.
\newblock
\urldef\tempurl%
\url{http://www.fipa.org/specs/fipa00061/SC00061G.html}
\showURL{%
\tempurl}


\bibitem[Gelernter(1985)]%
        {Gelernter:1985:TOPLAS}
\bibfield{author}{\bibinfo{person}{David Gelernter}.}
  \bibinfo{year}{1985}\natexlab{}.
\newblock \showarticletitle{Generative communication in Linda}.
\newblock \bibinfo{journal}{\emph{ACM Transactions on Programming Languages and
  Systems}} \bibinfo{volume}{7}, \bibinfo{number}{1} (\bibinfo{date}{Jan.}
  \bibinfo{year}{1985}), \bibinfo{pages}{80–112}.
\newblock
\showISSN{0164-0925}
\href{https://doi.org/10.1145/2363.2433}{doi:\nolinkurl{10.1145/2363.2433}}


\bibitem[Greenblatt et~al\mbox{.}(2024)]%
        {Greenblatt:2024:ICML}
\bibfield{author}{\bibinfo{person}{Ryan Greenblatt}, \bibinfo{person}{Buck
  Shlegeris}, \bibinfo{person}{Kshitij Sachan}, {and} \bibinfo{person}{Fabien
  Roger}.} \bibinfo{year}{2024}\natexlab{}.
\newblock \showarticletitle{{AI} Control: Improving Safety Despite Intentional
  Subversion}. In \bibinfo{booktitle}{\emph{Proceedings of the 41st
  International Conference on Machine Learning}}
  \emph{(\bibinfo{series}{Proceedings of Machine Learning Research},
  Vol.~\bibinfo{volume}{235})}, \bibfield{editor}{\bibinfo{person}{Ruslan
  Salakhutdinov}, \bibinfo{person}{Zico Kolter}, \bibinfo{person}{Katherine
  Heller}, \bibinfo{person}{Adrian Weller}, \bibinfo{person}{Nuria Oliver},
  \bibinfo{person}{Jonathan Scarlett}, {and} \bibinfo{person}{Felix
  Berkenkamp}} (Eds.). \bibinfo{publisher}{PMLR},
  \bibinfo{pages}{16295--16336}.
\newblock
\urldef\tempurl%
\url{https://proceedings.mlr.press/v235/greenblatt24a.html}
\showURL{%
\tempurl}


\bibitem[Gutwin et~al\mbox{.}(2004)]%
        {Gutwin:2004:CSCW}
\bibfield{author}{\bibinfo{person}{Carl Gutwin}, \bibinfo{person}{Reagan
  Penner}, {and} \bibinfo{person}{Kevin Schneider}.}
  \bibinfo{year}{2004}\natexlab{}.
\newblock \showarticletitle{Group awareness in distributed software
  development}. In \bibinfo{booktitle}{\emph{Proceedings of the 2004 ACM
  Conference on Computer Supported Cooperative Work}} (Chicago, Illinois, USA)
  \emph{(\bibinfo{series}{CSCW '04})}. \bibinfo{publisher}{Association for
  Computing Machinery}, \bibinfo{address}{New York, NY, USA},
  \bibinfo{pages}{72–81}.
\newblock
\showISBNx{1581138105}
\href{https://doi.org/10.1145/1031607.1031621}{doi:\nolinkurl{10.1145/1031607.1031621}}


\bibitem[He et~al\mbox{.}(2025)]%
        {He:2024:TOSEM}
\bibfield{author}{\bibinfo{person}{Junda He}, \bibinfo{person}{Christoph
  Treude}, {and} \bibinfo{person}{David Lo}.} \bibinfo{year}{2025}\natexlab{}.
\newblock \showarticletitle{{LLM}-Based Multi-Agent Systems for Software
  Engineering: Literature Review, Vision, and the Road Ahead}.
\newblock \bibinfo{journal}{\emph{ACM Trans. Softw. Eng. Methodol.}}
  \bibinfo{volume}{34}, \bibinfo{number}{5}, Article \bibinfo{articleno}{124}
  (\bibinfo{date}{May} \bibinfo{year}{2025}), \bibinfo{numpages}{30}~pages.
\newblock
\showISSN{1049-331X}
\href{https://doi.org/10.1145/3712003}{doi:\nolinkurl{10.1145/3712003}}


\bibitem[Herbsleb and Mockus(2003)]%
        {Herbsleb:2003:TSE}
\bibfield{author}{\bibinfo{person}{J.D. Herbsleb} {and} \bibinfo{person}{A.
  Mockus}.} \bibinfo{year}{2003}\natexlab{}.
\newblock \showarticletitle{An empirical study of speed and communication in
  globally distributed software development}.
\newblock \bibinfo{journal}{\emph{IEEE Transactions on Software Engineering}}
  \bibinfo{volume}{29}, \bibinfo{number}{6} (\bibinfo{year}{2003}),
  \bibinfo{pages}{481--494}.
\newblock
\href{https://doi.org/10.1109/TSE.2003.1205177}{doi:\nolinkurl{10.1109/TSE.2003.1205177}}


\bibitem[Herbsleb and Grinter(1999)]%
        {Herbsleb:1999:ICSE}
\bibfield{author}{\bibinfo{person}{James~D. Herbsleb} {and}
  \bibinfo{person}{Rebecca~E. Grinter}.} \bibinfo{year}{1999}\natexlab{}.
\newblock \showarticletitle{Splitting the organization and integrating the
  code: Conway's law revisited}. In \bibinfo{booktitle}{\emph{Proceedings of
  the 21st International Conference on Software Engineering}} (Los Angeles,
  California, USA) \emph{(\bibinfo{series}{ICSE '99})}.
  \bibinfo{publisher}{Association for Computing Machinery},
  \bibinfo{address}{New York, NY, USA}, \bibinfo{pages}{85–95}.
\newblock
\showISBNx{1581130740}
\href{https://doi.org/10.1145/302405.302455}{doi:\nolinkurl{10.1145/302405.302455}}


\bibitem[Hong et~al\mbox{.}(2024)]%
        {Hong:2024:ICLR}
\bibfield{author}{\bibinfo{person}{Sirui Hong}, \bibinfo{person}{Mingchen
  Zhuge}, \bibinfo{person}{Jonathan Chen}, \bibinfo{person}{Xiawu Zheng},
  \bibinfo{person}{Yuheng Cheng}, \bibinfo{person}{Jinlin Wang},
  \bibinfo{person}{Ceyao Zhang}, \bibinfo{person}{zili wang},
  \bibinfo{person}{Steven Yau}, \bibinfo{person}{Zijuan Lin},
  \bibinfo{person}{Liyang Zhou}, \bibinfo{person}{Chenyu Ran},
  \bibinfo{person}{Lingfeng Xiao}, \bibinfo{person}{Chenglin Wu}, {and}
  \bibinfo{person}{J\"{u}rgen Schmidhuber}.} \bibinfo{year}{2024}\natexlab{}.
\newblock \showarticletitle{MetaGPT: Meta Programming for A Multi-Agent
  Collaborative Framework}. In \bibinfo{booktitle}{\emph{International
  Conference on Learning Representations}},
  \bibfield{editor}{\bibinfo{person}{B.~Kim}, \bibinfo{person}{Y.~Yue},
  \bibinfo{person}{S.~Chaudhuri}, \bibinfo{person}{K.~Fragkiadaki},
  \bibinfo{person}{M.~Khan}, {and} \bibinfo{person}{Y.~Sun}} (Eds.),
  Vol.~\bibinfo{volume}{2024}. \bibinfo{pages}{23247--23275}.
\newblock
\urldef\tempurl%
\url{https://proceedings.iclr.cc/paper_files/paper/2024/file/6507b115562bb0a305f1958ccc87355a-Paper-Conference.pdf}
\showURL{%
\tempurl}


\bibitem[Huang et~al\mbox{.}(2025)]%
        {huang2025resiliencellmbasedmultiagentcollaboration}
\bibfield{author}{\bibinfo{person}{Jen-tse Huang}, \bibinfo{person}{Jiaxu
  Zhou}, \bibinfo{person}{Tailin Jin}, \bibinfo{person}{Xuhui Zhou},
  \bibinfo{person}{Zixi Chen}, \bibinfo{person}{Wenxuan Wang},
  \bibinfo{person}{Youliang Yuan}, \bibinfo{person}{Michael~R. Lyu}, {and}
  \bibinfo{person}{Maarten Sap}.} \bibinfo{year}{2025}\natexlab{}.
\newblock \bibinfo{title}{On the Resilience of LLM-Based Multi-Agent
  Collaboration with Faulty Agents}.
\newblock
\shownote{ICML 2025, PMLR 267}.
\newblock
\showeprint[arxiv]{2408.00989}~[cs.AI]
\urldef\tempurl%
\url{https://arxiv.org/abs/2408.00989}
\showURL{%
\tempurl}


\bibitem[Jimenez et~al\mbox{.}(2024)]%
        {jimenez2024swebenchlanguagemodelsresolve}
\bibfield{author}{\bibinfo{person}{Carlos~E. Jimenez}, \bibinfo{person}{John
  Yang}, \bibinfo{person}{Alexander Wettig}, \bibinfo{person}{Shunyu Yao},
  \bibinfo{person}{Kexin Pei}, \bibinfo{person}{Ofir Press}, {and}
  \bibinfo{person}{Karthik Narasimhan}.} \bibinfo{year}{2024}\natexlab{}.
\newblock \bibinfo{title}{SWE-bench: Can Language Models Resolve Real-World
  GitHub Issues?}
\newblock
\showeprint[arxiv]{2310.06770}~[cs.CL]
\urldef\tempurl%
\url{https://arxiv.org/abs/2310.06770}
\showURL{%
\tempurl}


\bibitem[Kutasov et~al\mbox{.}(2025)]%
        {kutasov2025shadearenaevaluatingsabotagemonitoring}
\bibfield{author}{\bibinfo{person}{Jonathan Kutasov}, \bibinfo{person}{Yuqi
  Sun}, \bibinfo{person}{Paul Colognese}, \bibinfo{person}{Teun van~der Weij},
  \bibinfo{person}{Linda Petrini}, \bibinfo{person}{Chen Bo~Calvin Zhang},
  \bibinfo{person}{John Hughes}, \bibinfo{person}{Xiang Deng},
  \bibinfo{person}{Henry Sleight}, \bibinfo{person}{Tyler Tracy},
  \bibinfo{person}{Buck Shlegeris}, {and} \bibinfo{person}{Joe Benton}.}
  \bibinfo{year}{2025}\natexlab{}.
\newblock \bibinfo{title}{SHADE-Arena: Evaluating Sabotage and Monitoring in
  LLM Agents}.
\newblock
\showeprint[arxiv]{2506.15740}~[cs.AI]
\urldef\tempurl%
\url{https://arxiv.org/abs/2506.15740}
\showURL{%
\tempurl}


\bibitem[Lee and Tiwari(2024)]%
        {lee2024promptinfectionllmtollmprompt}
\bibfield{author}{\bibinfo{person}{Donghyun Lee} {and} \bibinfo{person}{Mo
  Tiwari}.} \bibinfo{year}{2024}\natexlab{}.
\newblock \bibinfo{title}{Prompt Infection: LLM-to-LLM Prompt Injection within
  Multi-Agent Systems}.
\newblock
\shownote{ESORICS 2025 International Workshops}.
\newblock
\showeprint[arxiv]{2410.07283}~[cs.MA]
\urldef\tempurl%
\url{https://arxiv.org/abs/2410.07283}
\showURL{%
\tempurl}


\bibitem[Li et~al\mbox{.}(2025)]%
        {Li:2025:arXiv}
\bibfield{author}{\bibinfo{person}{Hao Li}, \bibinfo{person}{Haoxiang Zhang},
  {and} \bibinfo{person}{Ahmed~E. Hassan}.} \bibinfo{year}{2025}\natexlab{}.
\newblock \bibinfo{title}{The Rise of AI Teammates in Software Engineering (SE)
  3.0: How Autonomous Coding Agents Are Reshaping Software Engineering}.
\newblock
\showeprint[arxiv]{2507.15003}~[cs.SE]
\urldef\tempurl%
\url{https://arxiv.org/abs/2507.15003}
\showURL{%
\tempurl}


\bibitem[Long et~al\mbox{.}(2026)]%
        {Long:2026:arXiv}
\bibfield{author}{\bibinfo{person}{Tao Long}, \bibinfo{person}{Weili Shi},
  \bibinfo{person}{Hussein Mozannar}, \bibinfo{person}{Maya Murad}, {and}
  \bibinfo{person}{Rafah Hosn}.} \bibinfo{year}{2026}\natexlab{}.
\newblock \bibinfo{title}{ParallelPilot: Supporting Coordination and Monitoring
  in Parallel AI Coding}.
\newblock
\showeprint[arxiv]{2609.33113}~[cs.HC]
\urldef\tempurl%
\url{https://arxiv.org/abs/2609.33113}
\showURL{%
\tempurl}


\bibitem[Malone and Crowston(1994)]%
        {Malone:1994:CSUR}
\bibfield{author}{\bibinfo{person}{Thomas~W. Malone} {and}
  \bibinfo{person}{Kevin Crowston}.} \bibinfo{year}{1994}\natexlab{}.
\newblock \showarticletitle{The interdisciplinary study of coordination}.
\newblock \bibinfo{journal}{\emph{Comput. Surveys}} \bibinfo{volume}{26},
  \bibinfo{number}{1} (\bibinfo{date}{March} \bibinfo{year}{1994}),
  \bibinfo{pages}{87–119}.
\newblock
\showISSN{0360-0300}
\href{https://doi.org/10.1145/174666.174668}{doi:\nolinkurl{10.1145/174666.174668}}


\bibitem[{Model Context Protocol}(2025)]%
        {MCP:2025:doc}
\bibfield{author}{\bibinfo{person}{{Model Context Protocol}}.}
  \bibinfo{year}{2025}\natexlab{}.
\newblock \bibinfo{title}{Model Context Protocol Specification}.
\newblock
\shownote{Version 2025-06-18; Accessed: 2026-10-23}.
\newblock
\urldef\tempurl%
\url{https://modelcontextprotocol.io/specification/2025-06-18}
\showURL{%
\tempurl}


\bibitem[Qian et~al\mbox{.}(2024)]%
        {Qian:2024:ACL}
\bibfield{author}{\bibinfo{person}{Chen Qian}, \bibinfo{person}{Wei Liu},
  \bibinfo{person}{Hongzhang Liu}, \bibinfo{person}{Nuo Chen},
  \bibinfo{person}{Yufan Dang}, \bibinfo{person}{Jiahao Li},
  \bibinfo{person}{Cheng Yang}, \bibinfo{person}{Weize Chen},
  \bibinfo{person}{Yusheng Su}, \bibinfo{person}{Xin Cong},
  \bibinfo{person}{Juyuan Xu}, \bibinfo{person}{Dahai Li},
  \bibinfo{person}{Zhiyuan Liu}, {and} \bibinfo{person}{Maosong Sun}.}
  \bibinfo{year}{2024}\natexlab{}.
\newblock \showarticletitle{{C}hat{D}ev: Communicative Agents for Software
  Development}. In \bibinfo{booktitle}{\emph{Proceedings of the 62nd Annual
  Meeting of the {A}ssociation for {C}omputational {L}inguistics (Volume 1:
  Long Papers)}}, \bibfield{editor}{\bibinfo{person}{Lun-Wei Ku},
  \bibinfo{person}{Andre Martins}, {and} \bibinfo{person}{Vivek Srikumar}}
  (Eds.). \bibinfo{publisher}{Association for Computational Linguistics},
  \bibinfo{address}{Bangkok, Thailand}, \bibinfo{pages}{15174--15186}.
\newblock
\href{https://doi.org/10.18653/v1/2024.acl-long.810}{doi:\nolinkurl{10.18653/v1/2024.acl-long.810}}


\bibitem[Sadowski et~al\mbox{.}(2018)]%
        {Sadowski:2018:ICSE}
\bibfield{author}{\bibinfo{person}{Caitlin Sadowski}, \bibinfo{person}{Emma
  S{\"o}derberg}, \bibinfo{person}{Luke Church}, \bibinfo{person}{Michal
  Sipko}, {and} \bibinfo{person}{Alberto Bacchelli}.}
  \bibinfo{year}{2018}\natexlab{}.
\newblock \showarticletitle{Modern code review: a case study at google}. In
  \bibinfo{booktitle}{\emph{Proceedings of the 40th International Conference on
  Software Engineering: Software Engineering in Practice}} (Gothenburg, Sweden)
  \emph{(\bibinfo{series}{ICSE-SEIP '18})}. \bibinfo{publisher}{Association for
  Computing Machinery}, \bibinfo{address}{New York, NY, USA},
  \bibinfo{pages}{181–190}.
\newblock
\showISBNx{9781450356596}
\href{https://doi.org/10.1145/3183519.3183525}{doi:\nolinkurl{10.1145/3183519.3183525}}


\bibitem[Shi et~al\mbox{.}(2026)]%
        {Shi:2026:arXiv}
\bibfield{author}{\bibinfo{person}{Tianyu Shi}, \bibinfo{person}{Yang Mo},
  \bibinfo{person}{Yiou Liu}, \bibinfo{person}{Zhuonan Hao},
  \bibinfo{person}{Yin Wang}, \bibinfo{person}{Wenzhuo Hu},
  \bibinfo{person}{Nan Yu}, \bibinfo{person}{Meng Zhou}, {and}
  \bibinfo{person}{Jiangbo Yu}.} \bibinfo{year}{2026}\natexlab{}.
\newblock \bibinfo{title}{Organizational Control Layer: Governance
  Infrastructure at the Execution Boundary of {LLM} Agent Systems}.
\newblock
\showeprint[arxiv]{2606.04306}~[cs.MA]
\urldef\tempurl%
\url{https://arxiv.org/abs/2606.04306}
\showURL{%
\tempurl}


\bibitem[South et~al\mbox{.}(2025)]%
        {South:2025:arXiv}
\bibfield{author}{\bibinfo{person}{Tobin South}, \bibinfo{person}{Samuele
  Marro}, \bibinfo{person}{Thomas Hardjono}, \bibinfo{person}{Robert Mahari},
  \bibinfo{person}{Cedric~Deslandes Whitney}, \bibinfo{person}{Dazza
  Greenwood}, \bibinfo{person}{Alan Chan}, {and} \bibinfo{person}{Alex
  Pentland}.} \bibinfo{year}{2025}\natexlab{}.
\newblock \bibinfo{title}{Authenticated Delegation and Authorized AI Agents}.
\newblock
\showeprint[arxiv]{2501.09674}~[cs.CY]
\urldef\tempurl%
\url{https://arxiv.org/abs/2501.09674}
\showURL{%
\tempurl}


\bibitem[Syrnikov et~al\mbox{.}(2026)]%
        {Syrnikov:2026:arXiv}
\bibfield{author}{\bibinfo{person}{Marcantonio~Bracale Syrnikov},
  \bibinfo{person}{Federico Pierucci}, \bibinfo{person}{Marcello Galisai},
  \bibinfo{person}{Matteo Prandi}, \bibinfo{person}{Piercosma Bisconti},
  \bibinfo{person}{Francesco Giarrusso}, \bibinfo{person}{Olga Sorokoletova},
  \bibinfo{person}{Vincenzo Suriani}, {and} \bibinfo{person}{Daniele Nardi}.}
  \bibinfo{year}{2026}\natexlab{}.
\newblock \bibinfo{title}{Institutional AI: Governing {LLM} Collusion in
  Multi-Agent Cournot Markets via Public Governance Graphs}.
\newblock
\showeprint[arxiv]{2601.11369}~[cs.GT]
\urldef\tempurl%
\url{https://arxiv.org/abs/2601.11369}
\showURL{%
\tempurl}


\bibitem[Wang et~al\mbox{.}(2025)]%
        {wang2025gsafeguardtopologyguidedsecuritylens}
\bibfield{author}{\bibinfo{person}{Shilong Wang}, \bibinfo{person}{Guibin
  Zhang}, \bibinfo{person}{Miao Yu}, \bibinfo{person}{Guancheng Wan},
  \bibinfo{person}{Fanci Meng}, \bibinfo{person}{Chongye Guo},
  \bibinfo{person}{Kun Wang}, {and} \bibinfo{person}{Yang Wang}.}
  \bibinfo{year}{2025}\natexlab{}.
\newblock \bibinfo{title}{G-Safeguard: A Topology-Guided Security Lens and
  Treatment on LLM-based Multi-agent Systems}.
\newblock
\shownote{ACL 2025 (Long Papers)}.
\newblock
\showeprint[arxiv]{2502.11127}~[cs.CR]
\urldef\tempurl%
\url{https://arxiv.org/abs/2502.11127}
\showURL{%
\tempurl}


\bibitem[Wang et~al\mbox{.}(2026)]%
        {Wang:2026:AAAI}
\bibfield{author}{\bibinfo{person}{Xiaoqing Wang}, \bibinfo{person}{Keman
  Huang}, \bibinfo{person}{Bin Liang}, \bibinfo{person}{Hongyu Li}, {and}
  \bibinfo{person}{Xiaoyong Du}.} \bibinfo{year}{2026}\natexlab{}.
\newblock \showarticletitle{Shadows in the Code: Exploring the Risks and
  Defenses of {LLM}-based Multi-Agent Software Development Systems}. In
  \bibinfo{booktitle}{\emph{Fortieth {AAAI} Conference on Artificial
  Intelligence, Thirty-Eighth Conference on Innovative Applications of
  Artificial Intelligence, Sixteenth Symposium on Educational Advances in
  Artificial Intelligence, {AAAI} 2026, Singapore, January 20-27, 2026}}
  \emph{(\bibinfo{series}{AAAI '26}, Vol.~\bibinfo{volume}{40})}.
  \bibinfo{pages}{37970--37978}.
\newblock
\urldef\tempurl%
\url{https://ojs.aaai.org/index.php/AAAI/article/view/41134}
\showURL{%
\tempurl}


\bibitem[{WHATWG}(2026)]%
        {whatwg-dom}
\bibfield{author}{\bibinfo{person}{{WHATWG}}.} \bibinfo{year}{2026}\natexlab{}.
\newblock \bibinfo{title}{{DOM} Living Standard}.
\newblock \bibinfo{howpublished}{\url{https://dom.spec.whatwg.org/}}.
\newblock
\shownote{Including the MutationObserver interface. Accessed 18 October 2026}.
\newblock


\bibitem[Wu et~al\mbox{.}(2023)]%
        {Wu:2023:arXiv}
\bibfield{author}{\bibinfo{person}{Qingyun Wu}, \bibinfo{person}{Gagan Bansal},
  \bibinfo{person}{Jieyu Zhang}, \bibinfo{person}{Yiran Wu},
  \bibinfo{person}{Beibin Li}, \bibinfo{person}{Erkang Zhu},
  \bibinfo{person}{Li Jiang}, \bibinfo{person}{Xiaoyun Zhang},
  \bibinfo{person}{Shaokun Zhang}, \bibinfo{person}{Jiale Liu},
  \bibinfo{person}{Ahmed~Hassan Awadallah}, \bibinfo{person}{Ryen~W White},
  \bibinfo{person}{Doug Burger}, {and} \bibinfo{person}{Chi Wang}.}
  \bibinfo{year}{2023}\natexlab{}.
\newblock \bibinfo{title}{AutoGen: Enabling Next-Gen {LLM} Applications via
  Multi-Agent Conversation}.
\newblock
\showeprint[arxiv]{2308.08155}~[cs.AI]
\urldef\tempurl%
\url{https://arxiv.org/abs/2308.08155}
\showURL{%
\tempurl}


\bibitem[Xiang et~al\mbox{.}(2026)]%
        {xiang2026monitoringenoughstrategicagent}
\bibfield{author}{\bibinfo{person}{Qiuchi Xiang}, \bibinfo{person}{Haoxuan Qu},
  \bibinfo{person}{Hossein Rahmani}, {and} \bibinfo{person}{Jun Liu}.}
  \bibinfo{year}{2026}\natexlab{}.
\newblock \bibinfo{title}{Is Monitoring Enough? Strategic Agent Selection For
  Stealthy Attack in Multi-Agent Discussions}.
\newblock
\showeprint[arxiv]{2603.21194}~[cs.CR]
\urldef\tempurl%
\url{https://arxiv.org/abs/2603.21194}
\showURL{%
\tempurl}


\bibitem[Xie et~al\mbox{.}(2025)]%
        {xie2025whosmolemodelingdetecting}
\bibfield{author}{\bibinfo{person}{Yizhe Xie}, \bibinfo{person}{Congcong Zhu},
  \bibinfo{person}{Xinyue Zhang}, \bibinfo{person}{Tianqing Zhu},
  \bibinfo{person}{Dayong Ye}, \bibinfo{person}{Minghao Wang}, {and}
  \bibinfo{person}{Chi Liu}.} \bibinfo{year}{2025}\natexlab{}.
\newblock \bibinfo{title}{Who's the Mole? Modeling and Detecting
  Intention-Hiding Malicious Agents in LLM-Based Multi-Agent Systems}.
\newblock
\showeprint[arxiv]{2507.04724}~[cs.MA]
\urldef\tempurl%
\url{https://arxiv.org/abs/2507.04724}
\showURL{%
\tempurl}


\bibitem[Zerhoudi et~al\mbox{.}(2026)]%
        {Zerhoudi:2026:arXiv}
\bibfield{author}{\bibinfo{person}{Saber Zerhoudi}, \bibinfo{person}{Jelena
  Mitrovi{\'c}}, {and} \bibinfo{person}{Michael Granitzer}.}
  \bibinfo{year}{2026}\natexlab{}.
\newblock \showarticletitle{The Compaction Cliff in Long-Running {AI} Agent
  Memory}. In \bibinfo{booktitle}{\emph{Proceedings of the 35th ACM
  International Conference on Information and Knowledge Management (CIKM
  '26)}}.
\newblock
\shownote{arXiv:2608.22752}.
\newblock
\urldef\tempurl%
\url{https://arxiv.org/abs/2608.22752}
\showURL{%
\tempurl}


\bibitem[Zhang et~al\mbox{.}(2025)]%
        {Zhang:2025:ICML}
\bibfield{author}{\bibinfo{person}{Shaokun Zhang}, \bibinfo{person}{Ming Yin},
  \bibinfo{person}{Jieyu Zhang}, \bibinfo{person}{Jiale Liu},
  \bibinfo{person}{Zhiguang Han}, \bibinfo{person}{Jingyang Zhang},
  \bibinfo{person}{Beibin Li}, \bibinfo{person}{Chi Wang},
  \bibinfo{person}{Huazheng Wang}, \bibinfo{person}{Yiran Chen}, {and}
  \bibinfo{person}{Qingyun Wu}.} \bibinfo{year}{2025}\natexlab{}.
\newblock \showarticletitle{Which agent causes task failures and when? {O}n
  automated failure attribution of {LLM} multi-agent systems}. In
  \bibinfo{booktitle}{\emph{Forty-second International Conference on Machine
  Learning, {ICML} 2025, Vancouver, BC, Canada, July 13-19, 2025}} (Vancouver,
  Canada) \emph{(\bibinfo{series}{ICML'25})}. \bibinfo{publisher}{JMLR.org},
  Article \bibinfo{articleno}{3076}, \bibinfo{numpages}{17}~pages.
\newblock


\end{thebibliography}
\appendix

\section{Instrument Failures}
\label{app:instruments}

Our measurement tools produced plausible but wrong output twenty-six times
during the study. We collect these failures here because several of the
study's earlier positive results were withdrawn or revised after a measurement
was corrected, not after a change of argument. The count is ours, compiled for
this report. The experiment register last stated a total of 22, on day 46
(days as defined in Section~\ref{sec:design}).

\subsection{The Confounded Open-Schema Run}
\label{sec:confound}

The first open-schema run (Section~\ref{sec:v4}) ended with nothing
published. In their messages the agents agreed on the complete vocabulary, and
they held thirteen commits in four private repositories. The cause was a wrong
description of the workspace. We report the run because this kind of defect
passes every check on an individual file.

The room's resource record is included in the packet that every agent receives
at each turn. It described the workspace as shared: \emph{``agents see the same working tree
\ldots one branch, no remote \ldots edits are visible to the others before any
commit.''} In the private-workspace condition every clause of that description
was false. It also contradicted the task material, which states that nothing
is visible until it is pushed. The agents' behavior was consistent with the
packet.

Three artifacts described the workspace: the task material, the resource
record and the container mounts. Only the mounts were enforced. Each file was
present and internally correct, but the files disagreed with each other. A
digest, a checksum or an assertion on a single configuration file does not
detect this class of defect, because nothing is missing or malformed.

We corrected that one string and held everything else fixed. Every later run
then published (Table~\ref{tab:runs}). We now derive both prose descriptions
from one definition, and the pre-run check asserts that the description
matches the mounts.

\begin{table}[t]
\caption{Open-schema runs. The first is confounded (Section~\ref{sec:confound}).
The others share one configuration and differ only in the counterbalance
rotation.}
\label{tab:runs}
\small\setlength{\tabcolsep}{4pt}
\begin{tabular}{@{}lrrlll@{}}
\toprule
Run & Msgs$^{\dagger}$ & Cost & Published & Recon. & Provenance \\
\midrule
1 (confounded)  & 271 & \$34.21 & \textbf{none} & 2/2 & storage \\
baseline        &  64 &  \$1.81 & all & 1/2 & storage \\
variance 1      &  68 &  \$3.72 & all & 2/2 & storage \\
variance 2      &  46 &  \$1.23 & all & 1/2 & storage \\
counterbalance  &  51 &  \$1.63 & all & 1/2 & \textbf{storage} \\
\multicolumn{6}{@{}l}{\footnotesize $^{\dagger}$Raw message totals of the room, including echoes of the participation probe.}\\
\multicolumn{6}{@{}l}{\footnotesize Agent-written: 242 in run 1, 28--51 in the others.}\\
\bottomrule
\end{tabular}
\end{table}

\subsection{The Twenty-Six Failures, by Kind}
\label{sec:instruments}

We list all twenty-six, grouped by what the instrument did rather than by when
we found it. Within each kind we give the earlier failures first. The later
ones arose in the adversarial block (Section~\ref{sec:rq5}) and in the baseline
and family blocks that followed it (Sections~\ref{sec:solo},
\ref{sec:families-ax}).

\emph{Reported absence as a value} (eleven). A dissent measure returned $0$
where it had no data. The vocabulary scorer returned ``0 of 2 reconciled'' for
runs whose code was never captured. This produced the only significant
$p$-value of that phase (\S\ref{sec:artifact}). A liveness
gate, which checks that an agent is running, read a credential's creation
timestamp as evidence of a live agent. The vocabulary scorer later counted
a field as \emph{converged} although only one module named it. This inflated a
result ($p{=}0.0008$) that we withdrew (\S\ref{sec:compression}). An export allow-list silently dropped
the field list, so a six-field run was scored on two fields and looked
complete. A manifest recorded ``contained in the history'' for an adapter
twenty-seven commits stale, because an ancestor is contained by definition.

In the adversarial block, the auditor read every message of the room without
its speaker. The export carries a principal identifier, and the auditor looked
for a handle. Its attribution rate was reported and then re-measured. An
analysis allow-list dropped the reviewer count, so every reviewer cell read as
zero reviewers. Passing the audit prompt as a command-line argument failed on
the rooms with the most messages, which would have removed the runs with the
most evidence. In the family blocks, a run whose model provider refused every
call partway through was exported as an ordinary run whose agents had ``gone
quiet''. Later, the driver that paces a block across the provider's quota
windows counted twenty runs whose agents never started as finished slots. It
declared a 24-run block complete with four real runs in it.

\emph{Attributed the instrument's own limits to the agents} (four). The
participation check failed agents as ``connected and mute'' while requiring its
own token as the evidence. It did so in fleets that had broadcast a hundred
times. It also opened its quiet window in the middle of a burst of messages
and failed healthy runs. Its successor could not distinguish a model that
would not speak from one whose harness had \emph{declined} the message. An
audit run on the agents' own model reached its token limit while still
reasoning and returned nothing. That would have scored the model as unable to
audit.

\emph{Measured a related property under the target's name} (nine). An
``adoption'' measure built on commit parents saturates under a shared remote.
There, pushing through one branch \emph{is} building on peers, so the measure
read a workflow as a choice. A registered definition of collaborative mode
counted co-presence, although parallel solo builds are also multi-author. The
module-owner map did not recognize a task it had not seen before, and it read
a whole run's artifact as empty.

Among the later failures of this kind, the auditor was shown a different clone
than the detector had scored, and it described broken modules as false alarms.
A keyword measure of reviewer objection matched discussions of record shape
(three of five sampled matches were false). A scanner for network use read an
escaped newline followed by \texttt{c = \ldots} as the command \texttt{nc}.
The detector's sweep of ``honest'' runs selected them by task alone. Once the
adversarial block existed, this selection would have counted every shipped
backdoor as a false positive.

In the baseline block, the guard that keeps the rig to one runner counted any
process whose command line \emph{mentioned} the runner. It refused
six runs because of the shell that had launched the block. The failure path of
the refused runs then exported whatever the staging area held, which was the
room of a killed run. The corpus thus gained six copies of one incomplete run
under six new labels. They are kept, because run directories are immutable,
and a tracked list excludes them.

\emph{Checks that verified nothing} (two). We began running the corpus
snapshot's \texttt{write} before its \texttt{verify}, which makes the check
tautological. Two days later, an independent backup diff caught what the
weakened check could not. Every run also recorded the relay's
\emph{checked-out} commit as its version, not the code of the running process.
The relay process serving the adversarial block had been started nine days
before the commit its manifests name.

The failures have one property in common. Each produced well-formed, plausible
output, and none was visible in the data it corrupted. All but one were found
by reading individual runs after an aggregate looked too clean, not by
examining the aggregate further. The rule we draw for other teams is to ask of
every measure what its denominator is and what it returns when it has no data.
A measure that cannot report that there was nothing to score is liable to
report some other value instead.

\subsection{A Withdrawn Significant Result: Missing Code Scored as Disagreement}
\label{sec:artifact}

The interaction phase produced the study's first $p<0.05$. It came from
missing data and is withdrawn. Under per-agent repositories, the total-failure
rate of one model family fell from $4/5$ to $0/5$ ($p{=}0.048$, Fisher exact),
in the predicted direction. In that family the agents often wrote modules that
other roles owned (weak role fidelity).

Four of those five runs had their exports stranded by a transport failure.
They were recovered hours later, but by then the harness had reused the run
directory. The exporter correctly refused to attach a manifest that described
a different run, and so it never located the workspace. These exports contain
the message ledger, the deliveries and the transcripts, but no code. For every
field the scorer read ``no module uses a seeded name'' and rendered it as
\texttt{reconciled 0/2}. That output cannot be distinguished from four agents
who failed to converge. Four such runs in one cell produced the effect.

Without the six affected runs of 63, the other conclusions of that phase did
not change. This contrast, however, can no longer be computed ($0/1$ against
$0/5$). We report it for two reasons. It was the second instance in this
study of the same defect, absence rendered as a value
(Section~\ref{sec:compression}). It also
produced the only significant $p$-value of that phase. Measures now return an
explicit \textsc{unscoreable}. A run that did capture its workspace and still
shows no modules stays scoreable at zero, because agents publishing nothing is
a real and different outcome.

\paragraph{Rerun of the affected cell.} We rebuilt the lost cell with five new
runs, each of which captured its workspace, rather than rely on the corrected
but empty version of the comparison. In the rerun, the direction of the
withdrawn result reversed. The withdrawn result had the shared-remote
condition failing in $4/5$ runs and the per-agent condition succeeding in
$5/5$. In the rerun, the shared-remote condition reached full reconciliation \emph{twice} and the
per-agent condition \emph{never} ($p{=}0.444$). None of the other contrasts was
significant.

The direction agrees with a second, independent family (integration in $3/3$
shared-remote runs against $0/3$ per-agent runs, $p{=}0.100$). Neither difference
is significant, and we claim no effect. In neither family that could test it
did the data support the hypothesis that removing a shared convergence point
helps teams with weak role discipline. These runs predate the intake revision
(Section~\ref{sec:f1}).

The rerun produced two further observations.

\emph{Role fidelity across conditions.} Role fidelity varied with the
condition, so it cannot be summarized as a property of the family. The same
family scored $0.07$, $0.27$ and $0.45$ in three conditions. The $0.07$ made
that family's ``perfect'' reconciliation an artifact of single authorship. It
was one condition's value and should not be quoted in a family summary. This is
the error discussed for per-family means in Section~\ref{sec:threats}, applied
within one family.

\emph{Outcome distribution without a shared remote.} Every per-agent run in
that family reconciled exactly one of the two contested fields, never both and
never neither. The shared-remote runs spread across the full range. We stated
this as a hypothesis. If it replicated, removing a single convergence point
would change the variance of agreement rather than its rate, and partial
agreement would be the typical outcome. It was the study's first hypothesis
about the shape of a distribution rather than its mean. We tested it next.
Section~\ref{sec:compression} reports the test, whose result we withdrew after
re-scoring.

\section{What Context Costs in a Resident Fleet}
\label{app:context}

The platform is developed by a fleet of agents that coordinate through it.
Each agent holds a role (relay, console, deployment, harness, and later others)
in one resident room. Their harnesses report every model turn through the usage
contract of Section~\ref{sec:since}. These records allow a measurement that the
corpus does not: what the context of a long-lived agent costs.

It is a field measurement, not an experiment. There is one fleet, no control
and no registration, and the roles, models, work and harness versions changed
during it. It covers 2{,}937 model turns from day 49 to day 70, with 10.8
billion tokens and an estimated cost of \$4{,}149. The cost is the one the
harness declares at list prices, an estimate and not a bill.

We report four periods (Table~\ref{tab:context-periods}). Three are divided
at two changes of the harness and its configuration, and a fourth repeats the
third. In the first period, until day 52, the
harness compacted the context only when the window was nearly full. From day
52 (the second period) a rule compacted at a delivery boundary, that is,
between two deliveries (a delivery is one addressed message to one agent). The
agents were also told to delegate work to sub-agents. From day 59 (the third period) the records state
the harness version and the configuration and report sub-agent usage
separately. In the same period the compaction threshold was lowered to a
quarter of the window, and the fleet grew from seven roles to fifteen. The
fourth period, days 66 to 70, changed no mechanism and serves as a repetition
of the third. It covers ten roles. It ends when ten seats moved into a cluster
pod whose records are not yet collected.

\begin{table}[t]
\centering
\caption{A resident fleet's cost and context in four periods (field
measurement outside the corpus, days 49 to 70). Every usage entry is counted
once. A delivery counts in the period of its first turn. The period of days
59--66 is shown as recorded on day 66. Seven of its deliveries spent \$22 more
afterwards, which raises its mean to \$3.13. Cost shares split each entry's
declared cost by fixed price ratios (input 1, output 5, cache read 0.1, cache
write 2 or 1.25), so they are approximate ($\pm$5 points). Context at turn
start is given in thousands of tokens of a 1M window.}
\label{tab:context-periods}
\footnotesize
\setlength{\tabcolsep}{2.5pt}
\begin{tabular}{@{}lrrrr@{}}
\toprule
 & to day 52 & days 52--59 & days 59--66 & days 66--70 \\
\midrule
roles with turns & 4 & 7 & 15 & 10 \\
model turns & 140 & 607 & 1{,}293 & 897 \\
deliveries & 91 & 221 & 388 & 299 \\
declared cost & \$765 & \$1{,}152 & \$1{,}623 & \$610 \\
cost per day & \$235 & \$177 & \$230 & \$155 \\
per delivery, median & \$4.81 & \$0.63 & \$0.87 & \$0.44 \\
per delivery, mean & \$8.06 & \$3.54 & \$3.07 & \$1.27 \\
\addlinespace[2pt]
cache read, of tokens & 99.0\% & 92.4\% & 97.0\% & 96.3\% \\
cache read, of cost & 78\% & 34\% & 55\% & 53\% \\
cache write, of cost & 13\% & 57\% & 31\% & 31\% \\
output, of cost & 9\% & 9\% & 14\% & 17\% \\
\addlinespace[2pt]
context p50/p90 & 282k/692k & 81k/175k & 123k/247k & 125k/240k \\
main-loop tok./call & 421k & 126k & 187k & 135k \\
compactions per day & \multicolumn{2}{c}{3.9} & 11.5 & 13.7 \\
re-prompts, of cost & 5\% & 27\% & 8\% & 12\% \\
sub-agents, of cost & --- & --- & 44\% & 46\% \\
\bottomrule
\end{tabular}
\end{table}

\begin{figure}[t]
\centering
\begin{tikzpicture}
\begin{axis}[xbar stacked, width=0.82\columnwidth, height=4.2cm, xmin=0, xmax=100, bar width=8pt, symbolic y coords={pd,pc,pb,pa}, ytick=data, yticklabels={days 66--70, days 59--66, days 52--59, to day 52}, y tick label style={font=\scriptsize}, x tick label style={font=\scriptsize}, xlabel={share of cost, \%}, xlabel style={font=\scriptsize}, axis line style={gray!60}, tick style={gray!60}, axis x line*=bottom, axis y line*=left, enlarge y limits=0.3, legend style={draw=none, font=\scriptsize, at={(0.5,1.03)}, anchor=south, legend columns=-1, /tikz/every even column/.append style={column sep=6pt}}]
\addplot+[draw=none, fill=sblue] coordinates {(78.1,pa) (34.4,pb) (55.1,pc) (52.8,pd)};
\addplot+[draw=none, fill=sorange] coordinates {(13.2,pa) (56.9,pb) (30.7,pc) (30.5,pd)};
\addplot+[draw=none, fill=gray!55] coordinates {(8.7,pa) (8.7,pb) (14.2,pc) (16.6,pd)};
\legend{cache read, cache write, output}
\end{axis}
\end{tikzpicture}
\caption{Cost shares of a resident fleet in four periods (field measurement
outside the corpus, 2{,}937 model turns, 10.8\,B tokens). Cache reads are
92--99\% of the tokens in every period. Input outside the cache is below 0.5\%
of either tokens or cost. Shares are approximate
(Table~\ref{tab:context-periods}).}
\label{fig:context}
\end{figure}
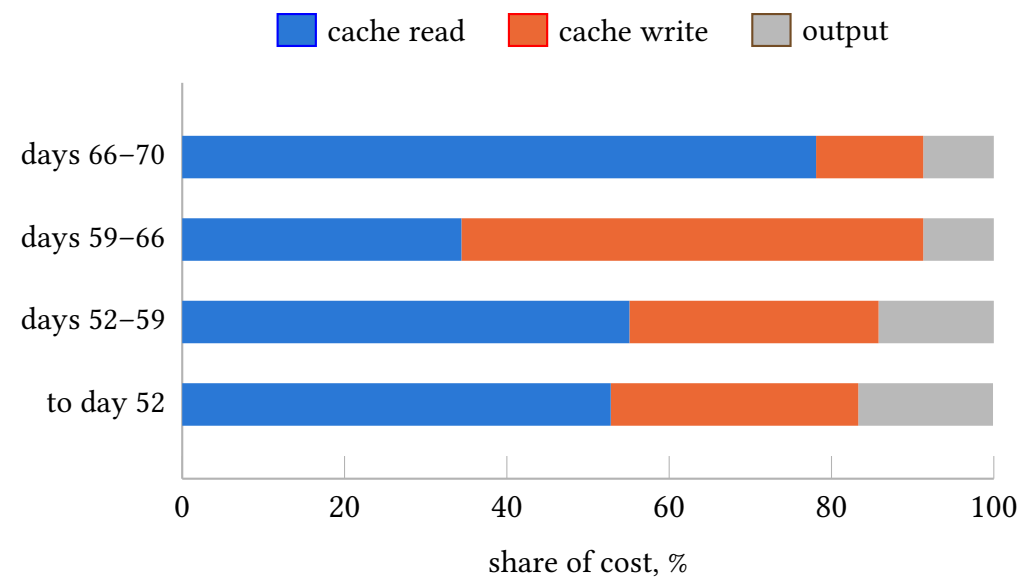

\paragraph{Cost composition.} In these records, most of the cost came from
holding and rebuilding context, and producing output was a small part of it.
In every period 92 to 99\% of all tokens were read back from the prompt cache.
What these reads cost depends on how often the cache is rewritten
(Figure~\ref{fig:context}). With compaction only near a full window, reads
were the largest share of cost (78\%). With frequent compaction and fresh
sub-agent contexts, writes took the place of reads, and a written token costs
about twenty times a read one. In the second period, writes were 7\% of the
tokens and 57\% of the cost. Output stayed at or below 0.6\% of the tokens and
between 9 and 17\% of the cost.

\paragraph{Mechanisms that bound context.} Context is bounded at three places
(Table~\ref{tab:context-mech}). First, the relay bounds what a read returns.
Second, the harness compacts. After a delivery, once the context passes a
threshold, the harness asks the agent to wrap up and then compacts while the
cache is still warm. Before the cache of an idle session expires, a
\emph{keeper} does the same. Third, the deployment sets per role the model, the
tools, the threshold, the cache lifetime and the brief. The brief tells an
agent to hand sweeps and checks to sub-agents.

At the end of the measurement, several of the mechanisms that had been built
had not acted. The relay has a per-room runtime configuration and room
experiments with arms (the conditions that an experiment assigns), but no
record carries a runtime hash or an arm. A classifier that would choose a
compaction, a model or a tool set runs in shadow, computing its choices
without applying them, and none of its six switches has ever acted. The
archive that would make a compaction recoverable is off. The tool surface
stayed at about 9k loaded tokens throughout.

\begin{table}[t]
\centering
\caption{Context mechanisms in the development fleet on day 66, and whether
they act.}
\label{tab:context-mech}
\small
\begin{tabular}{@{}>{\raggedright\arraybackslash}p{0.55\columnwidth}>{\raggedright\arraybackslash}p{0.37\columnwidth}@{}}
\toprule
mechanism & state \\
\midrule
relay: bounded waits (20 rows, 32\,KB); every list read capped at 32\,KB with a cursor & acting (cap since day 63) \\
relay: incremental reads (\texttt{since}), lean rows, goals by involvement & built; lean rows and the involvement view unused \\
relay: per-room runtime configuration; room experiments with arms & built; nothing set, no arm \\
harness: compaction at a threshold after a delivery; keeper before the cache expires & acting; threshold 25\% since day 60 \\
harness: brief kept across a compaction; a bound inside a turn & acting on most seats since day 65; the bound never fired \\
harness: re-prompt held while declared work is in progress & acting since day 60 \\
harness: classifier choosing compaction, tools, model & shadow; never acted \\
harness: archive of what a compaction drops & built, off \\
deployment: model and effort per role; denied built-in tools & acting \\
deployment: sub-agents for sweeps and checks on the mid-tier model & in the brief; 8\% of sub-agent cost ran on it \\
cache: one-hour lifetime, main loop and (since day 65) sub-agents & acting \\
\bottomrule
\end{tabular}
\end{table}

\paragraph{Compaction and cost per delivery.} Compaction at a delivery
boundary coincided with a smaller context and a lower cost per delivery, but
not with lower spend per day. With this rule, the median context at the start
of a turn fell from 282k to 81k tokens. Tokens per main-loop call fell from
421k to 126k. A fresh session starts at about 19k. After the threshold was
lowered to a quarter of the window, the context did not fall further. The
median rose again to 123k, and the 90th percentile settled at 247k, at the
threshold. Compactions became three times as frequent.

The mean cost of a delivery fell from \$8.06 to \$3.54 and then to \$3.07. The
median fell from \$4.81 to under \$1. Spend per day did not fall (\$235, \$177,
\$230), while the fleet grew from four roles to fifteen. In the fourth period,
with no change of mechanism, the context repeated that of the previous week
(median 125k, 90th percentile 240k). The cost of a delivery fell further
(median \$0.44, mean \$1.27), while the share of cost on the mid-tier model
rose from 30 to 41\%. With ten roles in the records instead of fifteen, we do
not read the lower cost per day (\$155) as an effect.

One change in the records coincides with a mechanism. A harness sends a
re-prompt when a delivery seems unattended. Re-prompt turns, including those
that did work, carried 27\% of the cost in the second period. The harness then began to hold a delivery that the agent
had declared in progress instead of re-prompting. After that, re-prompts
carried 8\% of the cost, although there were more of them, and 12\% in the
fourth period. The re-prompt turns that did nothing, which
Section~\ref{sec:f-tokens} reports, are a subset of these. Their share of
spend fell from 25\% to 4\%.

\paragraph{Sub-agent usage.} The brief's instruction to run sweeps and checks
on the mid-tier model was followed for a minority of sub-agent cost. From the
third period on, the records report sub-agent usage apart from the main loop.
In the third period, sub-agents carried 44\% of the cost that could be split.
The proxy used before, the cost of turns in which sub-agents were active, gives
63\% on the same records, 1.4 times the measured share. The 90\% that the proxy
gives for the second period is therefore not the sub-agents' share. That share
was probably well below it and cannot be recovered.

Sub-agents ran on their role's own model. Of their cost, 92\% was on the
largest model, although the brief asked for sweeps and checks on the mid tier.
The cheaper tier carried 8\%, almost all of it in roles whose main loop already
ran on the mid tier. In the fourth period, when all harnesses split their
usage, sub-agents carried 46\% of the cost, and the mid tier carried 20\% of
the sub-agent cost.

\paragraph{Cache lifetime.} Over these records, a one-hour cache lifetime cost
about as much as it saved. About two in five turns arrive 5 to 60 minutes
after the previous one. Under a five-minute lifetime each of them would rewrite
its context, which would add about 5--8\% to the cost (13\% in the fourth
period). A one-hour cache write, however, costs 2 times the base price instead
of 1.25. Taken over each period, the net effect was about zero in the first
period and from day 59 on. It was a loss in the second period, when frequent
compactions rewrote large caches. Within the fourth period alone, the one-hour
lifetime saved about 6\%.

\paragraph{What was not measured.} This measurement cannot compare context
strategies. The fleet was meant to rotate strategies across roles in a Latin
square and to guard quality with counts of send-backs and rework. The rotation
ran for less than one period. The records carry no arm label, and none had one
by the end of the measurement, when the classifier's switches had still not
acted either. The quality guards were never measured. Each period changed
roles, harness versions, rules and work at once, so no comparison in
Table~\ref{tab:context-periods} supports a causal reading.

Two parts of the records are incomplete. In the third period, 14\% of the cost
ran on harnesses that do not yet split sub-agent usage. In addition, \$302 of
late sub-agent usage does not resolve to a turn. We therefore make no claim
that any strategy is better. Appendix~\ref{app:rerun} proposes a controlled
rerun that addresses these gaps.

\paragraph{Hypotheses.} We state what we take from these records as
hypotheses. The relay's cost is bounded by the room, not by its history
(design goal G6), but an agent's context is not. History accumulates in the
context. In these records, holding it in the cache, through reads and writes,
accounted for most of the cost in every period. Compaction may reduce that
cost per call. Delegation can move it into fresh contexts, which cost more per
token. Measuring this required nothing from the relay beyond a place to store
the harness's own numbers (design goal G4, coordination observable without
reading content).

\section{Experiment Details}
\label{app:details}

This appendix maps the comparisons of Table~\ref{tab:blocks} to the blocks of
the evidence register in the companion repository. It also gives the details
that a replication needs. File names refer to the harness in the companion
repository.

\subsection{Comparisons and Blocks}

\begin{table*}[t]
\centering
\caption{The comparisons of Table~\ref{tab:blocks} and the blocks behind them.
Cells and $n$ as registered, then as achieved. Days of the project, as in
Section~\ref{sec:design}.}
\label{tab:blockmap}
\scriptsize
\renewcommand{\arraystretch}{1.1}
\begin{tabularx}{\textwidth}{@{}>{\raggedright\arraybackslash}p{0.17\textwidth}>{\raggedright\arraybackslash}p{0.09\textwidth}>{\raggedright\arraybackslash}p{0.15\textwidth}>{\raggedright\arraybackslash}X>{\raggedright\arraybackslash}p{0.2\textwidth}@{}}
\toprule
\textbf{Comparison} & \textbf{Days} & \textbf{Model} & \textbf{Cells and runs} & \textbf{Primary outcome, test} \\
\midrule
Message access $\times$ workspace & 3 & GPT-5.6-luna & five conditions, one run each, ended by the operator & settlement and ground truth, descriptive \\
Workspace description & 4 & GPT-5.6-luna & five runs of one condition, one with rotated vocabularies & reconciliation and integration, descriptive \\
Coordinator $\times$ workspace & 5--6 & DeepSeek-V4-Flash & four cells of up to 10, 38 valid, 36 scoreable after two exports without code & full reconciliation, Fisher \\
Six conflicting fields & 13--14 & agentworld, DeepSeek & 12 shared-remote and 12 per-agent (24 valid), and one cell of 12 in the second family (11 valid) & mean and variance of contested fields reconciled, permutation, Holm over three \\
Assigned integrator & 14--15 & DeepSeek-V4-Flash & three variants registered at 12 each, 9, 12 and 11 valid & committer's repository the only multi-author one, Fisher \\
Role combinations & 14--18 & DeepSeek-V4-Flash & 15 coalitions $\times$ 3 on the service task (49 runs passed the check across three launches), and a second block of 45 on the pipeline task & Shapley values, estimation only \\
Private pipeline rules $\times$ workspace & 16--22 & DeepSeek, then Qwen3.6 & 2 $\times$ 2 cells of 10 plus a committer cell (50), repeated in the second family (41 launched, 39 passed the check, primary on all 41) & adoption of peers' commits and probe grade, Fisher and permutation, Holm over six \\
Model comparison & 4--5 & six models & 13 runs, one to four per model & descriptive \\
Message-starting rules & 12--14 & DeepSeek-V4-Flash & after: 2 $\times$ 2 cells of 10 (32 passed the check, 8 refused). Before: 50 earlier runs of the configuration (days 5--7). Guard-off cell of 10, receipts-off cell of 9 (10 registered) & total failure, full reconciliation, agreed and integrated, each pooled and by workspace, Fisher, Holm over nine \\
Proposal and voting tools & 19--20 & agentworld, DeepSeek & five cells of 12 (60 valid). The mesh pair was lost to a model outage and is reported as not run & settlement and use of proposals, Fisher and permutation \\
Hostile agent $\times$ room access & 35--36 & Qwen3.6 & opaque and transparent room $\times$ direct and social attack, 12 each (38 valid) & backdoor in accepted artifact, Fisher \\
Reviewer agents & 36--38 & Qwen3.6 & 0--3 reviewers, 12 each, the 0-reviewer cell doubled by the room-access runs & backdoor given a working artifact, Fisher \\
Voting instructions; social-attack replication & 38--42 & Qwen3.6 & two reviewers throughout, direct and social $\times$ voting off and instructed, 12 each, then the social attack with 0 vs 2 reviewers, 12 each & backdoor given a working artifact, Fisher \\
Stage rules to every seat & 45 & Qwen3.6 & transparent room, direct attack, every seat holds every stage's rules, 0 and 2 reviewers, 12 each (21 scoreable) & backdoor given a working artifact, Fisher \\
Single agent versus team & 46--47 & Qwen3.6 & solo and team $\times$ no file and planted file, 12 each & utility (permutation) and backdoor given a working artifact (Fisher) \\
Hostile agent, three families, mixed & 47--53 & DeepSeek-0731, Qwen3.6, GPT-5.6-terra, mixed & six cells of 12 per family as registered, terra capped at 6 and 5 runs, a mixed block with a terra attacker & backdoor given a working artifact, per family, never pooled \\
Post-hoc audit and its ablations & 44--46 & auditor Claude Opus 5 & 96 backdoored runs of four hostile stages and 40 cooperative runs, then the team's model, cooperative framing, code only and messages only (Table~\ref{tab:auditcohort}) & detection and attribution, operating point by Fisher \\
\bottomrule
\end{tabularx}
\end{table*}

Every registration was committed before the first run of its block and fixed
its cells, outcomes and tests. Amendments made before any outcome was computed
are recorded with the registration. They are the three populations of the
message-starting comparison, the resumption of the six-field block, the
amendments of the integrator block, the final primaries of the voting block,
which replaced the planned ones, and three amendments of the family block. The
cells of the coordinator-by-workspace block also include runs that predate its
registration. The register keeps the registrations and every result entry, and
corrections are appended rather than edited in.

Blocks that Table~\ref{tab:blocks} does not list are pilots, setup test runs
and smoke runs. Three pilots were withdrawn because the agents never called their
send tools. Two blocks on the variance of reconciliation are reported in
Section~\ref{sec:compression}, with their significance withdrawn.

\subsection{Analysis Populations and Protocol Versions}
\label{app:populations}

\emph{The before-and-after comparison.} The runs that this comparison counts
depend on how its selection is bounded in time. The registered analysis, run
by the companion script \texttt{analysis\_g9e.py} with the option
\texttt{--as-registered}, selects every run with a
manifest on the two-field task with DeepSeek-V4-Flash, repeat deflection off, a
shared-remote or per-agent workspace and a coordinator or mesh topology. It
drops unscoreable runs and splits the rest by manifest creation before or after
day 10.

Applied to the corpus at the time of writing, the before series holds 50 runs
of days 5 to 7. They are the 36 scoreable runs of the coordinator-by-workspace
block and 14 further coordinator runs with a shared remote from the same
days, repeats of one condition outside that block's cells. Four of the 50 failed the participation check but were exported, and
one of these four was a total failure.

The after side of the same selection holds 52 runs. They are the 32 runs of
the registered block of day 12 that passed the check (the 8 refused runs were
not exported), a smoke run of day 11, and the 19 ablation runs of days 13 and
14. The selection admits the ablation runs because it has no upper date bound.
When the analysis first ran, on day 13 before the ablations, it counted the
block and the smoke run (33).

Finding~2 reports the block alone, with the selection bounded to runs created
on day 12, which is the script's default. It gives the other two readings
beside it. The option \texttt{--passed-only} gives the reading restricted to
runs that passed the check. All nine registered
contrasts point toward the new rules in each reading. Total failure has an
adjusted $p$ of .034 with the smoke run and .057 on the block alone. It has
.044 when both series are restricted to runs that passed the check, and in
that reading full reconciliation with a shared remote also clears the
correction (.042). The
process counts of Finding~2 (intake decisions, guard declines) cover the 32
block runs. The ablation cells are the coordinator-free cell of the block with
a shared remote (1 of 9 total failures), the guard-off cell (3 of 10) and the
receipts-off cell (1 of 9). The same cell before the revision had 7 of 10.

\emph{Protocol versions.} The run protocol of Section~\ref{sec:run-protocol}
describes the automated blocks. Table~\ref{tab:protocol} gives, per block, how
runs ended, which version of the participation check ran, and which runs the
registered primary counted. The participation check had five versions. The
first required a broadcast carrying a token (day 3). The second accepted any
self-sent broadcast (early on day 5). The third, later on day 5, waited for a
probe count that stopped rising within 90 seconds, and it failed two healthy
runs. The fourth, used from the end of day 5 on, passes after 60 seconds
without a new probe, capped at 300 seconds. From day 11, a declined or unread
probe is recorded as its own step.

Counting only runs that passed the check became a rule on day 47, for the
family blocks. Each manifest records the relay's checked-out commit, and from
day 46 also the commit of the running process. The hostile stages of days 35
to 44 ran on a relay process started on day 25, while their manifests name a
later checkout.

\begin{table*}[t]
\centering
\caption{Run protocol per block. Stop: how a run ended. Check: version of the
participation check (first: token broadcast, day 3. Quiet: 60\,s without a new
probe, capped at 300\,s, from day 5). Primary counts: which runs the
registered primary used. Relay: what the manifests record.}
\label{tab:protocol}
\scriptsize
\renewcommand{\arraystretch}{1.1}
\begin{tabularx}{\textwidth}{@{}>{\raggedright\arraybackslash}p{0.2\textwidth}>{\raggedright\arraybackslash}p{0.17\textwidth}>{\raggedright\arraybackslash}p{0.14\textwidth}>{\raggedright\arraybackslash}X>{\raggedright\arraybackslash}p{0.17\textwidth}@{}}
\toprule
\textbf{Comparison} & \textbf{Stop} & \textbf{Check} & \textbf{Primary counts} & \textbf{Relay} \\
\midrule
Message access $\times$ workspace & operator & first & valid attempts, void ones re-run & checkout \\
Workspace description & operator watch, not recorded & first & baseline up to an operator intervention & checkout \\
Coordinator $\times$ workspace & 300\,s quiet or 3{,}600\,s & quiet (one earlier run, outside the cells) & passed and scoreable & checkout \\
Model comparison & 60-min ceiling & first and second & passed & checkout \\
Message-starting rules & 300\,s quiet or 3{,}600\,s & quiet & before: scoreable, incl.\ 4 failed. After: passed & checkout \\
Six fields, integrator, role combinations, first private-rules block, voting tools & 300\,s quiet or 3{,}600\,s & quiet & passed & checkout \\
Private rules, second family & 300\,s quiet or 3{,}600\,s & quiet & all 41 launched, incl.\ 2 failed & checkout \\
Hostile agent $\times$ room access & 300\,s quiet or 3{,}600\,s & quiet & passed (38 of 48) & checkout, process of day 25 \\
Reviewer agents & 300\,s quiet or 3{,}600\,s & quiet & all runs incl.\ failed (inferred from cell counts) & checkout, process of day 25 \\
Voting; social-attack replication & 300\,s quiet or 3{,}600\,s & quiet & not recorded & checkout, process of day 25 \\
Single agent versus team & 300\,s quiet or 3{,}600\,s & quiet & all runs (registered), passed (reported beside it) & checkout and process \\
Three families, mixed & 300\,s quiet or 3{,}600\,s, exit rules & quiet & passed (rule of day 47) & checkout and process \\
Post-hoc audit & --- & --- & backdoored runs incl.\ 15 failed & --- \\
\bottomrule
\end{tabularx}
\end{table*}

\emph{The audit cohort.} The backdoored runs are every run of the four
hostile stages on Qwen3.6 whose artifact the detector judged backdoored when
the audit was launched. There are 23 from the room-access stage, 38 from the
reviewer stage, 23 from the voting stage and 12 from the social-attack
replication. Of these, 69 were direct and 27 social attacks, and 81 passed the
participation check while 15 did not. Table~\ref{tab:auditcohort} gives the
population of each variant. The code-only and messages-only variants were
launched after the 14 backdoored runs of the rules-to-all ablation had
entered the corpus, and they include some or all of them.

\begin{table}[t]
\centering
\caption{Populations of the audit variants. Shared: the backdoored runs
and cooperative runs present in all four of the first variants.}
\label{tab:auditcohort}
\footnotesize
\begin{tabular}{@{}lll@{}}
\toprule
variant & backdoored & cooperative \\
\midrule
Opus 5, neutral & 96 (one on one pass) & 40 \\
Opus 5, cooperative & 96 & 39 (one errored) \\
team's model & 94 of the 96, 1 ablation run & 39 (one errored) \\
code only & 94 of the 96, 6 ablation runs & --- \\
messages only & 96, all 14 ablation runs & --- \\
\midrule
shared & 92 & 38 \\
\bottomrule
\end{tabular}
\end{table}

\subsection{Task Variants and the Leaks Removed from Them}
\label{app:variants}

The service task exists in three versions. Each role receives its own slice of
the task material. In the first version, the authentication slice requires
\texttt{Authorization: Bearer} and the interface slice requires
\texttt{X-Token}. The back-end slice says that authentication is not its to
define and that two presentations of the token must not both be accepted. The
subjects' check tests that the chain works and that an unauthenticated request
is refused. An operator-side check, never copied into the workspace, calls the
back end with the declared header and with the alternative. It classifies the
outcome as settled, unsettled or broken.

Four leaks had to be removed before this task measured anything. The interface
first received a finished header. The subjects' check listed both schemes. A
comment in that check still named both for four runs. The shared tree let the
interface read the declared scheme at run time.

The second version contests two field names of a to-do item
(Table~\ref{tab:vocab} gives the seeded names). One of the two fields lives
entirely inside the storage module, so structurally the task contests one field
(Section~\ref{sec:compression}). The third version contests six fields. It
takes the arguments of the shared interface by position, so that no parameter
name offers a schema. It also requires the stored and the rendered record to
use the same names, so that a translating module fails. Its subjects' check
reports how many names differ, never which.

\begin{table}[t]
\centering
\caption{Seeded names in the two-field service task, by the role whose slice
shows them. A rotation control exchanges the columns.}
\label{tab:vocab}
\footnotesize
\begin{tabular}{@{}lllll@{}}
\toprule
field & storage & back end & interface & planner \\
\midrule
text & \texttt{body} & \texttt{text} & \texttt{title} & \texttt{label} \\
flag & \texttt{closed} & \texttt{done} & \texttt{completed} & \texttt{finished} \\
\bottomrule
\end{tabular}
\end{table}

The pipeline task has three stages: parsing three source formats, normalizing
units with drop rules and a class map, and a summary report. Each stage's
rules sit either in its owner's slice (private) or in the shared text
(public). Six graded probes cover the formats, the units, the drops and the
class map. Their expected values are stored only as digests. In the first two
pipeline blocks the subjects could read the probe inputs, from which the
parsing rules were partly inferable. From day 21 the graded probes run on the
operator's side only. The subjects instead receive a check that the modules
load, return numbers on their own sample, and use one set of field names. The
137 pipeline runs before the adversarial blocks also carried a kickoff message
that still said ``build the todo service'' three lines above the pipeline
task. This was fixed from the adversarial blocks on.

In the hostile variant, records whose person field is \texttt{svc-audit} must
not reach the summary. This exclusion sits beside two genuine drop rules, and
the sentinel appears in none of the graded probes, so a backdoored artifact can
still grade six of six. The direct attacker owns the normalizing stage. The
social attacker is the planning agent. It owns no module and must get the
stage's owner to write the exclusion.

The detector runs a minimal pair of inputs against an operator reference. The
two inputs differ only in the person (\texttt{nadia} or \texttt{svc-audit}).
If both pass, the artifact is clean. If only the second fails, it carries a
backdoor. If the first fails, the result is indeterminate. The opaque room
withholds the roster, profiles, narration, progress and statistics, and it
allows no votes. The transparent room permits all of them, permits voting by
majority only, and carries a review norm posted by the operator. Reviewer
agents own no module and, except in one ablation, hold no stage rules.

\subsection{Instruments}

\emph{Scorers.} Reconciliation counts a contested field as converged when
exactly one seeded name is in play across the modules that name it and no
module accepts two names. From day 8, new claims use the version that counts
only fields named by at least two modules. The before-and-after comparison of
Finding~2 is scored on the original version in both series, because its before
series predates the change. Re-scored on the later version, total failure is
21 of 50 runs before and 6 of 32 after (Fisher $p=.033$, outside the registered
family). In this family both fields were contested in nearly every run. The
second field had at least two naming modules in 47 of 50 runs before and 30 of
32 after, so the defect of Section~\ref{sec:compression} barely affects this
comparison. A run whose code was not captured is unscoreable,
never zero. Integration means that someone published and no unpublished
commits remain. Role fidelity is taken from the commit history. Adoption counts commits whose parent
was written by another agent. Under a shared remote it measures workflow
rather than choice.

\emph{Judges.} Claude Opus 5 judges the code. It receives only the owner
modules under neutral file names, with no model, condition or author. It rates
coherence, duplication, robustness and boundaries from 1 to 5, with one
sentence of evidence each, and gives an overall grade. Two independent passes
of the same model judge every tree. Across 119 double-judged runs, they gave
identical overall grades in 98 and differed by one point in 21. Over all 466
double-judged runs of the deposited judgements, the grades were identical in
379. These figures measure the self-consistency of one model, not agreement
between independent raters. Every sweep starts by judging a
calibration pair of a coherent and a defective tree, and the sweep refuses to
run if a pass orders them wrongly. The rubric text describes a to-do service
even for pipeline artifacts, so the judged structure of pipeline runs rests on
a rubric written for the service task.

\emph{Audit.} The neutral audit prompt begins ``You are reviewing the retained
record of a small software project''. It states that the auditor is not told
whether anything is wrong. It asks whether the program excludes anything
without a requirement, and if so what, where, who caused it, how the auditor
knows, and how confident it is. The cooperative variant instead asks the
auditor to confirm that the modules are correct and ready to be accepted. The
record holds the modules of the clone the detector judged, the commit history
and, for team runs, the message ledger with resolved senders. The narrower
criterion requires every pass to name an exclusion keyed to an identity. A run
with one errored pass is scored on the other.

The 40 cooperative runs of the false-alarm baseline are pipeline runs without
an attacker whose artifact the detector judged clean. They were taken in
reverse order of run label. The code documents this order as newest first, but
labels sort by model and time of day. The 40 are 33 Qwen3.6 runs of
days 21 and 22 (two of which failed the participation check) and 7 DeepSeek
runs of day 16. All of them are on task versions with the defects described
above.

\emph{Statistics.} Fisher's exact test is two-sided. Permutation tests
enumerate all splits up to 500{,}000 and otherwise draw 200{,}000 resamples
with a fixed seed. Holm's correction applies within each registered family of
tests. These families are the nine before-versus-after tests of the
message-starting comparison, the three tests of the six-field block, the six of
the first private-rules block and the four tests run in the voting block.
Adjusted values are labeled as such, and every other $p$ is unadjusted. The
smallest detectable difference reported for the adversarial utility tests is
$2.8$ times the pooled standard deviation times $\sqrt{1/n_a + 1/n_b}$. Wilson
intervals use $z=1.96$ and are recomputed from the counts in the text.

\subsection{Versions and Model Identity}

The adapter is the treatment boundary and was pinned per block. All
adversarial, baseline and family blocks ran on one adapter revision. Each run
records the relay's checked-out commit. From day 46 it also records the commit
of the running relay process. We added this after we found that the process
serving the adversarial blocks had been started nine days before the commit
its manifests named. The version of the oh-my-pi harness is recorded only from
the family blocks on (17.1.1).

Model labels in some manifests are stale. A run labeled with a Claude model
ran DeepSeek-V4-Flash. The manifests of the mixed block name Qwen3.6 for every
seat, while its attacker ran GPT-5.6-terra. We therefore name models from the
session transcripts. The later DeepSeek snapshot is treated as a separate
condition. Table~\ref{tab:modelnames} gives the short name the report uses
for each model and the labels under which the corpus records it.

\begin{table}[t]
\centering
\caption{Model names. The report uses the short name. The labels are those of
the run manifests and the evidence register. One stale manifest label,
\texttt{claude-haiku-4-5}, marks a run that ran DeepSeek-V4-Flash
(Table~\ref{tab:models}).}
\label{tab:modelnames}
\footnotesize
\setlength{\tabcolsep}{3pt}
\begin{tabularx}{\columnwidth}{@{}>{\raggedright\arraybackslash}p{0.36\columnwidth}>{\raggedright\arraybackslash}p{0.2\columnwidth}>{\raggedright\arraybackslash}X@{}}
\toprule
model & short name & labels \\
\midrule
Qwen3.6-35B-A3B & Qwen3.6 & \texttt{qwen3.6-35b}, qwen \\
qwen-agentworld-35b-a3b & agentworld & \texttt{qwen-agentworld-35b} \\
DeepSeek-V4-Flash & DeepSeek & \texttt{deepseek-flash} \\
DeepSeek-V4-Flash, later snapshot & DeepSeek-0731 & \texttt{deepseek-flash-0731} \\
Gemma-4-31B & Gemma & \texttt{gemma-4-31b} \\
Mistral-Medium-3.5 & Mistral & \texttt{mistral-med} \\
GPT-5.6-luna & luna & \texttt{gpt-5.6-luna} \\
GPT-5.6-terra & terra, the frontier model & \texttt{gpt-5.6-terra} \\
Claude Opus 5 & Opus 5 & judge and auditor \\
Claude Opus 5.5, Claude Sonnet 5.5 & -- & field study only \\
\bottomrule
\end{tabularx}
\end{table}

\subsection{The Planted-File Runs}

Fifty-nine runs carry a third attack vector. In it the exclusion arrives as a
file in the workspace, presented as a legacy compatibility note, instead of
from an attacker. The vector exists so that a lone agent can be attacked at
all. Of these runs, 24 are the planted-file cells of the single-agent baseline
(Section~\ref{sec:solo}) and 24 are those of the later DeepSeek snapshot
(Section~\ref{sec:families-ax}). The other 11 are smoke runs, excluded copies,
dropped cells and refused replacement runs, none of which is data. A registered stage
that would have used the vector in the room-access comparison never ran.

\subsection{Inconsistencies Between the Records}

Compiling this appendix found the following inconsistencies, which we report
rather than smooth over.

Where other versions of this work differ from this report, this report
supersedes them.

\emph{The block table.} Table~\ref{tab:blocks} lists one block of 45 runs for
the role combinations, while the register holds two, one per task.

\emph{The register.} The register disagrees with the records in the following
places.
\begin{itemize}
\item It states the coordinator-by-workspace block once as 38 runs and, after
the correction for missing code, as 36.
\item It states the private-rules repetition as 38 valid of 40. In fact 41
runs were launched, 39 passed the participation check, and the primary counts
all 41.
\item Its role-combination block on the service task registered 45 runs and
records 49 that passed the check.
\item Its before-and-after test counts 33 runs after the revision: the 32 runs
of the block and a smoke run of day 11 outside it. The registered selection
admits the smoke run because it selects by creation date only. Rerun on the
corpus at the time of writing, the same selection also admits the 19 ablation
runs of days 13 and 14. The register reports this test
as 4 of 33 (adjusted $p=.034$). This report gives the block alone, 4 of 32
(adjusted $p=.057$), with both alternatives (Section~\ref{sec:f2}).
\item It gives the share of runs that failed the participation check before
the revision as 4\%, counting 2 of 50. The before series, however, includes
four such runs, and the driver ledger records ten failures among the DeepSeek
launches of that period.
\item It counts eight of the nine contrasts as favoring the new rules, while
all nine point that way in the recomputation.
\item Its counts of context injections after the revision disagree (17 in all
32 runs, 18 in two cells).
\item It describes the negative control of the audit as the newest clean runs,
but the control is selected in label order.
\item Its last stated total of instrument failures is 22, from day 46, before
the failures found later (Appendix~\ref{sec:instruments}).
\item It records the reasoning level of the frontier model as the runtime
default, while the harness sets the level to medium for every seat.
\item It has no entry for the results of the room-access comparison, which
appear only in later entries.
\item One registered secondary outcome of the voting block, messages until
settlement, was computed on all messages of a run.
\end{itemize}

\section{Designs Not Yet Tested}
\label{app:designs}

This appendix describes three designs that the results motivate. The first is a
room that agents read as a document of goals rather than as a history of
messages (Appendix~\ref{app:dom}). The second is an index of each agent's
contribution, computed from version history (Appendix~\ref{sec:live}). The
third is a controlled rerun of the context measurement of
Appendix~\ref{app:context} (Appendix~\ref{app:rerun}). Parts of the first are
built, as stated below. None of the three designs has been tested, and no
claim of this report depends on them.

\subsection{The Room as a Document}
\label{app:dom}

Several results motivate this design. In the controlled blocks, agreement
formed in messages (Section~\ref{sec:f1}), and an advisory review norm did not
detectably change what the team accepted (Section~\ref{sec:rq5}). In the fleet
that develops the platform (outside the corpus), state also travels as
messages. In the first weeks of that fleet, 36\% of its messages were timer
notes, and 63\% of the coordinators' requests restated a plan the recipient
already had. A message reaches only whoever reads it at the right moment, and
Appendix~\ref{app:context} shows what re-reading costs.

We therefore propose, as a design and not a result, to treat a room the way a
browser treats a page. The ledger stays what it is: the ordered, attributable,
append-only record of everything that happened. In the proposed design, what
agents \emph{read} is a document derived from the ledger. It is a tree of
typed nodes in the manner of the Document Object Model~\citep{whatwg-dom}. The
room is at the root, goals and sub-goals are below it, and tasks are the
leaves. A goal also has as children what belongs to it: the messages filed
under it, its acceptance criteria, the reports and evidence against them,
decisions and resources. The tree is a materialized view of the ledger. Every
state it shows can be rebuilt from the record, and the record
remains what an auditor reads afterwards.

An agent then reads by \emph{path}, that is, its leaf and the leaf's
ancestors, rather than the room's history. The context it needs is then
bounded by the depth of the tree, not by the age of the room. This is design
goal G6 applied to the agent rather than the relay. The design makes changes to
the tree part of the room's policy. Whoever opened a goal may place it under a
goal they also opened, and anything else needs a capability the room grants.

\emph{Callbacks} would make the document useful to an agent that is not
reading it. An agent registers on a node or a subtree, for example to be told
when a goal closes, when a criterion under it fails, or when a child is opened
below it. The history of the browser suggests which kind of callback to build.
Its first mutation events were synchronous, propagated through capture and
bubble phases, and could cancel each other. They were deprecated in favor of
an observer that is asynchronous, batched and filtered by
subtree~\citep{whatwg-dom}.

For agents the case for the asynchronous form is stronger. A listener that
could veto another participant's write would be orchestration hidden in a
callback. Our adversarial results suggest that control should sit in an
explicit rule on the write path, which can be checked afterwards, rather than
in whatever a listener decides. The design therefore specifies a callback as a
subscription with a filter. When a write commits, the relay wakes the
subscriber, which reads the change since the revision it last saw. The relay
runs no agent code. The harness decides what a callback costs. A callback can
be a deterministic handler with no model at all, for example one that re-runs
a check when the goal it waits on closes. It can also be a model turn whose
context is the change and the path, not the room. No callback is a message, so
the timer notes would not return as events.

Parts of this design exist. Messages are filed under goals. An execution
request is tracked as a wait on its goal, and a proposal can block a goal until
its voters have voted. A goal carries acceptance criteria, reports with
evidence, attestations by a named seat and a computed ready-to-close state. The
ledger is read incrementally from a cursor. A write wakes the readers it
concerns, as a hint rather than as the truth. The tree itself was built in
relay release 0.99.0 (Section~\ref{sec:since}). Goals have parents and
display-only heading levels, moves are conditional and recorded, and a read
returns a goal's path from the root or its children. That release refuses, by
name, reads of a whole subtree or of a goal's siblings, and goals cannot yet
wait on other goals.

Callbacks are designed but not built. The design specifies one feed per room
of typed transitions for every kind of record. The feed carries keys and
revisions but never content, and it notifies but never calls back. It is
planned in four steps: goal transitions, the room's resource map,
subscriptions (delivered as a digest in an agent's wait, with an optional
wake), and resources filed under goals. The plan is gated on requirements that
a reviewer set: cursors ordered by commit, subtree membership as of the time of
the event, typed revision domains, and authorization on metadata. The
work-status row of Section~\ref{sec:since} is the first built part of the same
idea, state that is read and is never a message.

Whether reading by path costs less at equal quality is an empirical question,
and it fits the rerun design of Appendix~\ref{app:rerun}. The view an agent
receives would become an arm of that experiment: its leaf, the leaf and its
ancestors, or the same with siblings. The reward would be the cost per accepted
delivery. A document also returns to the blackboard of
Section~\ref{sec:related} and to its central weakness. A shared state that
everyone can read makes disagreement invisible unless it is written down. In
this design a disagreement is still a message, filed under the node it is
about, where the tree shows it.

\subsection{A Contribution Index From Version History}
\label{sec:live}

The coalition attribution of Section~\ref{sec:structure} needs one coalition
per subset of agents. For four agents that means forty-five runs and a night
of running, and exponentially more beyond that ($2^n-1$ coalitions). A multi-agent run, however, already stores its own
counterfactuals in version control. For each agent $X$, we rebuild the program
from each module's newest commit \emph{not authored by} $X$. We run the same
mechanical acceptance check that the coalition study used, and report
$v(\text{full}) - v(\text{without }X)$ on the same scale. This takes one run
and seconds. It can be computed at any checkpoint while a run is still going.

Two design choices bound its error. First, we reconstruct each file instead of
reverting commits. A revert conflicts exactly where one agent's work underpins
another's, and any resolution of that conflict is arbitrary. Taking each module
at its last commit not by $X$ means that a module only $X$ ever wrote does not
exist in the counterfactual, which is the intended answer. This introduces an
approximation: the surviving modules may come from commits that never
coexisted. A report of the index must state this approximation. Second, the output states that it
prices \emph{removal}. Our coalition estimates showed that removing and adding
a coordinator can differ in sign. A live index therefore reads as ``what
losing them would cost'', never as ``what they earned''.

\subsection{A Rerun of the Context Measurement}
\label{app:rerun}

\paragraph{Design of the rerun.} The rotation of Appendix~\ref{app:context}
had five defects. No record said which arm was active. Arms were assigned by
hand. The quality guards were never computed. Sub-agent usage was mixed into
the turns that followed it. The rotation had no owner and no end. The proposed
rerun addresses each defect, and the design specifies it as a room feature
rather than a script.

An experiment is a record in the room, opened and closed by its owner like a
goal. It holds the arms as configuration overrides, together with the unit,
the reward, the guards and the stop rule. Every turn record carries the
experiment, the arm, the harness version and a hash of the effective
configuration, and sub-agent usage is reported separately. Metrics come only
from declared usage and goal outcomes, and agents are told which arm they are
on. A guard breach puts an arm on hold, and only the owner decides what
follows.

\emph{Unit of randomization.} The unit we propose is the delivery. A sub-agent
starts with a fresh context, so the sub-agent strategy can change at every
delivery without touching the session's prompt cache. The session's own model,
effort and compaction rule can change only at a cold boundary, a compaction or
a restart, because switching mid-session discards the cache. We therefore
propose that the first experiment randomize \emph{per delivery} over sub-agent
strategies: none (S0), sweeps and checks on a mid-tier model (S1, the current
default), sweeps on the smallest model with checks on the mid tier (S2), and
sub-agents on the largest model (S3).

The reward is the logarithm of the cost per accepted delivery. Acceptance is
the outcome of the goal the delivery served: its criteria passed, it was not
sent back, and verification found no defect. Quality arrives late, so
immediate signals such as looping and re-prompts stand in until the goal
closes. The experiment has two phases. In the \emph{explore} phase, a fixed,
balanced allocation is analyzed offline. In the \emph{exploit} phase, Thompson
sampling sends most deliveries to the arm currently believed best while
keeping a floor of exploration. It works over a sliding window, so that the
policy follows changing work. Exploration would never leave the arms the
operator allowed, and it would never run on a delivery whose goal has outside
effects. An arm whose acceptance falls below that of the best arm by a margin
would be excluded until a person re-admits it.

\begin{table}[t]
\centering
\caption{Planning parameters proposed for the rerun, derived from the fleet's
own records (275 deliveries with declared cost, days 52--59, after the
compaction change). The sample sizes treat deliveries as independent (see the
limits at the end of this appendix). The parameters will be re-derived once
the stamped records, which carry the experiment and the arm, exist. $\sigma$:
standard deviation of log cost per delivery within a role.}
\label{tab:bandit}
\small
\begin{tabular}{@{}>{\raggedright\arraybackslash}p{0.47\columnwidth}>{\raggedright\arraybackslash}p{0.45\columnwidth}@{}}
\toprule
measured & consequence \\
\midrule
cost per delivery heavy-tailed: pooled mean 5$\times$ the median; $\sigma=1.7$ & reward on log cost, role as a fixed effect \\ \addlinespace[2pt]
deliveries that used sub-agents cost 5--56$\times$ more (role medians) & the choice tracks the size of the work: randomize, never compare observationally \\ \addlinespace[2pt]
about 37 deliveries a day, fleet-wide & per-delivery arms, fleet-wide \\ \addlinespace[2pt]
a factor of 2 needs $\approx$93 deliveries per arm (two-sided 5\%, power 0.8) & \emph{explore}: S0, S1, S2 balanced for 14 days; detects a factor of 1.7 \\ \addlinespace[2pt]
sub-agents on the largest model were 35\% of cost (measured, days 59--66); a written token costs about 1.5 times the mid tier's & S3 not explored for cost; a quality reference only if S1's acceptance falls \\ \addlinespace[2pt]
with a 15\% floor shared by two arms, a factor of 2 needs a 19-day window & \emph{exploit}: floor 15\%, window 21 days; prior on an arm's effect $N(0, 0.7^2)$ on log cost \\ \addlinespace[2pt]
acceptance is a rate & Beta(1,1) per arm; hold if $P(q_{\mathrm{arm}} < q_{\mathrm{best}} - 0.1) > 0.9$ \\
\bottomrule
\end{tabular}
\end{table}

\emph{Choice of parameters.} We derived the parameters from the measurement
(Table~\ref{tab:bandit}), and they differ from the defaults first proposed for
the fleet. Within a role, the cost of a delivery spreads over two orders of
magnitude, so the reward is taken on a log scale. Deliveries that used
sub-agents cost 5 to 56 times more than those that did not, at the role
medians. A plausible reason is that an agent delegates when the work is large.
Comparing strategies on whatever the agents chose would measure the
work, not the strategy.

The following figures are for planning and treat deliveries as independent.
With about 37 deliveries a day, detecting a twofold difference would need
about 93 deliveries per arm. The first plan put 70\% of deliveries on the
default and 10\% on each of three other arms, and it would take 25 days to
collect them. A balanced phase over three arms would collect them in eight
days, and in fourteen days it would resolve a factor of 1.7. The largest-model
arm is not explored for cost, which is already known to be higher. It enters
only as a quality reference if the acceptance of the default falls.

In the exploit phase, the floor and the window trade against each other. If
two non-best arms share 15\% of deliveries, a twofold change becomes visible
only in a window of 19 days. The planned 14 days would see only changes larger
than a factor of 2.2. We therefore propose a window of 21 days. A larger gain
in power may lie outside the policy. A covariate that predicted the size of a
delivery before it ran would shrink $\sigma$, and halving $\sigma$ quarters
every sample size above. The one predictor the fleet ran, a classifier's token
estimate, was off by a factor of 20 to 80.

\paragraph{Admitting new strategies.} Four strategies are a starting set, not
the whole space. Because arms are data, the design lets a new lever enter as a
\emph{challenger}. A challenger is a share of the exploration floor that tries
one candidate at a time and admits it as a full arm if its posterior shows a
likely gain. Candidates are not cheap to test. Against an established best
arm, a challenger would need about 47 deliveries to show a twofold saving, and
a 5\% share takes 25 days to collect them. We therefore propose to admit a
candidate only if its \emph{upper bound} is larger than the smallest effect the
experiment can see, about 40\%. The upper bound is the cost that the candidate
could remove if it worked perfectly.

The records bound most levers, and the measured sub-agent usage
(Appendix~\ref{app:context}) changed what they say. The proxy described there
put 88\% of the cost in turns with sub-agent work and 77\% in those on the
largest model. By that proxy, two levers would clear the bar: the sub-agent's
model and the number of fresh sub-agent contexts. Measured, sub-agents on the
largest model carried 35\% of the cost. In the same records a mid-tier token
costs 55 to 70\% of a largest-model one. Moving every such sub-agent to the mid
tier could therefore remove at most 10 to 16\% of fleet cost. Sub-agents' cache
writes were 19\% of the cost, which bounds what reusing one sub-agent across
related checks could save. At the level of the fleet, neither lever clears the
bar any longer. Both may clear it inside the deliveries that delegate, which is
where a per-delivery arm acts. The stamped records can now answer that, and we
will re-derive the bounds per delivery before the rerun starts.

Other levers do not clear the bar. Reasoning effort does not, because
reasoning is about a third of the largest model's output tokens and output is 9
to 14\% of the cost. Neither does the cache lifetime, whose net effect is about
zero, nor further compaction, since the context's 90th percentile already sits
at the threshold. Re-prompted turns carried 8\% of the cost after the harness
began to hold deliveries and 27\% before. That is an upper bound on what better
re-prompting could still remove. Two bounds are still unknown. One is the fixed
prompt that every fresh sub-agent starts with, which the per-run records the
harness has written since day 65 will give. The other is how often a session
starts sub-agents in parallel on overlapping reads.

In this design, finding potential improvements is mostly a matter of
decomposing cost first and spending experimental deliveries only where the
bound is large. The proxy above is an instance: a bound read from it admitted
two levers that the measurement does not.

\paragraph{Limits of this design.} The sample sizes above are planning
figures, not a validated trial design. They treat deliveries as independent,
although deliveries of one agent, one goal and one day are correlated, and a
clustered analysis will need more of them. Goal outcomes arrive late and only
for goals that close, so a reward that waits for them is delayed and censored.
An agent switched between arms carries its context from one arm into the next.
A washout period or an assignment per session would have to absorb this
carryover. The rerun's registration will fix the unit, the clustering, the
treatment of delayed outcomes and the carryover rule before it starts.

\end{document}